\documentclass[prb,preprint,showpacs, floatfix,superscriptaddress]{revtex4}

\pdfoutput=1
\usepackage{graphicx}
\usepackage{dcolumn}
\usepackage{bm,amsmath,verbatim}

\def\sgn{{\text{sgn\,}}}
\def\be{\begin{equation}}
\def\ee{\end{equation}}
\def\bea{\begin{eqnarray}}
\def\eea{\end{eqnarray}}
\def\bse{\begin{subequations}}
\def\ese{\end{subequations}}

\def\sgn{{\text{sgn\,}}}
\renewcommand{\v}[1]{{\bf #1}}

\def\be{\begin{eqnarray}}
\def\ee{\end{eqnarray}}

\newcommand{\ua}{\uparrow}
\newcommand{\da}{\downarrow}

\begin{document}

\title{Non-Abelian quantum order in spin-orbit-coupled semiconductors: The search for topological Majorana particles in solid state systems}

\author{Jay D. Sau}
\affiliation{Condensed Matter Theory Center and Joint Quantum Institute, Department of Physics, University of
Maryland, College Park, MD 20742}
\author{Sumanta Tewari}
\affiliation{Department of Physics and Astronomy, Clemson University, Clemson, SC 29634}
\affiliation{Condensed Matter Theory Center and Joint Quantum Institute, Department of Physics, University of
Maryland, College Park, MD 20742}
\author{Roman Lutchyn}
\affiliation{Condensed Matter Theory Center and Joint Quantum Institute, Department of Physics, University of
Maryland, College Park, MD 20742}
\author{Tudor Stanescu}
\affiliation{Department of Physics, West Virginia University, Morgantown, WV 26506}
\affiliation{Condensed Matter Theory Center and Joint Quantum Institute, Department of Physics, University of
Maryland, College Park, MD 20742}
\author{S. Das~Sarma}
\affiliation{Condensed Matter Theory Center and Joint Quantum Institute, Department of Physics, University of
Maryland, College Park, MD 20742}
\date{\today}

\begin{abstract}
We show that an ordinary semiconducting thin film with spin-orbit coupling
can, under appropriate circumstances, be in a quantum topologically ordered state supporting exotic Majorana excitations which follow non-Abelian statistics. The key to the quantum topological order is
the coexistence of spin-orbit coupling with proximity-induced $s$-wave superconductivity and an externally-induced Zeeman coupling of the spins. For the Zeeman coupling below a critical value, the system is a non-topological (proximity-induced) $s$-wave superconductor.
However, for a range of Zeeman coupling above the critical value, the lowest energy excited state inside a vortex is a zero-energy Majorana fermion state. The system, thus, has entered into a non-Abelian $s$-wave superconducting state via a topological quantum phase transition (TQPT) tuned by the Zeeman coupling.
In the topological phase, since the time reversal symmetry is explicitly broken by the Zeeman term in the Hamiltonian, the edge of the film constitutes a \emph{chiral} Majorana wire. Just like the $s$-wave superconductivity, the Zeeman coupling can also be
 proximity-induced in the film by an adjacent magnetic insulator. We show this by an explicit model tight-binding calculation for both types of proximity effects in the heterostructure geometry. Here we show that the same TQPT can be accessed by varying the interface transparency between the film and the superconductor. For the transparency below (above) a critical value, the system is a topological (regular) $s$-wave superconductor.
 In the one-dimensional version of the same structure and for the Zeeman coupling above the critical value, there are localized Majorana zero-energy modes at the two ends of a semiconducting quantum nanowire. In this case, the Zeeman coupling can be induced more easily by an external magnetic field parallel to the wire, obviating the need for a magnetic insulator.  We show that, despite the fact that the superconducting pair potential in the nanowire is explicitly $s$-wave, tunneling of electrons to the ends of the wire reveals a pronounced zero-bias peak. Such a peak is absent when the Zeeman coupling is below its critical value, \emph{i.e.,} the nanowire is in the non-topological $s$-wave superconducting state.
 We argue that the observation of this zero-bias tunneling peak in the semiconductor nanowire is possibly the simplest and clearest experiment proposed so far to unambiguously detect a Majorana fermion mode in a condensed matter system.
\end{abstract}

\pacs{03.67.Lx, 71.10.Pm, 74.45.+c}
\maketitle

\section{Introduction}
Particle statistics of a collection of indistinguishable particles is a genuinely quantum mechanical concept without any classical analog.
In spatial dimensions three and above, pairwise interchange of particle coordinates in a many-body system is equivalent to a simple permutation
of the coordinates. Consequently, each interchange has the effect of either a change of sign (fermion) or no change at all (boson) on the
many-body quantum wave function. In $(2+1)$ dimensions, however, exchanges and permutations are not necessarily equivalent. ~\cite{Leinaas, Wilczek, Wilczek2} In this case, under simple interchange of the particle coordinates, the corresponding space-time trajectories can form non-trivial braids in the $(2+1)$-dimensional space-time. ~\cite{nayak_RevModPhys'08} Consequently,
  in ($2+1$) dimensions,
  particles can have quantum statistics strikingly different from the
statistics of bosons and fermions.

A straightforward extension of the statistics of bosons and fermions is the Abelian anyonic statistics,
in which the many body wave-function, under pairwise exchange of the particle coordinates,
picks up a phase $\theta$, which can take any value between $0$ (bosons) and $\pi$ (fermions).
Since a phase factor is only a one-dimensional representation of the braid group in $2D$, the statistics is still Abelian.
On the other hand, if the many body ground state wave function happens to be a linear combination of states from a degenerate subspace,
a pairwise exchange of the particles can unitarily \emph{rotate} the wave function in the ground state subspace.
In this case, the effect of exchanging the particle positions is an operation on the wave function vector by a unitary matrix representing
this rotation.
Consequently, the statistics is non-Abelian,~\cite{Kitaev, nayak_RevModPhys'08} and the corresponding system is a non-Abelian quantum system.
It has been proposed that such systems, if the ground state subspace is concurrently protected by an energy gap,
can be used as a fault-tolerant platform for topological quantum computation (TQC).~~\cite{Kitaev, nayak_RevModPhys'08, Freedman}

 One important class of non-Abelian quantum systems, sometimes referred to as the Ising topological class, \cite{nayak_RevModPhys'08}
   is characterized by quasiparticle excitations called Majorana fermions, which involve no energy cost (when the mutual separation among the excitations is large). The second quantized operators, $\gamma_i$, corresponding to these zero energy excitations are self-hermitian, $\gamma_i^{\dagger}=\gamma_i$, which is in striking contrast to ordinary fermionic (or bosonic) operators for which $c_i \neq c_i^{\dagger}$. However, since $\gamma_i$ and $\gamma_j$ anticommute when $i \neq j$, they retain some properties of ordinary fermion operators as well. The Majorana fermions, which are actually more like half-fermions, were envisioned \cite{Majorana} by E. Majorana in 1935 as fundamental constituents of nature (\emph{e.g.} neutrinos are thought to be Majorana, rather than Dirac, fermions).  Majorana modes are intriguing \cite{Wilczek-3} because each Majorana particle is its own anti-particle unlike Dirac fermions where electrons and positrons (or holes) are distinct. Although the
   emergence of Majorana excitations, which are
   effectively fractionalized objects (anyons)
   obeying non-Abelian anyonic statistics,
   in solid state systems is by itself an extraordinary
   phenomenon, a great deal of
   attention has also been focused on them because of the possibility of carrying out fault
   tolerant TQC in two
   dimensional systems using these Majorana particles.
    TQC, in contrast
   to ordinary quantum computation, would not require
   any quantum error correction since the Majorana
   excitations are immune to local noise by virtue of
   their non-local topological nature. \cite{Stern, nayak_RevModPhys'08}  The direct experimental
   observation of Majorana modes in solid state systems
   would therefore be a remarkable breakthrough both from the
   perspective of fundamental physics of fractional
   statistics in nature and the technological
   perspective of building a working quantum computer.  It is
   therefore not surprising that there has been
   recent resurgence of immense interest for the
   experimental realization (and detection) of Majorana
   fermions in solid state systems.
Recently, some exotic ordered states in condensed matter systems, such as the Pfaffian states in
fractional quantum Hall (FQH) systems,~\cite{Moore, Nayak-Wilczek, Read, dassarma_prl'05, Halperin, Bonderson, Sling, Rosenow}
 $p$-wave superconductors/superfluids,~\cite{Read, volovik1, Ivanov, Stern-1, DasSarma_PRB'06, Sumanta, tewari_prl'2007, dunghai, Yip, Rad1, Rad2, volovik2, volovik3, nishida, Beenakker-defect} theoretical models that can be potentially simulated in cold atom optical lattice systems, \cite{Kitaev-1, Kivelson, Trebst, Duan, Chuanwei1}
as well as the surface state of a topological insulator (TI) or related systems~\cite{fu_prl'08, Jackiw1, fu_prl'09, akhmerov_prl'09, robustness, SSLdS, Nilsson, Law, Nilsson-1, Tanaka, fu_paper,  Palee, Qi}
have been discussed
as systems which can support Majorana fermions as the lowest energy excitations. In the context of optical lattice systems, it has also been proposed that a 2D $p_x+ip_y$ superfluid can be realized using only $s$-wave Feshbach resonance modified by the topological Berry phases arising from artificially generated spin-orbit coupling.~\cite{Zhang-pwave}

It has been shown recently ~\cite{Sau} that even a regular semiconducting film with a sizable Rashba-type spin-orbit coupling,
such as InGaAs thin films, can host, under suitable conditions, Majorana fermions as low energy excitations. Since the basic effects behind the emergence of the Majorana fermion excitations -- spin-orbit coupling, $s$-wave superconductivity, and Zeeman splitting -- are physically well-understood and experimentally known to occur in many solid state materials, the proposed semiconductor heterostructure ~\cite{Sau} is possibly one of the simplest condensed matter systems supporting Majorana quasiparticles and non-Abelian quantum order. By an analysis of the real-space Bogoliubov-de Gennes (BdG) equations for a vortex in the semiconductor thin film, in which $s$-wave superconductivity and a Zeeman splitting are proximity-induced (Fig. (1a)), it has been shown that the lowest energy quasiparticle excitation in the vortex core is a zero-energy Majorana fermion mode. This real space analysis has also been supported by a momentum space analysis in the form of an index theorem \cite{index-sm} analogous to such a treatment in the context of one-dimensional Dirac theory.~\cite{Jackiw2, Sch}. Here a comment about the various means to induce a Zeeman splitting in the semiconductor thin film is in order. Note that when the spin-orbit coupling is of the Rashba type, we require a Zeeman splitting which is 
 perpendicular to the plane of the film (Zeeman splitting parallel to the film does not produce a gap in the one-electron band-structure, a firm requirement of our non-Abelian state). \cite{Sau, index-sm} Inducing such a splitting by applying a strong perpendicular magnetic field is not convenient, because the magnetic field will give rise to unwanted order parameter defects such as vortices. It is for this reason that we propose to induce the Zeeman splitting by the exchange proximity effect of an adjacent magnetic insulator (we ignore the small coupling of the spins in the semiconductor with the actual magnetic field of the magnetic insulator).  More recently, it has been shown that, when the spin-orbit coupling also has a component which is of the Dresselhaus type, the appropriate Zeeman splitting can also be induced by applying an in-plane magnetic field.~\cite{alicea}  The Majorana mode is separated by a finite energy gap (so-called mini-gap) from the other conventional fermionic excited states in the vortex core. Thus, for a collection of well-separated vortices, the resulting degenerate ground state subspace is protected from the environment by the mini-gap. This enables the potential use of the semiconductor heterostructure in Fig. (1a) in TQC.

One of the main goals of the present paper is to provide the important mathematical details relevant to our solutions of the BdG equations in the semiconductor heterostructure. These mathematical details, which were left out in Ref.~[\onlinecite{Sau}], are given in Sec.~[II] and Sec.~[III] below. It is important to note that, unlike the case of the surface of a $3D$ strong topological insulator adjacent to an $s$-wave superconductor,~\cite{fu_prl'08} the BdG equations in the spin-orbit coupled semiconductor are not exactly solvable. We therefore only show that, in a specified region of the parameter space, a single non-degenerate solution of the BdG equations, which is spatially localized around the vortex core, is allowed. We also show that the second quantized operator corresponding to such a solution is indeed a Majorana fermion operator. (In a subsequent section (Sec.~[VI]) we confirm the existence of such zero-energy Majorana fermion states localized at the vortex cores by a full numerical solution of the BdG equations set up on a sphere.) In the next few sections (Secs.~[IV-IX]) we provide a comprehensive discussion of the interesting physics of non-Abelian topological order arising via the complex interplay of spin-orbit coupling, Zeeman splitting, and $s$-wave superconductivity externally induced in a host system. We also deduce the parameter space needed for the establishment of the non-Abelian order, as well as the associated TQPT accessed by tuning the system in or out of this parameter space.
 In Sec.~[X] we study the superconducting and magnetic proximity effects in the host semiconductor thin film by a microscopic model tight-binding calculation. In the last part of the paper (Sec.~[XI]), we consider a one-dimensional version of our proposed
structure-- a semiconducting nanowire with proximity induced $s$-wave
superconductivity and a Zeeman splitting. We emphasize that the Zeeman splitting can now be induced by a
 magnetic field parallel to the length of the wire, because such a field does produce a gap in the one-electron band structure without producing unwanted excitations in the adjacent superconductor. This obviates the need for a nearby magnetic insulator. For the Zeeman
splitting above a critical value, the wire is in a non-Abelian topological
 phase with zero-energy Majorana excitations at the ends.
We propose a scanning tunneling experiment from the ends of the
 semiconducting nanowire as possibly the most realistic experiment proposed so far to detect a Majorana fermion state in a condensed matter system.

 We use the terminology Majorana particle or Majorana fermion or Majorana state or
Majorana excitation or Majorana mode interchangeably in this article, all of them
meaning precisely the same entity, namely, the non-degenerate zero-energy eigenstate
(\emph{i.e.} a solution of the BdG equations) at the vortex core of a
spin-less chiral $p$-wave or other such topological superconductor.  We emphasize that this object obeys the
non-Abelian braiding statistics rather than ordinary fermionic statistics, and the
Majorana particle is its own anti-particle in contrast to the ordinary Dirac
fermions where electrons and holes (positrons) are distinct particle-hole conjugates
of each other. A part of the results presented in this paper -- approximate solutions of the BdG equations in the semiconductor heterostructure -- has been published elsewhere. ~\cite{Sau} In Secs.~[II,III] we provide all the mathematical details relevant to the solutions of the BdG equation which were left out in Ref.~[\onlinecite{Sau}]. Most of the results contained in the subsequent sections are new. Some additional mathematical details related to the solution of the BdG equations are relegated to the appendix.

\begin{figure}[tbp]
\begin{center}
\includegraphics[width=0.47\textwidth]{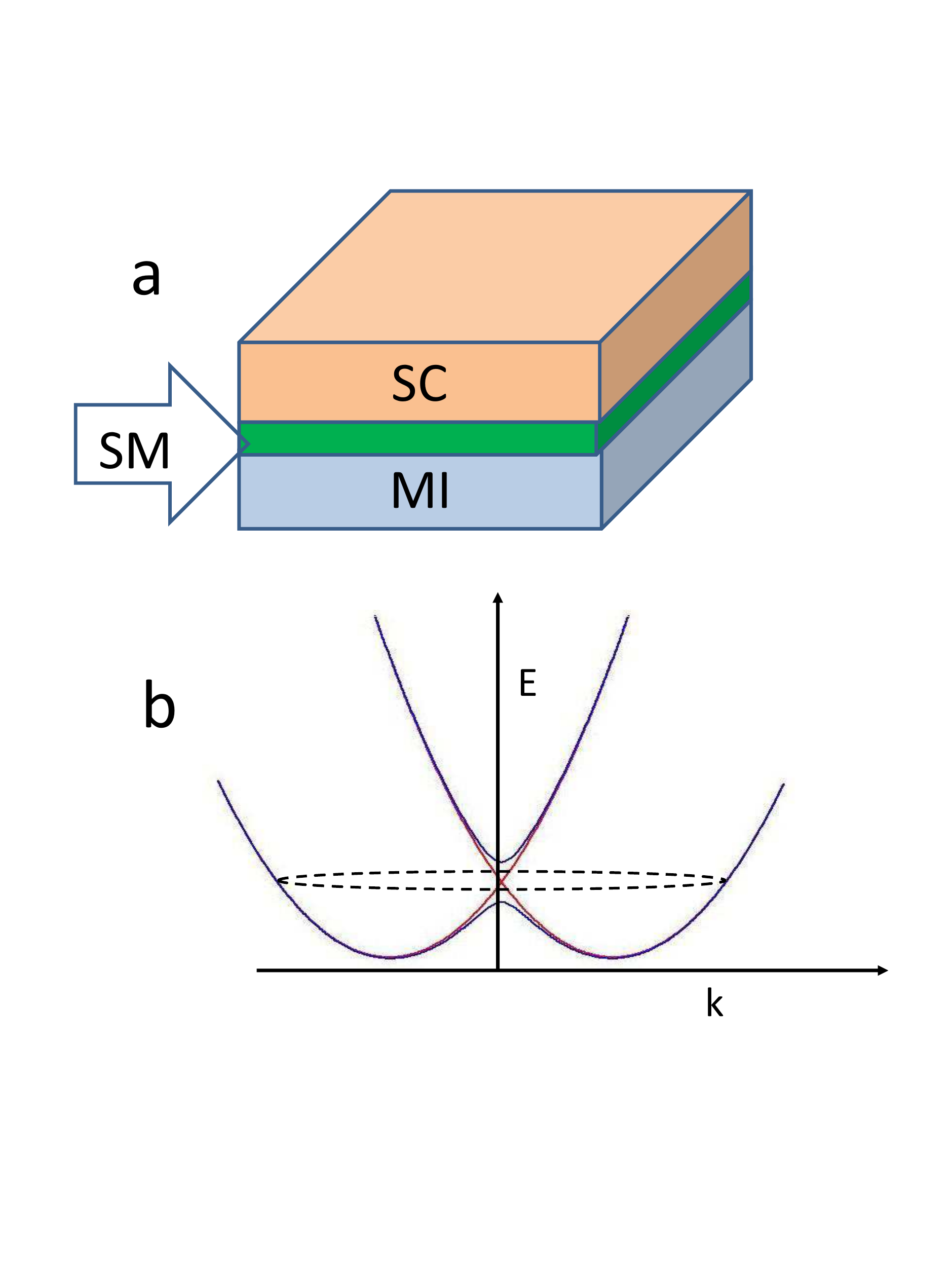}
\end{center}
\caption{(a): The proposed heterostructure of a semiconductor (SM) sandwiched between an $s$-wave superconductor (SC) and a magnetic insulator (MI).
In this geometry, the semiconducting film can support non-Abelian topological order. (b): Single-particle band-structure in the semiconducting film with and without the Zeeman splitting induced by the MI. Without the Zeeman splitting, the two spin-orbit shifted bands touch at $k_x=k_y=|k|=0$ (red lines). Then, for
 any value of the chemical potential, the system has two Fermi surfaces.
 With a finite Zeeman splitting, the bands have an energy gap near $k_x=k_y=|k|=0$ (blue lines).
If the chemical potential lies in the gap, the system just has one Fermi surface (indicated by the dotted circle).}
\label{Fig1}
\end{figure}

\section{Hamiltonian.}
\label{sec:Hamiltonian}

The single-particle effective Hamiltonian $H_0$ for the conduction band of a spin-orbit coupled semiconductor in contact with a magnetic insulator is given by (we set $\hbar=1$ henceforth)
\begin{align}\label{eq:H0}
H_0\!=\! \frac{p^2}{2 m^*}\!-\!\mu\!+\!V_z \sigma_z\!+\!\alpha (\vec \sigma \!\times \! \vec p)\!\cdot\! \hat{z}.
\end{align}
Here, $m^*$, $V_z$ and $\mu$ are the conduction-band effective mass of an electron, effective Zeeman coupling induced
by proximity to a magnetic insulator (we neglect the direct coupling of the electrons with the magnetic field from the magnetic insulator), and chemical potential, respectively.  The coefficient
 $\alpha$ describes the strength of the Rashba spin-orbit coupling and $\sigma_{\alpha}$ are the Pauli matrices.

The proximity-induced superconductivity in the semiconductor can be described by the Hamiltonian,
\begin{equation}
\hat{H}_{p}=\int d\mathbf{r}\,\{\Delta(\mathbf{r})\hat{c}^{\dagger}_{\uparrow}(\mathbf{r})\hat{c}^{\dagger}_{\downarrow}(\mathbf{r})+\rm H.c\},
\end{equation}
where $\hat{c}_{\sigma}^{\dagger}(\mathbf{r})$ are the creation operators for electrons with spin $\sigma$ and $\Delta(\mathbf{r})$ is the proximity-induced gap. The pairing term $\hat{H}_{p}$ and the non-interacting
part $H_0$ can be combined to obtain the BCS mean-field Hamiltonian $H_{BCS}=H_0+H_p$.
The excitation spectrum of this Hamiltonian is defined in terms
of the Bogoliubov quasiparticle operators
\begin{equation}
\hat{\gamma}^\dagger=\int d\mathbf{r}\,\sum_{\sigma}u_{\sigma}(\mathbf{r})\hat{c}_{\sigma}^{\dagger}(\mathbf{r})+v_{\sigma}(\mathbf{r})\hat{c}_{\sigma}(\mathbf{r})
\label{eq:BCSqp}
\end{equation}
 which satisfy
\begin{equation}\label{eq:qpeqn}
[\hat{H}_{BCS},\hat{\gamma}^\dagger]=E\hat{\gamma}^\dagger.\end{equation} Such a quasiparticle
operator $\hat{\gamma}$ can be used to construct excited states $\hat{\gamma}^{\dagger}|\Psi_0\rangle$ with energy $E+E_0$ from the ground state $|\Psi_0\rangle$ with energy $E_0$. The ground state $|\Psi_0\rangle$ is defined as the
lowest energy state of the BCS Hamiltonian satisfying $\hat{\gamma}|\Psi_0\rangle=0$.
The equation for the quasiparticle operator, Eq.~\ref{eq:qpeqn}, can be re-written
 as the BdG equations in the Nambu basis,
\begin{equation}
\left(\begin{array}{cc}H_0&\Delta(\mathbf{r})\\\Delta^*(\mathbf{r})&-\sigma_y H_0^* \sigma_y\end{array}\right)\Psi(\mathbf{r})=E\Psi(\mathbf{r})
\label{eq:H5}.
\end{equation}
 Here, $\Psi(\mathbf{r})$ is the wave function in the Nambu spinor basis, $\Psi(\mathbf{r})=(u_{\uparrow}(\mathbf{r}),u_{\downarrow}(\mathbf{r}),v_{\downarrow}(\mathbf{r}),-v_{\uparrow}(\mathbf{r}))^T$.
Introducing the Pauli matrices $\tau_{\alpha}$ in the Nambu space the
  Hamiltonian on the left hand side in Eq.~(\ref{eq:H5}) can be written as
\begin{equation}
H_{BdG}=[\frac{p^2}{2 m^*}\!-\!\mu\!+\!V_z \sigma_z\!+\!\alpha (\vec \sigma \!\times \! \vec p)\!\cdot\! \hat{z}]\tau_z+[\Delta(\bm r)\tau_++h.c]
\end{equation}
where $\tau_+=\tau_-^\dagger=\frac{\tau_x+\imath\tau_y}{2}$.

\section{BdG equations for a vortex.}
\label{sec:BdG}
The single-particle Hamiltonian $H_0$ can be written in polar coordinates as
\begin{align}
H_0&=\eta p^2-\mu+V_z \sigma_z +\alpha (\sigma\times p)\cdot \hat{z}\nonumber\\
&=-\eta\nabla^2-\mu + V_z\sigma_z+\imath\frac{\alpha}{2} (\sigma_+p_--\sigma_-p_+)
\end{align}
where $\eta=\frac{\hbar^2}{2 m^*}$, $\sigma_+=\sigma_-^\dagger=\sigma_x+\imath \sigma_y$
and $p_+=p_x+\imath p_y=e^{\imath\theta}(-\imath\partial_r+\frac{1}{r}\partial_\theta)$ and $p_-=p_x-\imath p_y=e^{-\imath\theta}(-\imath\partial_r-\frac{1}{r}\partial_\theta)$.
The full BdG Hamiltonian for an $n$-fold vortex can be written conveniently in the Nambu space as
\begin{equation}
H_{BdG}=(-\eta\nabla^2-\mu)\tau_z + V_z\sigma_z+\imath\frac{\alpha}{2} (\sigma_+p_--\sigma_-p_+)\tau_z+\Delta(r)[\cos{(n\theta)}\tau_x+\sin{(n\theta)}\tau_y].
\end{equation}

In order to diagonalize the above Hamiltonian it is convenient to note
that  the BdG Hamiltonian has a combined
spin-orbit-pseudospin rotational symmetry. This symmetry can be expressed
compactly by noting that $H_{BdG}$ commutes with the operator
\begin{equation}J_{z}=L_z+\frac{1}{2}(\sigma_z-n\tau_z).
\label{eq:Jz}
\end{equation}
Therefore, the eigenspinors of the BdG
Hamiltonian can be taken to be $J_z$ eigenstates with eigenvalue
$J_z=m_J$ of the form
\begin{equation}
\Psi_{m_J}(r,\theta)=e^{\imath L_z \theta}\Psi_{m_J}(r)=e^{\imath (m_J-\sigma_z/2+n\tau_z/2) \theta}\Psi_{m_J}(r)=\left (\begin{array}{c}u_{\uparrow,m_J}(r)e^{\imath (m_J+\frac{n-1}{2}) \theta}\\u_{\downarrow,m_J}(r)e^{\imath (m_J+\frac{n+1}{2}) \theta}\\v_{\downarrow,m_J}(r)e^{\imath (m_J-\frac{n+1}{2}) \theta}\\-v_{\uparrow,m_J}(r)e^{\imath (m_J-\frac{n-1}{2}) \theta} \end{array}\right)\label{eq:theta}.
\end{equation}

The above equation can be used
to eliminate the angular degree of freedom $\theta$ from the
 BdG equations as follows:
\begin{align}
&H_{BdG}\Psi_{m_J}(r,\theta)=E_{m_J}\Psi_{m_J}(r,\theta)\\
&\tilde{H}_{BdG,m_J}\Psi_{m_J}(r)=E_{m_J}\Psi_{m_J}(r).
\end{align}
Here
$\tilde{H}_{BdG,m_J}=e^{-\imath (m_J-\sigma_z/2+n\tau_z/2)\theta}H_{BdG}e^{\imath (m_J-\sigma_z/2+n\tau_z/2)\theta}$ is $\theta$ independent.
More specifically
\begin{align}
&\tilde{H}_{BdG,m_J}=-\{\eta(\partial_r^2+\frac{1}{r}\partial_r+\frac{(2 m_J-\sigma_z+n\tau_z)^2}{4 r^2})+\mu\}\tau_z + V_z\sigma_z\nonumber\\
&-\frac{\imath\alpha}{2} \{\sigma_+-\sigma_-\}\tau_z\partial_r-\imath\frac{\alpha}{2 r} \{\sigma_+\frac{2 m_J+n\tau_z+1}{2}+\sigma_-\frac{2 m_J+n\tau_z-1}{2}\}\tau_z+\Delta(r)\tau_x.
\end{align}

Under the action of the particle-hole transformation operator,
 $\Xi=\sigma_y\tau_y K$, the $m_J$ spinor eigenstate with energy $E$
 transforms into a $-m_J$ eigenstate with energy $-E$ because
 $$\Xi e^{\imath (m_J-\sigma_z/2+n\tau_z/2) \theta}\Psi_{m_J}(r)= e^{\imath (-m_J-\sigma_z/2+n\tau_z/2) \theta}\Xi\Psi_{m_J}(r).$$
 Therefore, a necessary condition for a non-degenerate, $E=0$, Majorana
 state solution is that $m_J=0$. From here onwards we will write $\tilde{H}_{BdG, m_J =0}=\tilde{H}_{BdG}$ and $\Psi_{m_J=0}(r)=\Psi(r)$. Single-valuedness of the spinor wave-functions in Eq.~(\ref{eq:theta})
 requires that
 $(n-1)/2$ must be an integer. Therefore, only
 odd vortices can have non-degenerate Majorana eigenstates. From here
 onwards, for the simplicity of discussion, we will
consider only zero energy solutions at the cores of single-flux-quantum vortices $(n=1)$ .

The BdG matrix  $\tilde{H}_{BdG}$ may be reduced to a real Hamiltonian by
applying the $\sigma_z$ rotation $U=e^{\imath\sigma_z\pi/4}$
as $\tilde{H}_{BdG}\rightarrow U^\dagger \tilde{H}_{BdG}U$.
 The solutions of the resulting  $E=0$
BdG equation $\tilde{H}_{BdG}\Psi(r)=0$ must come in complex conjugate
pairs $\Psi(r)$ and $\Psi^*(r)$. Therefore the solutions $\Psi(r)$ can
be required to be real without loss of generality.
 For such real solutions, it follows from the particle-hole symmetry of
 the BdG equations that $\sigma_y\tau_y \Psi(r)$ is also a solution.
 Thus, any non-degenerate  $E=0$ solution must be real and satisfy
 the property $\sigma_y\tau_y \Psi(r)=\lambda \Psi(r)$.
Moreover, because $(\sigma_y\tau_y)^2=1$, the possible values of
 $\lambda$ are $\lambda=\pm 1$.

Using the relation $\tau_x=\imath\lambda\sigma_y\tau_z$, which follows from $\sigma_y\tau_y=\lambda$,
 the BdG Hamiltonian for a given value of $\lambda$ is of the form
 \begin{align}
&\tilde{H}_{BdG}=-\{\eta(\partial_r^2+\frac{1}{r}\partial_r+\frac{(-\sigma_z+\tau_z)^2}{4 r^2})+\mu\}\tau_z + V_z\sigma_z\nonumber\\
&-\frac{\alpha}{2} \{\sigma_++\sigma_-\}\tau_z\partial_r-\frac{\alpha}{2 r} \{\sigma_+\frac{\tau_z+1}{2}+\sigma_-\frac{\tau_z-1}{2}\}\tau_z+\imath \lambda\sigma_y\tau_z\Delta(r)\label{eq:decBdg}.
\end{align}
The Hamiltonian in this limit does not couple the $\tau_z=\pm 1$ sectors
(electron and hole). This allows one to write the BdG differential equation
in terms of only the electron-sector $(\tau_z=+1)$ of the spinor
$\Psi_0(r)=(u_{\uparrow}(r),u_{\downarrow}(r))^T$.
 The corresponding reduced BdG equations for a single vortex $(n=1)$
take the form of a $2\times 2$ matrix differential equation:
\begin{align}\label{eq:zeroenergy}
\!&\!\left(\!\begin{array}{cc}\!\!-\!\eta (\partial_r^2\!+\!\frac{1}{r}\partial_r)\!+\!V_z\!-\!\mu\!&\! \lambda\Delta(r)\!+\!\alpha (\partial_r\!+\!\frac{1}{r} )\\\\ -\lambda \Delta(r)\!-\!\alpha \partial_r  \!&\! -\!\eta (\partial_r^2\!+\!\frac{1}{r}\partial_r\!-\!\frac{1}{r^2}\!)\!-\!V_z\!-\!\mu\! \end{array}\!\right)\!\!\Psi_0(r)\!=\!0.
\end{align}

The hole part of the spinor is not independent and is constrained by
the value of $\lambda$ such that
$v_{\uparrow}(r)=\lambda u_{\uparrow}(r)$ and $v_{\downarrow}(r)=\lambda u_{\downarrow}(r)$ and the Majorana spinor has the form $\Psi(\bm r)=\left(\Psi_0(\bm r),\imath\sigma_y\Psi_0(\bm r)^*\right)^T$.

We now approximate the radial dependence of $\Delta(r)$ by $\Delta(r)=0$ for $r<R$ and $\Delta(r)=\Delta$ for $r\geq R$ where $R$ is of the order of the radius of a vortex core. In view of the  topological stability of the putative Majorana zero-energy solution to local changes in the Hamiltonian, ~\cite{Read} such an approximation can be made without any loss of generality.

\subsection{Solution inside the  vortex core.}
Inside the vortex core  $(r<R)$, which is the non-superconducting region $(\Delta(r)=0)$, it is possible to construct explicit analytic solutions $\Psi(r,z)$ to these equations in terms of the Bessel functions
$J_0(z)$ and $J_1(z)$ as,
\begin{equation}
\Psi(r,z)=  \left(\begin{array}{c}u_{\uparrow}J_0(z r)\\u_{\downarrow}J_1(z r)\end{array}\right).
\label{eq:Bessel}
\end{equation}
 By substituting Eq.~(\ref{eq:Bessel}) into Eq.~(\ref{eq:zeroenergy})  we find that $(u_\uparrow,u_\downarrow)$ and $z$ satisfy
\begin{align}
&\left(\begin{array}{cc}\eta  (-\partial_r^2-\frac{1}{r}\partial_r)+V_z-\mu & \alpha  (\partial_r+\frac{1}{r} )\\ -\alpha  (\partial_r)  & \eta (-\partial_r^2-\frac{1}{r}\partial_r+\frac{1}{r^2})-V_z-\mu \end{array}\right)\left(\begin{array}{c} u_{\uparrow} J_0 (z r )\\ u_{\downarrow} J_1(z r)\end{array}\right)\nonumber\\
&=\left(\begin{array}{c}(-\eta z^2+V_z-\mu) u_{\uparrow} J_0 (z r )+  z \alpha u_{\downarrow} J_0(z r) \\ z\alpha u_{\uparrow} J_1(z r)+(-\eta z^2 -V_z-\mu)u_{\downarrow} J_1(z r)\end{array}\right)=0,
\end{align}
which implies
\begin{equation}
\left(\begin{array}{cc}-\eta z^2+V_z-\mu & z\alpha  \\ \alpha z  & -\eta z^2 -V_z-\mu \end{array}\right)\left(\begin{array}{c} u_{\uparrow} \\ u_{\downarrow}\end{array}\right)=0.
\label{eq:matrix}
\end{equation}
Existence of solutions in terms of $(u_\uparrow,u_\downarrow)^T$ requires
that $z$ satisfies the characteristic equation of the matrix in
 Eq.~(\ref{eq:matrix}) ,
which is given by
\begin{equation}
(\eta z^2-\mu)^2-V_z^2-z^2\alpha^2=0.
\end{equation}
This is a quadratic equation in $z^2$ with real roots.
Therefore, the solutions for $z$ are either purely real or purely imaginary
and come in pairs with opposite signs.
 A real root $z=k$ of this equation corresponds to a crossing of
 some band at the Fermi level.
For purely real roots $z$, only the solution $\Psi(r,z)$ corresponding to
positive $z$ is normalizable
at the origin and therefore physically acceptable.
 On the other hand, the purely imaginary roots $z=\pm \imath k$
lead to a single real solution $\Psi(r,\imath k)=\Psi(r,-\imath k)$.
Thus we can see that in general there are $2$ linearly independent solutions. If all $4$ solutions of $z$ are real then these
correspond to the $2$ Fermi surfaces obtained from the intersection of the bands with the Fermi level. If only $1$ pair of solutions is real then
the imaginary pair corresponds to a decaying state.

The BdG equations describing the proximity-induced supercondutivity at a
 TI/SC interface~\cite{fu_prl'08} follow from the BdG equations for the
 present system, Eq.~(\ref{eq:zeroenergy}), by taking $\eta=0$. In this case the matrix equation reduces to
\begin{align}
&\left(\begin{array}{cc}V_z-\mu & z\alpha  \\ \alpha z  &  -V_z-\mu \end{array}\right)\left(\begin{array}{c} u_{\uparrow} \\ u_{\downarrow}\end{array}\right)=0
\end{align}
This equation only has $1$ pair of real solutions $z=\pm \sqrt{\mu^2-V_z^2}/\alpha$ and therefore has only $1$ linearly independent solution in the core
of the vortex.

\subsection{Solution outside the vortex core.}
The solution outside the vortex does not have a simple analytic form
as the solution inside.
Motivated by the large $r$ asymptotic expansion for Bessel functions, for $r>R$ we can consider a series expansion of the form
\begin{align}
\left(\begin{array}{c}u_{\uparrow}(r)\\u_{\downarrow}(r)\end{array}\right)=\frac{e^{-z r}}{r^{1/2}}\left(\begin{array}{c}\rho_{\uparrow}(1/r)\\\rho_{\downarrow}(1/r)\end{array}\right)
\end{align}
where $\rho(x)$ are analytic power series in $x$. We expect to be able to close such a series of equations since the matrix in Eq.~(\ref{eq:zeroenergy}) only has
derivatives and powers of $1/r$.
\begin{align}
&\left(\begin{array}{cc}\eta  (-\partial_r^2-\frac{1}{r}\partial_r)+V_z-\mu & \lambda\Delta+\alpha  (\partial_r+\frac{1}{r} )\\ -\lambda \Delta-\alpha  (\partial_r)  & \eta (-\partial_r^2-\frac{1}{r}\partial_r+\frac{1}{r^2})-V_z-\mu \end{array}\right)\frac{e^{-z r}}{r^{1/2}}\left(\begin{array}{c}\rho_{\uparrow}(1/r)\\\rho_{\downarrow}(1/r)\end{array}\right)=0\\
&\left(\begin{array}{cc}\eta  (-\partial_r^2-\frac{1}{4 r^2}+2 z\partial_r-z^2) +V_z-\mu & \lambda\Delta+\alpha  (\partial_r+\frac{1}{2 r} -z)\\ -\lambda\Delta-\alpha  (\partial_r-\frac{1}{2 r}-z)  & \eta  (-\partial_r^2+\frac{3}{4 r^2}+2 z\partial_r-z^2)-V_z-\mu \end{array}\right)\left(\begin{array}{c}\rho_{\uparrow}(1/r)\\\rho_{\downarrow}(1/r)\end{array}\right)=0.
\label{eq:64}
\end{align}
As shown in the appendix, the last equation has a simple solution as a power-series in $1/r$ which can be determined numerically. Moreover in this power series expansion we can determine the equation for the $0$-th order term by formally setting $1/r=0$ as below
\begin{align}
&\left(\begin{array}{cc}-\eta z^2 +V_z-\mu & \lambda\Delta-z\alpha  \\ -\lambda \Delta+z\alpha    & -\eta z^2-V_z-\mu \end{array}\right)\left(\begin{array}{c}\rho_{\uparrow}(0)\\\rho_{\downarrow}(0)\end{array}\right)=0\label{eq:largeRmode_vortex}.
\end{align}

Setting $z=\imath k$ one can see that the matrix appearing in the above equation is related to the one determining the quasiparticle bandstructure
from the BdG equations. The values of $z$ consistent with the above equation are determined by setting
\begin{align}
&Det\left(\begin{array}{cc}-\eta z^2 +V_z-\mu & \lambda\Delta-z\alpha  \\ -\lambda\Delta+z\alpha    & -\eta z^2-V_z-\mu \end{array}\right)
\nonumber\\
&=(\eta z^2-\mu)^2-V_z^2+(z\alpha\mp \Delta)^2=0\label{eq:char_eq}.
\end{align}
The two families of solution for $\lambda=\pm 1$ are related simply by flipping the sign of $z$.
The sign of the product of the roots $z_n$ of Eq.~\ref{eq:char_eq} is given by $S=\sgn{(\prod_n(z_n))}=\sgn{(C_0)}$ where $C_0=(\mu^2+\Delta^2-V_Z^2)$
is the polynomial evaluated at $z=0$. The parity of the number
of normalizable solutions with $Re(z_n)>0$ is given by $P=\sgn{(\prod_n Re(z_n))}$.
Since Eq.~\ref{eq:char_eq} is real, complex roots $z_n$ occur in conjugate pairs. Therefore, complex roots cannot affect the sign of either $S$ or $P$.
It follows that $S=P$.

\begin{figure}[tbp]
\begin{center}
\includegraphics[width=0.47\textwidth]{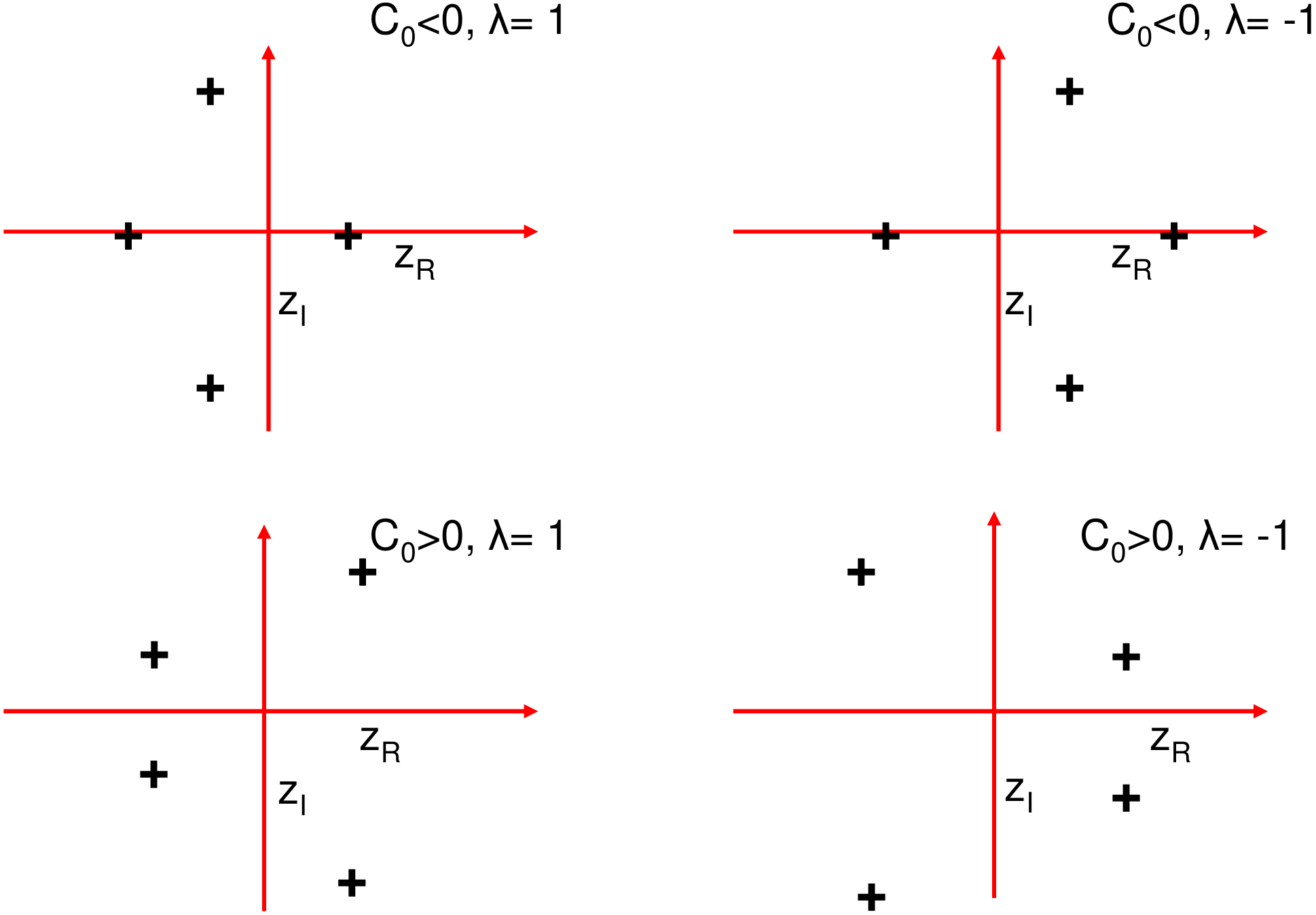}
\end{center}
\caption{Complex roots $z=z_R+\imath z_I$ of the characteristic equation
 Eq.~\ref{eq:char_eq} shown in the complex plane.  In the non-Abelian phase $(C_0=(\Delta^2+\mu^2-V_Z^2)<0)$, 3 roots with positive real parts and 1
root with negative real part for $\lambda=-1$ while there are only 2 roots on either side of the imaginary axis for $C_0>0$ for $\lambda=-1$. The roots in the $\lambda=1$ channel are the negative of the roots in the $\lambda=1$ channel.}\label{fig:roots}
\end{figure}

 Therefore, the condition $|V_Z|>\sqrt{\Delta^2+\mu^2}$ implies that there are an odd number of roots with positive real parts. Specifically, as seen in Fig.~(\ref{fig:roots}), if $C_0<0$, there are $3$ roots on one
side of the imaginary axis and $1$ root on the other side.
Similarly for $C_0>0$ we must have $2$ solutions on each side of the imaginary axis. A slightly different version of this argument has previously 
been presented. \cite{roman}

\subsection{Matching Boundary conditions at the edge of the vortex core.}
As discussed before, for $\Delta^2+\mu^2>V_Z^2$, one of the channels labeled by $\pm$ has a solution with
 $3$ decaying (negative real parts) solutions and $1$ growing solution. Out of these
$3$ decaying solutions, one is purely real and negative and the other $2$ are complex conjugate with negative real parts. An observation that
can be made by considering Eq.~(\ref{eq:64}) is that if $(\rho_{\uparrow}(1/r),\rho_{\downarrow}(1/r))^T$ corresponds to a value $z$ then
$(\rho^*_{\uparrow}(1/r),\rho^*_{\downarrow}(1/r))^T$ corresponds to an eigenvalue $z^*$. Thus from one pair of complex conjugate decaying
eignevalues we can construct a pair of real solutions
\begin{align}
\left(\begin{array}{c}u_{\uparrow,(1,2)}(r)\\u_{\downarrow,(1,2)}(r)\end{array}\right)&=s_{(1,2)}\{e^{-z r}\left(\begin{array}{c}\rho_{\uparrow}(1/r)\\\rho_{\downarrow}(1/r)\end{array}\right)\pm e^{-z^* r}\left(\begin{array}{c}\rho^*_{\uparrow}(1/r)\\\rho^*_{\downarrow}(1/r)\end{array}\right)\}
\end{align}
where $s_{1}=1$ and $s_{2}=\imath$ with the solution 1 corresponding to $+$ and 2 to $-$.

The non-degenerate real eigenvalue already corresponds to a real eigenvector
\begin{align}
\left(\begin{array}{c}u_{\uparrow,3}(r)\\u_{\downarrow,3}(r)\end{array}\right)&=e^{-z r}\left(\begin{array}{c}\rho_{\uparrow}(1/r)\\\rho_{\downarrow}(1/r)\end{array}\right).
\end{align}

On the other hand, for $r<R$ we expect a 2 parameter family with the general solution given by
\begin{align}
\left(\begin{array}{c}v_{\uparrow}(r)\\v_{\downarrow}(r)\end{array}\right)&= a_4 \left(\begin{array}{c}u_{\uparrow,4}(r)\\u_{\downarrow,4}(r)\end{array}\right)+a_5 \left(\begin{array}{c}u_{\uparrow,5}(r)\\u_{\downarrow,5}(r)\end{array}\right).
\end{align}

Matching the gradient and the wave-function at $r=R$ we get
\begin{align}
\left(\begin{array}{c}v_{\uparrow}(R)\\v_{\downarrow}(R)\\ \partial_r v_{\uparrow}(R)\\ \partial_r v_{\downarrow}(R)\end{array}\right)= a_4 \left(\begin{array}{c}u_{\uparrow,4}(R)\\u_{\downarrow,4}(R)\\ \partial_r u_{\uparrow,4}(R)\\ \partial_r u_{\downarrow,4}(R)\end{array}\right)+a_5 \left(\begin{array}{c}u_{\uparrow,5}(R)\\u_{\downarrow,5}(R)\\ \partial_r u_{\uparrow,5}(R)\\ \partial_r u_{\downarrow,5}(R)\end{array}\right)&=\sum_{j=1}^3 a_j \left(\begin{array}{c}u_{\uparrow,j}(R)\\u_{\downarrow,j}(R)\\ \partial_r u_{\uparrow,j}(R)\\ \partial_r u_{\downarrow,j}(R)\end{array}\right).
\end{align}

Together with the normalization constraint on the global wave-function,
this leads to 5 equations in 5 variables, which leads to a unique solution
for the Majorana mode in the case $C_0<0$. However, for the other case
with $C_0>0$, there are only 2 decaying modes outside the vortex core. The
existence of a Majorana mode would then require us to satisfy 5 equations
with 4 variables. Such a problem in general is over constrained and no
Majorana solutions exist in this case.

\section{Majorana solution for vortex in the spin-orbit coupling.}
For a planar system, the Rashba spin-orbit term
 $\alpha(\bm\sigma\times\bm p)\cdot\hat{z}$ in the Hamiltonian we
 considered can also be written as $\alpha\bm\sigma\cdot\bm p$. These
2 terms are simply related to each other by a $\sigma_z$ spin rotation and
a more general Rashba-type spin-orbit term can be written as
\begin{equation}
H_{SO}=\alpha[\cos{\zeta}(\bm\sigma\times\bm p)\cdot\hat{z}+\sin{\zeta}
\bm\sigma\cdot\bm p].
\end{equation}
A recent proposal~\cite{Sato} has considered a defect in such a spin-orbit
coupling where the angle of the spin-orbit $\zeta$ varies in space to
form a vortex $(\zeta(\theta)=\theta)$.
The full BdG Hamiltonian for such a vortex can be written
 in Nambu space as
\begin{equation}
H_{BdG}=(-\eta \nabla^2-\mu)\tau_z + V_z\sigma_z+\frac{1}{4} \left(\sigma_+\left\{\alpha(r)e^{\imath \theta},p_-\right\}+\sigma_-\left\{\alpha(r)e^{-\imath \theta},p_+\right\}\right)\tau_z+\Delta\tau_x
\end{equation}
where the anti-commutation must be introduced in the spin-orbit term to
preserve hermiticity.

Substituting the circular-polar form for the derivatives we note that the
Hamiltonian becomes
\begin{align}
&\tilde{H}_{BdG}=-\{\eta(\partial_r^2+\frac{1}{r}\partial_r-\frac{\partial_\theta^2}{4 r^2})+\mu\}\tau_z + V_z\sigma_z\nonumber\\
&-\frac{\imath}{2} \{\sigma_+(\{\partial_r,\alpha(r)\}+\frac{\alpha(r)}{r}\{e^{-\imath\theta},e^{\imath\theta}\partial_\theta\})+h.c\}+\Delta\tau_x
\end{align}
which in turn simplifies to a $\theta$ independent form
\begin{align}
&\tilde{H}_{BdG}=-\{\eta(\partial_r^2+\frac{1}{r}\partial_r-\frac{\partial_\theta^2}{4 r^2})+\mu\}\tau_z + V_z\sigma_z\nonumber\\
&-\frac{1}{2} \{\sigma_+(\{\alpha(r)\partial_r+\alpha'(r)\}-\frac{\imath\alpha(r)}{r}\{\partial_\theta+\imath\})+h.c\}+\Delta\tau_x.
\end{align}
Therefore, $J_z=L_z$ commutes with the above Hamiltonian and the
spinor form is
\begin{equation}
\psi_{m_J}(r,\theta)=e^{\imath L_z \theta}\psi_{m_J}(r)=e^{\imath m_J \theta}\psi_{m_J}(r)=e^{\imath m_J \theta}\left (\begin{array}{c}u_{\uparrow,m_J}(r)\\u_{\downarrow,m_J}(r)\\v_{\downarrow,m_J}(r)\\-v_{\uparrow,m_J}(r)\end{array}\right).
\end{equation}
As before only the $m_J=0$ channel can lead to a non-degenerate Majorana solution,
and the BdG equation in this channel is given by
\begin{align}
&\tilde{H}_{BdG}=-\{\eta(\partial_r^2+\frac{1}{r}\partial_r)+\mu\}\tau_z + V_z\sigma_z\nonumber\\
&-\frac{1}{2} \{\sigma_+(\{\alpha(r)\partial_r+\alpha'(r)\}+\frac{\alpha(r)}{r})+h.c\}+\Delta\tau_x.
\end{align}

Since the above BdG equation is real, it can be reduced to a  $2\times 2$ matrix differential equation:
\begin{align}\label{eq:zeroenergy_new}
\!&\!\left(\!\begin{array}{cc}\!\!-\!\eta (\partial_r^2\!+\!\frac{1}{r}\partial_r)\!+\!V_z\!-\!\mu\!&\lambda\Delta+\alpha(r)\partial_r+\alpha'(r)+\frac{\alpha(r)}{r}\\ -\lambda \Delta+\alpha(r)\partial_r+\alpha'(r)+\frac{\alpha(r)}{r}&\! -\!\eta (\partial_r^2\!+\!\frac{1}{r}\partial_r)\!-\!V_z\!-\!\mu\! \end{array}\!\right)\!\!\Psi_0(r)\!=\!0.
\end{align}

As before, considering a step function vortex profile $\alpha(r)=0$ for $r<R$ and
$\alpha(r)=\alpha$ for $r>R$, one notices that the reduced
BdG equation outside
the spin-orbit vortex core resembles the reduced BdG equation in the same region for a regular vortex in the large $r$ limit (Eq.~(\ref{eq:zeroenergy}) with $\Delta(r)=\Delta$).
 Inside the vortex core $\alpha(r)=0$, and the BdG equations, as before, are analytically solvable via Bessel functions as below
\begin{align}
&\left(\begin{array}{cc}\eta  (-\partial_r^2-\frac{1}{r}\partial_r)+V_z-\mu & \lambda\Delta\\ -\lambda\Delta  & \eta (-\partial_r^2-\frac{1}{r}\partial_r)-V_z-\mu \end{array}\right)\left(\begin{array}{c} u_{\uparrow} J_0 (z r )\\ u_{\downarrow} J_0(z r)\end{array}\right)=0
\end{align}
where
\begin{align}
&\left(\begin{array}{cc}-\eta z^2+V_z-\mu & \lambda\Delta  \\ -\lambda\Delta & -\eta z^2 -V_z-\mu \end{array}\right)\left(\begin{array}{c} u_{\uparrow} \\ u_{\downarrow}\end{array}\right)=0.
\end{align}
As before, this leads to 2 solutions inside and 3 solutions
outside the vortex core, with 5 constraints at the interface. This leads
to a single non-degenerate Majorana solution at the interface.

Contrary to the result in Ref.~[\onlinecite{Sato}] for the asymptotic $(r\rightarrow \infty)$ behavior of the zero energy wave function, $\psi_0(r)\propto e^{-V_Z r/\alpha}$, we find that the asymptotic zero energy wave function behaves as
$\psi_0(r)\propto e^{-\Delta r/\alpha}\frac{e^{-\imath \alpha r}}{\sqrt{r}}$ for $\mu=0$ and small $\Delta < V_Z$. Therefore, according to our
 result, the decay length of the zero energy wave function diverges in 
the limit of vanishingly weak superconductivity  $(\Delta \rightarrow 0)$
and the Majorana mode disappears by delocalizing over the entire system.
 This is in contrast to the result for the wave function in Ref.~[\onlinecite{Sato}] where the zero energy Majorana solution remains localized 
for arbitrarily small values of the superconducting gap.

\section{Majorana solution on the surface of a topological insulator}
Now we apply a similar argument to the vortex in proximity-induced $s$-wave superconductivity on a TI surface ~\cite{fu_prl'08} which is obtained from our Rashba model by setting $\eta=0$. The equation for the allowed
values of $z$ in the superconductor for $r>R$ are then
\begin{align}
&\mu^2-V_z^2+(z\alpha\mp \Delta)^2=0\\
&z=\pm (\Delta \pm \imath\sqrt{\mu^2-V_z^2})/\alpha
\end{align}
Therefore, in each  of the $\pm$ channels,  for small $V_z$, there are a pair of complex conjugate eigenvalues
 on the same side of the imaginary axis. For the $+$ channel both the eigenvalues are to the right of the imaginary
axis and therefore are acceptable decaying solutions. Thus there are 2 linearly independent solutions $(u_{\uparrow,1},u_{\downarrow,1})$ and $(u_{\uparrow,2},u_{\downarrow,2})$ for $r>R$.
From our previous discussion it is now clear that there is only one such solution $(u_{\uparrow,3},u_{\downarrow,3})$ for $r<R$.

Since the Hamiltonian is linear in the derivative, the boundary conditions only require us to match the wave-function $(u_{\uparrow}(r), u_{\downarrow}(r))^{T}$ at $r=R$ and not
the derivative. The boundary conditions that the zero energy solution must satisfy at $r=R$ are given by,
\begin{align}
\left(\begin{array}{c}u_{\uparrow}(R)\\u_{\downarrow}(R)\end{array}\right)= a_3 \left(\begin{array}{c}u_{\uparrow,3}(R)\\u_{\downarrow,3}(R)\end{array}\right)&=\sum_{j=1}^2 a_j \left(\begin{array}{c}u_{\uparrow,j}(R)\\u_{\downarrow,j}(R)\end{array}\right).
\end{align}
Together with the normalization condition for the zero energy wave function, the above equations provide $3$ constraints for the three variables $a_1, a_2, a_3$. This yields a unique zero energy Majorana wave-function for a
vortex on a TI surface.

\section{Numerical calculation of the vortex excitation spectrum in the spin-orbit coupled semiconductor.}
\begin{figure}[tbp]
\begin{center}
\includegraphics[scale=0.4, angle=270]{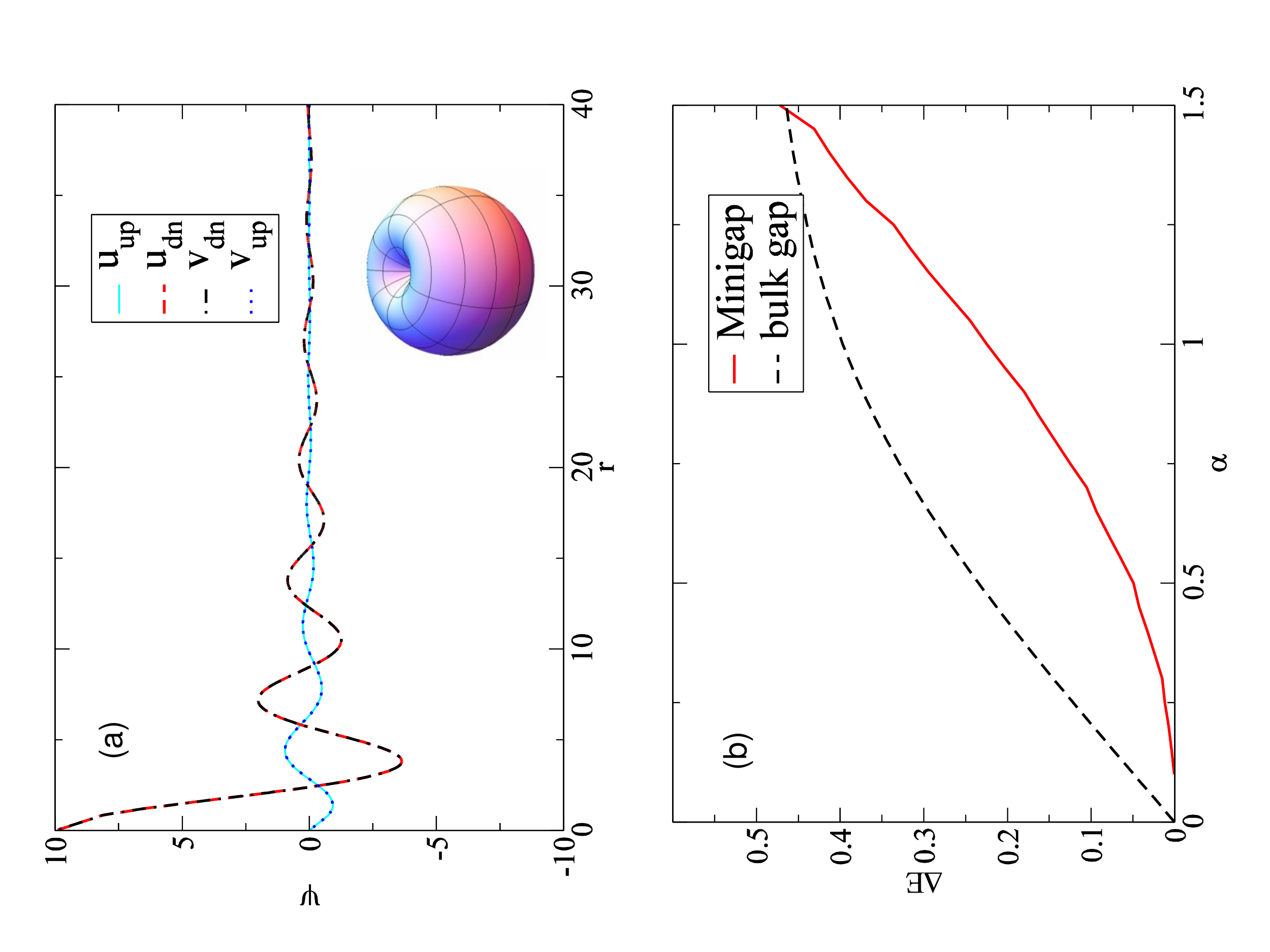}
\end{center}
\caption{(Color online) (a)
Plots for the individual components of the 4-component wave-function
$\Psi(r)=(u_{\ua}(r),u_{\da}(r),v_{\da}(r),-v_{\ua}(r))^T$ for the
zero energy Majorana state at the north pole. The components of $\Psi$
satisfy $u_\sigma=v^*_\sigma$ confirming the Majorana nature of the wave-function.
 We also show the semiconductor heterostructure on the surface of a
 sphere with a vortex and an antivortex (with reduced superconducting amplitudes at the vortex cores) situated at the north and the south poles.
 (b)Numerical results for the vortex mini-gap $\Delta E$ (solid line)
and bulk gap (dashed line) plotted against the spin-orbit coupling
strength $\alpha$ on the semiconductor.
In these plots we have used $\Delta=0.5$, $\mu=0.0$ and $\eta=1.0$ in the units where $V_z=1$.
The spin-orbit coupling strength $\alpha=0.3$ in (a) and varies for $(b)$.
In these units, the size of the vortex core has been taken to be unity.}\label{fig:minigap}
\end{figure}

In previous sections we calculate and show the existence of a Majorana mode in a vortex at the interface of an $s$-wave superconductor
and a spin-orbit coupled semiconductor. The most important information missing from these analytical calculations is the excitation gap  above the zero-energy Majorana state, the so-called minigap. A proper calculation
of this requires a numerical solution of the vortex problem which can be done by considering the system on a sphere with a vortex-antivortex pair~\cite{kraus'08, kraus'09} as shown in the inset of Fig. ~(\ref{fig:minigap}(a)).

The BdG Hamiltonian of this problem can be written as
\begin{equation}
H=[\eta \bm p^\dagger \bm p+\frac{\alpha}{2} \{(\sigma \times \hat{R})\cdot \bm p+\bm p^\dagger \cdot (\sigma \times \hat{R}) \}-\mu]\tau_z+ V_Z \sigma \cdot \hat{R}+\Delta(\mathbf{r})\tau_x
\end{equation}
where $\bm p=-\imath[\bm\nabla-\hat{R}(\hat{R}\cdot\bm\nabla)]$ is the
non-Hermitean gradient operator restricted to the surface of the sphere
and $\hat{R}=\frac{\bm r}{r}$.
The above Hamiltonian takes a more familiar form in angular coordinates as
\begin{align}
H&=[\frac{\eta}{R^2} L^2-\frac{\alpha}{R} L\cdot \sigma-\mu ]\tau_z+V_Z \sigma\cdot \hat{R}+\Delta(\theta)\{\cos{\phi}\tau_x+\sin{\phi}\tau_y\}\nonumber\\
&= [\frac{\eta}{R^2} L^2-\frac{\alpha}{R} \{L_z\sigma_z+\frac{1}{2}L_+\sigma_-+\frac{1}{2}L_-\sigma_+\}-\mu]\tau_z+V_Z \{R_z \sigma_z+\frac{1}{2}R_+\sigma_-+\frac{1}{2}R_-\sigma_+\}\nonumber\\
&+\frac{1}{2}\frac{\Delta(\theta)}{\sin{\theta}}\{R_+\tau_-+R_-\tau_+\}
\end{align}
where $R_z=\cos{\theta}$ and $R_\pm=\sin{\theta}e^{\pm \imath \phi}$. In these equations $R_-=R_+^\dagger$ and $L_-=L_+^\dagger$.
The spectrum of excitations of this system is found by solving the eigenvalue problem
\begin{equation}
H\Psi=E\Psi.\label{eq:sphereBdG}
\end{equation}

Similar to the vortex in the planar geometry, the BdG Hamiltonian has
 a combined spin-orbit-pseudospin rotational symmetry.
 This symmetry can be expressed
compactly by noting that the Hamiltonian commutes with
\begin{equation}
J_{z}=L_z+\frac{1}{2}(\sigma_z-\tau_z),
\end{equation}
 where we have used the
identity $[R_{\pm},L_z]=\pm R_{\pm}$.
The $\phi$ dependence of the eigenstates with $m_J=m$ can be written as $e^{\imath L_z\phi}\Psi_m(\theta)=e^{\imath(m-(\sigma_z-\tau_z)/2)\phi}\Psi_m(\theta)$. The $\phi$ independent part of the eigenstate $\Psi_m(\theta)$
then satisfies a 1 dimensional BdG equation $H_m\Psi_m(\theta)=E_m\Psi_m(\theta)$ where
\begin{equation}
H_m=U_m^\dagger(\phi)H U_m(\phi)
\label{eq:Hm}
\end{equation}
which can be explicitly checked to be $\phi$ independent, and $U_m(\phi)=e^{\imath(m-(\sigma_z-\tau_z)/2)\phi}$. To solve for
the $\theta$ dependent part of $\Psi_m(\theta)$ it is necessary to
convert $H_m$ to a discrete matrix by expanding $\Psi_m(\theta)=\sum_l c_{l,m}P_l^{(m)}(\cos{\theta})$ where $P_l^{(m)}(\cos{\theta})$ are the
associated Legendre polynomials which are the $\phi$ independent parts
of the spherical Harmonics. In the associated Legendre polynomial basis
the kinetic energy term $L^2$ has the simple diagonal form $l(l+1)$.
Under the transformation in Eq.~(\ref{eq:Hm}), the terms  $R_{z,\pm}$ in $H$ transform into
$P_1^{(0,\pm 1)}(\cos{\theta})$ in $H_m$.
 Therefore, its matrix elements in the associated Legendre polynomial
basis can be calculated using the spherical harmonic addition theorem.
A similar procedure can be used to calculate the matrix elements of
the $\theta$ dependent vortex.
For a vortex, we take $\Delta(\theta)=\Delta\tanh{R\sin{\theta}/\xi}$
where $\xi$ is taken to be the length-scale of the vortex.
From symmetry properties it is clear that $\Delta(\theta)/\sin{\theta}$
 is an even polynomial in $\sin{\theta}$ and  can be written
in terms of associated Legendre polynomials as
\begin{equation}
\frac{\Delta(\theta)}{\sin{\theta}}=\sum_{l}c_{(2 l+1)} P^1_{(2 l+1)}(\cos{\theta})
\end{equation}
where the associate Legendre polynomial can be written as
$P^1_l(\cos{\theta})=-\sin{\theta}P^{'}_l(\cos{\theta})$
and $c_l=\frac{(2 l+1)}{2 l(l+1)}\int_{-1}^1P^1_l(x)\Delta(x)dx$. As with the $R$ operators, the angular momentum matrix elements can be
 calculated from the
above expansion by using the spherical harmonic addition theorem.

As in the analytic solution for the vortex, the angular momentum index
 $m$ transforms from $m\rightarrow -m$ under the
particle-hole transformation $\Xi$ and we expect non-degenerate $E=0$
 Majorana solutions of Eq. ~\ref{eq:sphereBdG} only in the $m=0$ channel.
This is confirmed by our numerical solution of Eq. ~\ref{eq:sphereBdG}
 where we find that only in the topological phase $C_0<0$, are there a
 pair of states in the $m=0$ angular momentum channel whose eigenvalues
 approach 0 exponentially
 with increasing radius $R$. The non-zero
energy eigenvalue of the Majorana fermion is a result of the presence of
2 vortices in our calculation with a finite distance between them.
The wave-function of the $E=0$ eigenvalue of the $m=0$ angular 
momentum channel, that is localized at the north-pole  is plotted in 
Fig. ~(\ref{fig:minigap}(a)). The components of the wave-function are 
seen to satisfy $u_\sigma=v^*_\sigma$ confirming the Majorana character
of these states. In the figure, the wave-functions of the Majorana modes
 is seen to decay and oscillate away from the North pole.  The splitting
between the Majorana modes into a pair of exponentially small
oscillating eigevalues is a result of the overlap between the
 Majorana modes at the two poles~\cite{meng}.

Aside from the $E=0$ eigenvalue in the $m=0$ angular momentum
channel, an isolated vortex confines a set of non-zero
eigenvalues in other $m\neq 0$ angular momentum
 channels. Of these, the eigenvalue with the smallest absolute value occurs
in the $m=1$ angular momentum channel and has an eigenvalue equal
to the so-called mini-gap of the vortex. As mentioned before, 
the superconductivity in the non-Abelian superconducting phase is 
re-entrant with a bulk gap that is proportional to the spin-orbit coupling
strength. As seen from Fig. ~(\ref{fig:minigap}(b)) both the bulk and mini-gap
 are proportional to the  spin-orbit coupling strength $\alpha$.
For spin-orbit coupling $\alpha\sim 1$ and chemical potential $\mu=0$,
 both the mini-gap and the bulk gap are of order the induced pairing 
potential $\Delta$. Therefore for the semiconductor structure 
where superconductivity is proximity induced, the mini-gap of a 
vortex can be tuned to be of order $\Delta$ if the chemical potential 
$\mu$ can be tuned to be less than order $V_Z$, the Zeeman potential 
applied to the semiconductor. This is different from the conventional 
case of a regular (not proximity induced) where the chemical 
potential is of order $E_F$ and the mini-gap is of order $\Delta^2/E_F$
which is much smaller than $\Delta$.

Thus the ability to independently control the chemical potential
in the semiconductor heterostructure provides us with a powerful
tool that can increase the mini-gap of the vortex in the semiconductor
heterostructure shown in Fig. ~(\ref{Fig1}) by orders of magnitude from
the values in chiral p-wave superconductors to of the order of 1 K.
This leads to the possiblility of performing TQC with the
Majoranas trapped in vortices in the heterostructure at temperatures
which are as large as a fraction of a 1K.

\section{Bulk topological quantum phase transition.}
 We found that the Majorana modes exist for a spin-orbit coupled semiconductor system
only in the parameter regime $C_0=V_Z^2-(\Delta^2+\mu^2)>0$. This
in turn was related to the parity of the roots in one half
of the complex wave-vector plane of solutions outside the vortex core.
As pointed out before in the context of Eq.~(\ref{eq:char_eq}),
these roots are indeed properties of the reduced
bulk superconducting Hamiltonian in the absence of a vortex. We also
expect such a connection between the bulk properties and the
existence of Majorana modes on general topological grounds ~\cite{Read}.

To show explicitly the connection between the condition for the
 existence of
Majorana modes $(C_0<0)$ and the bulk properties, we note that even though
the gap in the bulk superconducting state prevents the existence of
propagating states at $E=0$, it allows evanescent
 states. Since the states at $E=0$ are particle-hole symmetric eigenstates
$\Psi_0$ of a real Hamiltonian, we can apply an argument analogous to Eq.(\ref{eq:zeroenergy}) to obtain a bulk BdG equation
 in a $\lambda$ channel
\begin{align}
\!&\!\left(\!\begin{array}{cc}\!\!-\!\eta \nabla^2+\!V_z\!-\!\mu\!&\! \lambda\Delta\!+\!\alpha (\partial_x+\imath\partial_y)\\\\ -\lambda \Delta\!-\!\alpha (\partial_x-\imath\partial_y)  & -\!\eta \nabla^2\!-\!V_z\!-\!\mu\! \end{array}\!\right)\!\!\Psi_0(x,y)\!=\!0.
\end{align}
Considering an evanescent state of the form $\Psi_0(x,y)=e^{-z(x\cos{\theta}+y\sin{\theta})}\Psi_0$ leads to a constraint on $z$ which was
previously written as Eq. (\ref{eq:char_eq}). Therefore, the condition
on $C_0$, which determines whether the phase supports a Majorana solution or
not is precisely related to the parity of decaying evanescent modes in
a given $\lambda$ channel in the bulk superconductor at $E=0$.

A change of the parity of the decaying evanescent modes requires an $E=0$
 mode
to become propagating, which can only exist if the bulk superconductor is
gapless. Therefore, a change of the sign of $C_0$, which determines the
topological nature of the phase, must be accompanied by a closing
of the bulk spectrum. This is determined by the full BdG Hamiltonian
for a state with momentum $k(\cos{\theta},\sin{\theta})$ and can be
written in the Nambu space as
\begin{equation}\label{eq:polar_bulk_H}
H_{BdG}=(\eta k^2-\mu)\tau_z + V_z\sigma_z+\frac{\imath\alpha k}{2} (e^{-\imath \theta}\sigma_+-e^{\imath \theta}\sigma_-)\tau_z+\Delta\tau_x.
\end{equation}
The spectrum is obtained by considering $Det(H_{BdG}-E_k)=0$ which can be
simplified to the equation,
\begin{equation}
E_k^2=V_z^2+\Delta^2+\tilde{\epsilon}^2+\alpha^2 k^2\pm 2 \sqrt{V_z^2\Delta^2+\tilde{\epsilon}^2(V_z^2+\alpha^2 k^2)}
\end{equation}
where $\tilde{\epsilon}=\eta k^2-\mu$. Setting $k=0$, it can be seen that
\begin{equation}
E_0^2=(V_Z\pm\sqrt{\Delta^2+\mu^2})^2
\end{equation}
which vanishes as $C_0$ becomes zero, as expected. Recent work \cite{parag} has shown that the quantity $C_0$ is the Pfaffian of the BdG Hamiltonian at $k=0$, $(C_0=Pf(H_{BdG}( k=0)\sigma_y\tau_y))$. The sign of $C_0$, which determines whether the phase of the superconductor is
non-Abelian or not has been shown to be related \cite{parag} to the parity of the first Chern number topological
invariant describing time-reversal broken topological superconductors. \cite{volovik, volovik-yakovenko, yakovenko, Roy}

\begin{figure}[tbp]
\begin{center}
\includegraphics[width=0.47\textwidth]{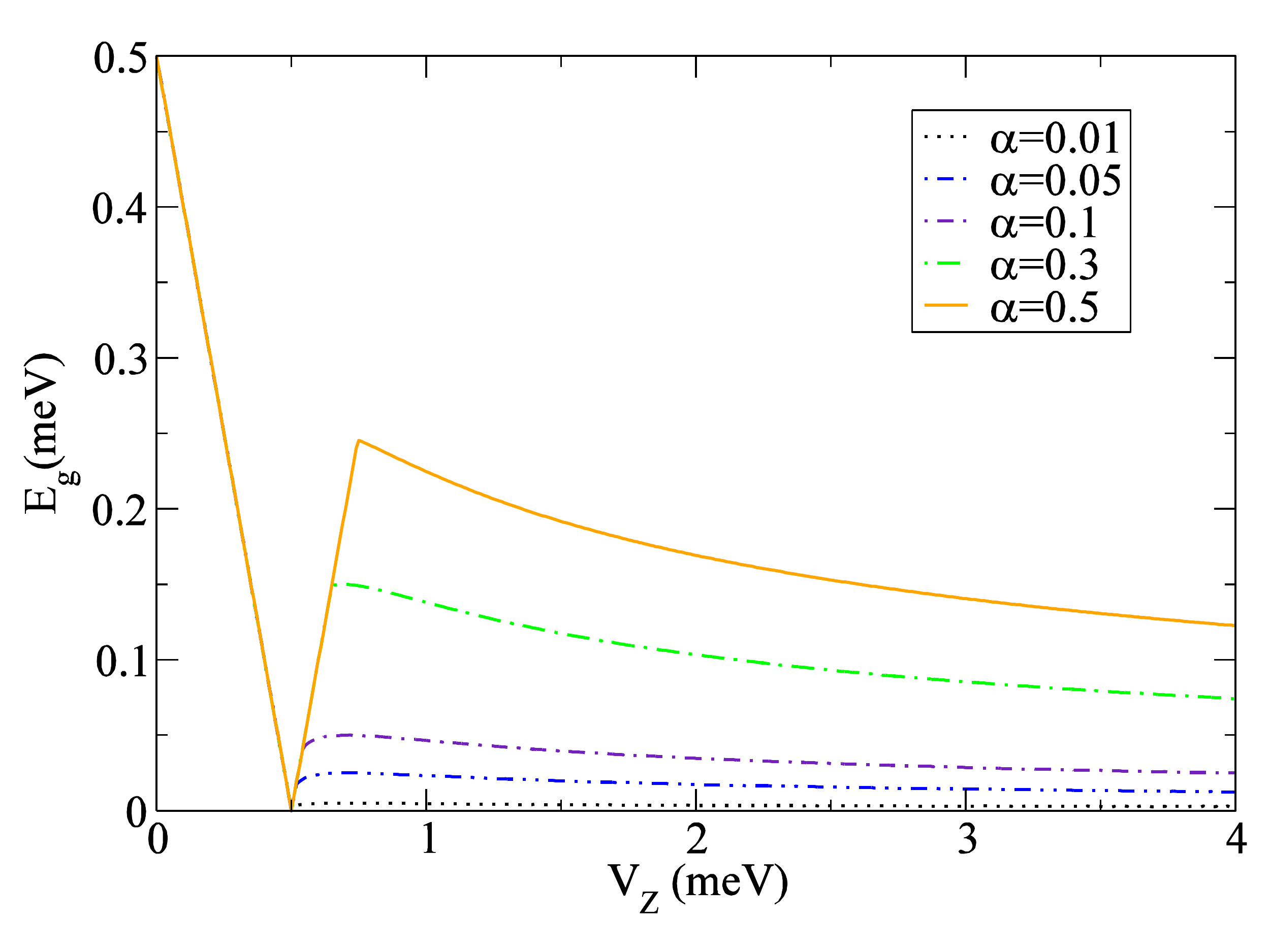}
\end{center}
\caption{Quasiparticle gap versus Zeeman coupling for various values of spin-orbit interaction $\alpha$. The strength of the spin-orbit coupling in the inset is such that $\alpha=0.3$ corresponds to $0.1$ eV-\AA.
 The gap vanishes at the critical value $V_z=\sqrt{\Delta^2+\mu^2}$. The
 spin-orbit coupling has negligible effect below this critical point
 and the superconducting gap is of a conventional
 $s$-wave type. Above the critical value and in the absence of spin-orbit
 coupling, the superconducting gap is destroyed by
the Zeeman coupling. Spin-orbit coupling opens up a gap in this phase
 leading to re-entrant superconductivity with is topological.}\label{fig:phasediag}
\end{figure}

The phase diagram of the spin-orbit coupled semiconductor system can be
 understood from Fig.~(\ref{fig:phasediag}), which gives the variation of
the quasiparticle gap versus the Zeeman splitting. One knows from topological stability of the Majorana fermion mode that,
due to its non-degeneracy, the Majorana state is protected as long as the bulk gap does not close as one moves
through the parameter space.  In Fig.~(\ref{fig:phasediag}), the gap closes (at the wave vector $k=0$) for the Zeeman splitting corresponding to
$V_z^2= V_{zc}^2=\Delta^2 + \mu^2$. The phase with $V_z > V_{zc}$ supports the non-degenerate Majorana state while the phase with $V_z < V_{zc}$ does not. These two regions are separated by a gapless point in the parameter space, which signifies a topological quantum phase transition. The quantum phase transition is topological since the superconducting order on both sides is explicitly $s$-wave and the two phases differ only by the topological properties such as Majorana modes in defects and boundaries.

\section{Competition between superconductivity and Zeeman splitting.}
The proposal to realize Majorana fermion modes in spin-orbit coupled
semiconductor system involves the introduction of a large Zeeman
potential. In general, a Zeeman splitting is known to compete with and eventually destroy
superconductivity. To understand better
the competition between the Zeeman splitting and superconductivity
in a spin-orbit coupled semiconductor, we first consider
the case without spin-orbit coupling. This case is described by the
BdG Hamiltonian
\begin{equation}
H_{BdG}=(\eta k^2-\mu)\tau_z + V_z\sigma_z+\Delta\tau_x.
\end{equation}
The dispersion relation of this Hamiltonian is
$E_k=\pm V_z \pm \sqrt{\Delta^2+\tilde{\epsilon}^2}$.
 In this case, with $V_z=0$, we obtain a conventional proximity
induced $s$-wave superconductor with no Majorana phase.
 As $V_z$ increases above $\sqrt{\mu^2+\Delta^2}$ the quasiparticle gap of the system closes and one obtains a metal with a
 Fermi momentum $k_F$ given by
$\eta k_F^2=\mu\pm\sqrt{V_z^2-\Delta^2}$.
This is the well-known Chandrasekhar-Clogston limit~\cite{chandrasekhar} where
 strong Zeeman splitting suppresses the superconducting quasiparticle gap.
This suppression is due to the fact that, in the spin-polarized regime $(|V_Z|>|\mu|)$, a small
pairing potential cannot open a $s$-wave superconducting gap since the latter couples opposite spins.

The BdG Hamiltonian at $k_F$ is doubly degenerate and is
given by
\begin{equation}
H_{BdG}=\sqrt{V_z^2-\Delta^2}\tau_z + V_z\sigma_z+\Delta\tau_x.
\end{equation}
The degeneracy of the above Hamiltonian at the gapless point,
 which arises from the particle-hole symmetry, is lifted by the Rashba
spin-orbit coupling term $\alpha k_F\sigma_x\tau_z$ in the semiconductor. This yields a topological superconductor with a gap given by
\begin{equation}
E_g\approx 2\alpha k_F\frac{\Delta}{V_z}.
\end{equation}

Considering the gap as a function of the Zeeman splitting (Fig.~(\ref{fig:phasediag})), it is clear
that, for Zeeman splitting  below the critical value $V_Z<\sqrt{\Delta^2+\mu^2}$, the superconductivity is non-topological in nature.
The topological
superconducting phase that supports Majorana fermions is created by
the application of a Zeeman
splitting to \emph{suppress the conventional pairing potential}. In this regime,
the spin-orbit coupling can open up a gap resulting in a
re-entrant superconducting phase.  However,
as is evident from the previous discussions,
the re-entrant superconductivity is unconventional (topological) in the sense
that it supports Majorana fermions.

\section{Topologically protected edge states at interfaces.}
One of the signatures of a topological phase is the existence of
gapless edge states which are inextricably linked to bulk topological
 properties such as Majorana modes in vortices. \cite{nayak_RevModPhys'08, kane_rmp, sandwich_interferometry, Beenakker-sandwich}
 The spin-orbit coupled
 semiconductor structure introduced in Sec. II can be shown to have
 gapless edge states using methods similar
to the ones described in the previous sections. Furthermore it turns out
that this approach to analyze the existence of Majorana edge modes does
not impose additional requirements such as rotational invariance that
were critical for the demonstration of a Majorana solution in a vortex.
Therefore, this method can be used to examine the question of the
existence of Majorana edge modes even in the heterostructures with more
general forms of spin-orbit coupling proposed by Alicea~\cite{alicea}
where the Zeeman splitting can be introduced by an in-plane magnetic
 field.

 \subsection{BdG Hamiltonians for edges.}
Edges can be created in the semiconductor heterostructure by varying a
 parameter of the Hamiltonian such as $\mu$, $V_Z$ or $\Delta$
 perpendicular to the edge of a surface. Without loss of generality we
 can consider an edge that is perpendicular to the direction $\hat{y}$.
Because of translational symmetry along the edge, the resulting edge BdG
 Hamiltonian has $k_x$ as a parameter.  The momentum parameter $k_x$
 transforms as $k_x\rightarrow -k_x$ under the particle-hole
 transformation.  Therefore, a non-degenerate Majorana mode can only
 exist for $k_x=0$.

Fixing $k_x=0$ reduces the two dimensional edge problem to a $1$-dimensional BdG Hamiltonian for a single band semiconductor
with spin-orbit coupling (assumed to be linear in the momentum $k_y$),
 which in general can be written as
\begin{equation}
H_{BdG}=(-\eta \partial_y^2-\mu(y))\tau_z + V_z\bm\sigma\cdot \hat{\bm B}+\imath\alpha\partial_y\hat{\bm \rho}\cdot\bm \sigma\tau_z+\Delta(y)\cos{\phi}\tau_x+\Delta(y)\sin{\phi}\tau_y
\end{equation}
where the unit vector $\hat{\bm B}$ is the direction of the effective Zeeman field
and the unit vector $\hat{\bm \rho}$ characterizes the  spin-orbit coupling. Using the
spin rotation transformations on $H_{BdG}$, we can choose $\hat{\bm \rho}=\hat{\bm y}$ without loss of generality. This yields the Hamiltonian
\begin{equation}
H_{BdG}=(-\eta \partial_y^2-\mu(y))\tau_z + V_z\bm\sigma\cdot \hat{\bm B}+\alpha(\imath \sigma_y)\tau_z\partial_y +\Delta(y)\cos{\phi}\tau_x+\Delta(y)\sin{\phi}\tau_y\label{eq:BdGedge}
\end{equation}
which is invariant under spin-rotations about the $y$-axis. Therefore
without loss of generality we can reduce the above Hamiltonian to
\begin{equation}
H_{BdG}=(-\eta \partial_y^2-\mu(y))\tau_z + V_Z(\cos{\nu}\sigma_z+\sin{\nu}\sigma_y)+\alpha(\imath \sigma_y)\tau_z\partial_y+\Delta(y)\cos{\phi}\tau_x+\Delta(y)\sin{\phi}\tau_y.
\end{equation}

Non-degenerate Majorana spinor solutions are of the form $\Psi=(u,\imath\sigma_y u^*)$ and are completely determined by the 2-spinor $u$.  This fact
was used to obtain the Majorana solutions for vortices to reduce the
BdG equation from a $4\times 4$ system of equations to a $2\times 2$
system of equation. However, this reduction procedure required the BdG Hamiltonian
to be real which is not the case for general forms of spin-orbit coupling
and Zeeman splitting.
The BdG equation for the zero energy mode $H_{BdG}\Psi=0$
may be reduced to an equation for $u$ as
\begin{equation}
\left[(-\partial_y^2-\mu(y)) + V_Z(\cos{\nu}\sigma_z+\sin{\nu}\sigma_y)+\alpha(\imath \sigma_y)\partial_y\right]u+\Delta(y)e^{\imath\phi}(\imath\sigma_y)u^*=0.
\end{equation}
This equation  is not real but may be reduced to a system of real
 equations by
writing $u=u_R+\imath u_I$ and taking the real and imaginary parts of
the resulting equation giving a pair of equations of the form
\begin{align}
&\left[(-\partial_y^2-\mu(y)) + V_Z\cos{\nu}\sigma_z+\alpha(\imath \sigma_y)\partial_y+\Delta(y)\cos{\phi}(\imath\sigma_y)\right]u_R-[\Delta(y)\sin{\phi}-V_Z\sin{\nu}](\imath\sigma_y)u_I=0\\
&\left[(-\partial_y^2-\mu(y)) + V_Z\cos{\nu}\sigma_z+\alpha(\imath \sigma_y)\partial_y-\Delta(y)\cos{\phi}(\imath\sigma_y)\right]u_I+[\Delta(y)\sin{\phi}-V_Z\sin{\nu}](\imath\sigma_y)u_R=0.
\end{align}
The above pair of equations is similar to the pair of equations obtained
for the two $\lambda=\pm 1$ channels except that earlier the two channels were
decoupled. For $\Delta(y)$ independent of $y$, the 2 channels can also
be decoupled by choosing $\phi$ such that $\Delta(y)\sin{\phi}=V_Z\sin{\nu}$. In what follows, we will make this choice and also replace
$V_Z\cos{\nu}\rightarrow V_Z$ and $\Delta\cos{\phi}\rightarrow \Delta$.
This results in a reduced BdG equation for the $E=0$ reduced spinor
 $\Psi_0(y)$
\begin{equation}
\left(\begin{array}{cc}-\eta \partial_y^2+V_z-\mu(y) & \lambda\Delta(y)+\alpha  \partial_y\\ -\lambda \Delta(y)-\alpha \partial_y  & -\eta \partial_y^2-V_z-\mu(y) \end{array}\right)\Psi_0(y)=0
\end{equation}
where $\lambda=\pm 1$.

An edge in a two dimensional system of the type considered above is
 defined by requiring some parameter of the Hamiltonian to vary across
 the edge situated at $y=0$. We take this parameter to be constant for
 $y<0$ and $y>0$. In this
 case, our
previous approach can be applied in a way even simpler than the application to the vortex problem, since the solutions on both sides of the
interface at $y=0$ can be approximated as a sum $\Psi_0(y)=\sum_n a_n e^{-z_n y}u_n$ where, as in the interior of the vortex (but far from the vortex core), (Eq.\ref{eq:largeRmode_vortex}),
\begin{align}
&\left(\begin{array}{cc}-\eta z_n^2 +V_z-\mu & \lambda\Delta-z_n\alpha  \\ -\lambda \Delta+z_n\alpha    & -\eta z_n^2-V_z-\mu \end{array}\right)u_n=0.\label{eq:largeRmode_edge}
\end{align}
 Similar to the vortex case, in the topological phase
$C_0=(\Delta^2+\mu^2-V_Z^2)<0$, there are 3 values of $z_n$ such that $Re(z_n)<0$, while in
the non-topological phase $C_0>0$, there are only 2 solutions in a given
$\lambda$ channel.
The coefficients $a_n$ in the solution are determined
 by matching the boundary conditions on $\Psi_0(y)$ at $y=0$.
The coefficient $C_0$, written in terms of the original parameters
of the wire, reduces to
\begin{equation}
C_0=\Delta^2\cos^2{\phi}+\mu^2-V_Z^2\cos^2{\nu}=\Delta^2+\mu^2-V_Z^2
\end{equation}
and is not affected by the $\phi$ and $\nu$ parameters that were introduced
to make the BdG Hamiltonian real. The procedure of reducing the BdG Hamiltonian to a
real Hamiltonian only introduces the additional constraint $|V_Z\sin{\nu}|<\Delta$.

\subsection{Chiral edge states.}
Based on analogy with FQHE and chiral $p$-wave superconductors, one
 expects a chiral gap-less state confined to the edge of the
 semiconductor heterostructure.
An edge can be created in such structures by raising the
chemical potential $\mu$ towards the edge such that electrons stay confined
inside the system. Therefore, an edge of a system confined to $y<0$
 is defined by $\mu(y)=0$ for $y<0$ and $\mu(y)=\infty>|V_z|$ for $y>0$.
In these structures we assume that $\Delta(y)=\Delta$ is independent of $y$.

The BdG equation now reduces to a $2\times2$ system of equations,
\begin{equation}
[-\eta \partial_x^2-\mu+V_z\sigma_z-\imath\alpha\sigma_y\partial_x+\imath\lambda\sigma_y\Delta]\psi(x)=0,
\end{equation}
where
\begin{equation}
\psi(x)=\left(\begin{array}{c}u_\uparrow(x)\\u_\downarrow(x)\end{array}\right).
\end{equation}
In order to solve a semi-infinite system we make a plane-wave trial solution ansatz
\begin{equation}\label{eq:psix}
\psi(x)=e^{z x}\left(\begin{array}{c}u_\uparrow\\u_\downarrow\end{array}\right)
\end{equation}
where $z$ must now satisfy
\begin{equation}\label{eq:z2}
\left(\begin{array}{cc}-\eta z^2+V_z-\mu & z\alpha  \\ \alpha z  & -\eta z^2 -V_z-\mu \end{array}\right)\left(\begin{array}{c} u_{\uparrow} \\ u_{\downarrow}\end{array}\right)=0.
\end{equation}

As in the case of the vortex, for $V_z^2>(\Delta^2+\mu^2)$ there are 3 solutions on the right half of the complex $z$ plane and 1 solution on the right half for $\lambda=-1$. The situation is opposite for $\lambda=1$.
Solutions with $Im(z)>0$ are physical on the left-edge of the system while
$Im(z)<0$ is physical on the right edge of the system. Thus for
 $\lambda=-1$ there are 3 physical solutions on the left-edge of the
 system which
is the exact number needed to make the 2-component spinor vanish at the
left-edge. Consequently, there is a localized zero mode at the left edge
of the system in the $\lambda=-1$ channel. Similarly there is a localized zero mode at the right edge of the system in
the $\lambda=+1$ channel.
Finally, for $V_z^2<(\Delta^2+\mu^2)$, there are no zero energy solutions at
either edge since there are only 2 physical solutions at each edge which
is insufficient to match the boundary conditions.
Since the wave-function is confined to the edge, we expect the boundary
conditions $\Psi_{0,\ua}(0)=\Psi_{0,\da}(0)=0$, which together with normalization lead to 3 constraints. As mentioned before, in
the topological phase we obtain 3 $a_n$ coefficients corresponding to the
3 normalizable solutions in the interior. Therefore, there is a unique
zero energy state resulting from a matching of the boundary conditions.
This state is a Majorana mode for the end-point of the nanowire in the topological phase$(C_0<0)$, which disappears when we tune the wire through a
phase transition to $C_0>0$.
The Majorana modes at the edges discussed above only occured at $k_x=0$.
The complete spectrum of the edge is obtained by considering the BdG
Hamiltonian at small $k_x\neq 0$ using the $k\cdot p$ perturbation theory. To lowest order in $k_x$, the Nambu Spinor wave-function can be approximated as $\Psi_{k_x}(x,y)\approx e^{i k_x x}\Psi_0(y)$
with an energy $E_{k_x}=v k_x$ where $v=\langle\Psi_0|\sigma_y\tau_z|\Psi_0\rangle$.

A similar chiral Majorana wire is obtained by considering an edge
between the topological phase $C_0<0$ and $C_0>0$ where
$\mu(y)$ is constant but $V_Z(y)=0$ for $y>0$. In that case there are 5
constraints to match as in the vortex case, and there are 5 coefficients,
3 arising from the topological phase at $y<0$ and 2 from the non-topological phase $y>0$.

\subsection{Non-chiral Majorana edge states.}
Now we consider the junction of a pair of topological
superconducting islands with phases $\phi$ and $\phi'=\pi-\phi$ which is a
geometry that is of particular interest to TQC architechtures~\cite{fu_prl'08,Sau}. For such a choice of phases, the effective
pairing potential in
 $\Delta \cos{\phi(x)}$ is a step function given by $\Delta(x)=\Delta\cos{\phi}$ for $x<0$ and $\Delta(x)=-\Delta\cos{\phi}$ for $x>0$. As before we then replace $\Delta\cos{\phi}\rightarrow \Delta$. Focusing on the
 $k_x=0$ particle-hole symmetric momentum for the edge ,solutions for $x<0$ and $x>0$
can be expanded in terms of spinor functions given in
 Eq.~\ref{eq:psix} which is written in terms of eigenvalues $z$ and
 eigenvectors  satisfying Eq.~\ref{eq:z2}. Normalizable solutions must
 now be composed of superpositions of exponentials with $Re(z)<0$ for $y<0$
and $Re(z)>0$ for $x>0$. We note at this point that the equations for $x>0$ and $x<0$ differ by a change in sign of $\Delta(x)$ across the
interface which corresponds to a change in sign of $z$. Thus, as before for
 $\lambda=-1$, in the topological phase we have 3 values of $z$
such that $Re(z)>0$ for $x>0$ and 3 values such that $Re(z)<0$ for $x<0$.
Following the boundary condition matching argument of the last section,
for the $\pi$ junction there are 6 states at $x=0$ to compose wave-functions at $x=0$ which need to satisfy 5 constraints. Therefore, generically there
will be a pair of zero energy modes satisfying these equations.

It might appear that, unlike the case for the chiral edge states, the pair
of Majorana states cannot be topologically protected. In the case of the
TI/SC interface, ~[\onlinecite{fu_prl'08}] the existence of such a pair of
non-chiral Majorana edge modes at a phase difference $\pi$ was
a consequence of time-reversal symmetry which is broken here.  In our
calculation we find that because of this time-reversal breaking, the
pair of degenerate zero modes may occur at a phase difference of $\pi-2 \phi$. In fact, by considering the evolution of the Andreev bound state spectrum in the junction as a function of phase difference,~~\cite{roman} it is possible to show that
even though the Majorana nature of the pair of non-chiral Majorana modes
is not topologically protected, it is not possible to eliminate the zero-modes all together. The zero-crossing of the non-chiral Majorana modes
may only be shifted to different values of phase by time-reversal breaking
perturbations. This is a result of the fact that
 the 2 particle-hole symmetry related branches of the Andreev bound state
spectrum differ by Fermion parity. Therefore, even though an infinitesimal
perturbation can lift the degeneracy of the 2 zero modes at
some value of phase difference $\delta\phi$,
it can only do so by shifting the crossing to a neighboring value of
$\delta\phi$.
Similar to the chiral edge modes one can use the $k \cdot p$ perturbation
theory to construct a pair of linearly dispersing modes from
the pair of zero energy states at $k_y=0$.

\section{Model calculations of proximity effects in superconductor - semiconductor - magnetic insulator heterostructures}

In this section we study, starting from a microscopic tight binding model, the excitation spectrum of a semiconductor thin film sandwiched between an s-wave superconductor (SC) and a ferromagnetic insulator (MI). We determine the dependence of the effective SC gap and Zeeman splitting induced by proximity effects on the parameters that characterize the heterostructure model. We also calculate the dynamical contributions to a low-energy effective theory of the proximity effect and identify parameter regimes suitable for the experimental implementation of the semiconductor-based proposal of a platform for topological quantum computation.

To study the proximity effect in the SC - semiconductor - MI heterostructure, we consider the minimal microscopic model defined by the Hamiltonian
\begin{equation}
H_{\rm tot} = H_0 + H_{SC} + H_{MI} + H_{\tilde{t}_S} + H_{\tilde{t}_M}. \label{Htot}
\end{equation}
The $H_0$ term describes the semiconductor thin film,
\begin{eqnarray}
H_0 &=& \sum_{i, j, \sigma} t_{ij}c_{i\sigma}^{\dagger}c_{j\sigma} + \frac{\alpha}{2}\sum_{i, \sigma, \sigma^{\prime}} \left[c_{i+\delta_x\sigma}^{\dagger}(i\hat{\tau}_y)_{\sigma, \sigma^{\prime}}c_{{i\sigma}^{\prime}}\right. \nonumber \\
&-& \left.c_{i+\delta_y\sigma}^{\dagger}(i\hat{\tau}_x)_{\sigma,
\sigma^{\prime}}c_{{i\sigma}^{\prime}} + {\rm h.c.}\right] + \sum_{i,\sigma} V(z_i)c_{i\sigma}^{\dagger}c_{i\sigma} , \label{H0}
\end{eqnarray}
where the first contribution describes hopping on a cubic lattice, while the second represents a lattice model of the Rashba spin-orbit interaction. The hopping matrix elements are non vanishing for nearest neighbors, $t_{ij}=-t_0$, and next nearest neighbors, $t_{ij}=-t_1$, and we also include an on-site contribution $t_{ii}=\epsilon_0$ that shifts the bottom of the semiconductor spectrum to zero energy. The parameter $\alpha$ represents the Rashba coupling constant and $\delta_{x(y)}$ are nearest neighbor displacements in the xy-plane of a cubic lattice with lattice parameter $a$. The system is assumed to be infinite in the x- and y-directions and contains N planes perpendicular to the z-direction. The quantities $\hat{\tau}_{x(y)}$ are Pauli matrices and $c_{i\sigma}^{\dagger}, c_{i\sigma}$ are electron creation and annihilation operators, respectively. The last term represents an external bias potential that modify the on-site energies along the z-direction.

For the superconductor we use a simple mean-field model defined by the Hamiltonian
\begin{eqnarray}
H_{SC} = -t_s\sum_{\langle i, j\rangle, \sigma} b_{i\sigma}^{\dagger}b_{j\sigma} &+&\epsilon_s\sum_{i,\sigma}b_{i\sigma}^{\dagger}b_{i\sigma}  \label{Hsc} \\
&+&\sum_{i}\left(\Delta b_{i\uparrow}^{\dagger}b_{i\downarrow}^{\dagger} + {\rm h.c.}\right),   \nonumber
\end{eqnarray}
where $t_s$ represents the nearest neighbor hopping on a cubic lattice and $\Delta$ is the mean-field s-wave SC order parameter. The SC and the semiconductor thin film have a planar interface perpendicular to the z-direction. For clarity, the electron creation and annihilation operators inside the SC were denoted $b_{i\sigma}^{\dagger}$ and $b_{i\sigma}$,  respectively.

The third term in Eq. (\ref{Htot}) represents the ferromagnetic insulator which, again, is modeled at the mean field level by the Hamiltonian
\begin{equation}
H_{MI} = -\sum_{\langle i, j\rangle, \sigma} t_{m\sigma} a_{i\sigma}^{\dagger}a_{j\sigma} +\sum_{i, \sigma}\left(\epsilon_{m\sigma} - \sigma\frac{\Gamma}{2}\right)a_{i\sigma}^{\dagger}a_{i\sigma}. \label{Hmi}
\end{equation}
The nearest neighbor hopping is spin dependent and we consider the case $t_{m\downarrow} = -t_{m\uparrow}>0$. The spin-dependent on-site energy was divided into a contribution $\epsilon_{m\sigma} = \epsilon_{m0}+6t_{m\sigma}$ that places the top of the valence band and the bottom of the conduction band at the same energy $\epsilon_{m0}$ and a term proportional to the insulating gap $\Gamma$. The last two terms in Eq. (\ref{Htot}) describe the coupling at the two interfaces, semiconductor-SC and semiconductor-MI, respectively:
\begin{eqnarray}
H_{\tilde{t}_S} &=& -\sum_{\langle i, j\rangle, \sigma} \tilde{t}_S (b_{i\sigma}^{\dagger} c_{j\sigma} + {\rm h.c.})  \label{HtilS}, \\
H_{\tilde{t}_M} &=& -\sum_{\langle i, j\rangle, \sigma} \tilde{t}_M (a_{i\sigma}^{\dagger} c_{j\sigma} + {\rm h.c.})  \label{HtilM}.
\end{eqnarray}
The parameters $\tilde{t}_S$ and $\tilde{t}_M$ characterize the transparencies of the two interfaces and provide the energy scales for the coupling between the semiconductor thin film and the SC and MI, respectively. These energy scales are crucial for determining the strength of the proximity effects.

We diagonalize Eq. (\ref{Htot}) numerically for a system with periodic boundary conditions along the x- and y-directions and a finite size along the z-direction. The semiconductor, SC and MI contain $N$, $N_s$ and $N_m$ planes, respectively. For the semiconductor thin film we consider $5\leq N\leq 20$, while $N_s$ and $N_m$ are typically of the order of several hundreds. For large systems, the proximity effects are independent on $N_s$ and $N_m$ and we explicitly checked that the values used in the calculations are within that regime. The hopping matrix elements for the semiconductor are $t_0=1.57eV$ and
$t_1=0.39eV$, which, for a cubic lattice with $a=5.5$\AA~  generate in the low wavelength limit a small effective mass $m^* = 0.04m_e$ characteristic of semiconductors such as InAs. The value of the Rashba coupling constant is $\alpha = 18$ meV, corresponding to a strong spin-orbit coupling of the order 100 meV \AA. For the SC, the hopping parameter is $t_s=0.35eV$ and $\Delta = 1.5meV$, while the magnetic insulator is described by  $t_{m\downarrow} = -t_{m\uparrow} = 0.36eV$ and a gap value $\Gamma=200 meV$. The coupling parameters $\tilde{t}_S$ and $\tilde{t}_M$ are varied and typically range within several hundreds meV.

\begin{figure}[tbp]
\begin{center}
\includegraphics[width=0.47\textwidth]{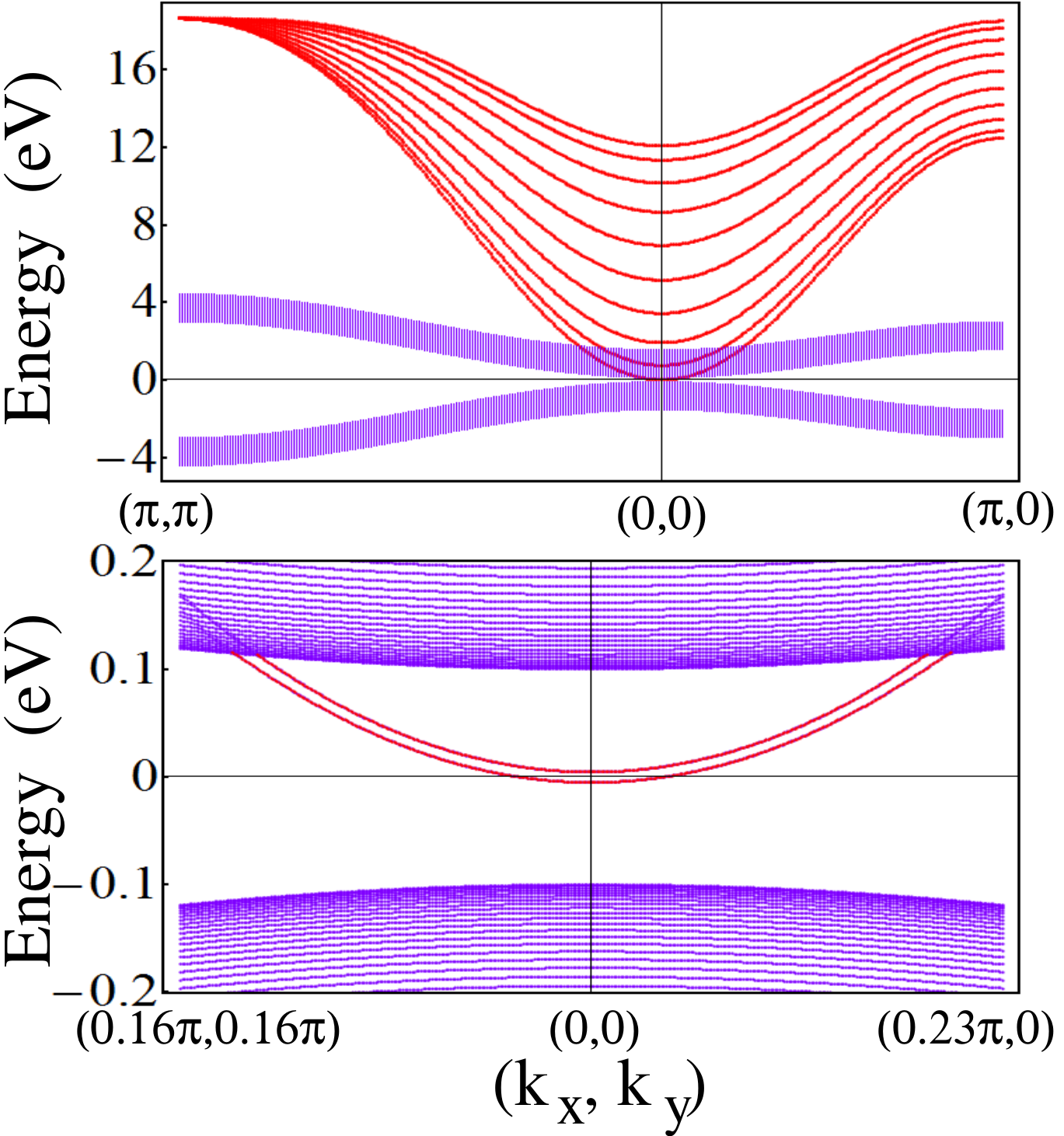}
\end{center}
\caption{(Color online) Band structure for a semiconductor-ferromagnetic insulator heterostructure described by equations (\ref{Htot}), (\ref{H0}), (\ref{Hmi}) and (\ref{HtilM}) with $N=10$ and $\tilde{t}_M=250$meV. The red points represent semiconductor states, while the magenta bands are ferromagnetic insulator states. The lower panel shows the low-energy behavior around ${\mathbf  k}=0$. Notice the spitting of the semiconductor band at ${\mathbf  k}=0$ due to the effective Zeeman term induced by ferromagnetic proximity effect.}
\label{FigSsM1}
\end{figure}

\subsection{The magnetic proximity effect}

First, we turn on the coupling at the interface between the semiconductor and the MI. Fig. \ref{FigSsM1} shows the spectrum of a slab containing  a semiconductor thin film with $N=10$ atomic layers in contact with a MI. The coupling at the interface is $\tilde{t}_m = 250$meV. The semiconductor spectrum is characterized by 10 strongly dispersive bands represented by red curves in the upper panel of Fig. \ref{FigSsM1}. For an isolated semiconductor, these bands are weakly split by the spin-orbit interaction and are double degenerate at ${\mathbf k}=0$. However, as a result of the magnetic proximity effect, an effective Zeeman splitting removes this degeneracy. This is shown in the lower panel of Fig. \ref{FigSsM1} for the lowest energy mode that has the minimum inside the insulating gap. The induced Zeeman splitting
is entirely a result of the exchange interaction between the MI and
the SM layers and is not related to the magnetic field produced by the
ferromagnetic MI. In fact, effective Zeeman splittings have been known
to induced by anti-ferromagnetic insulators as well through a phenomenon
commonly known as exchange bias~\cite{exchangebias}.

\begin{figure}[tbp]
\begin{center}
\includegraphics[scale=0.5,angle=0]{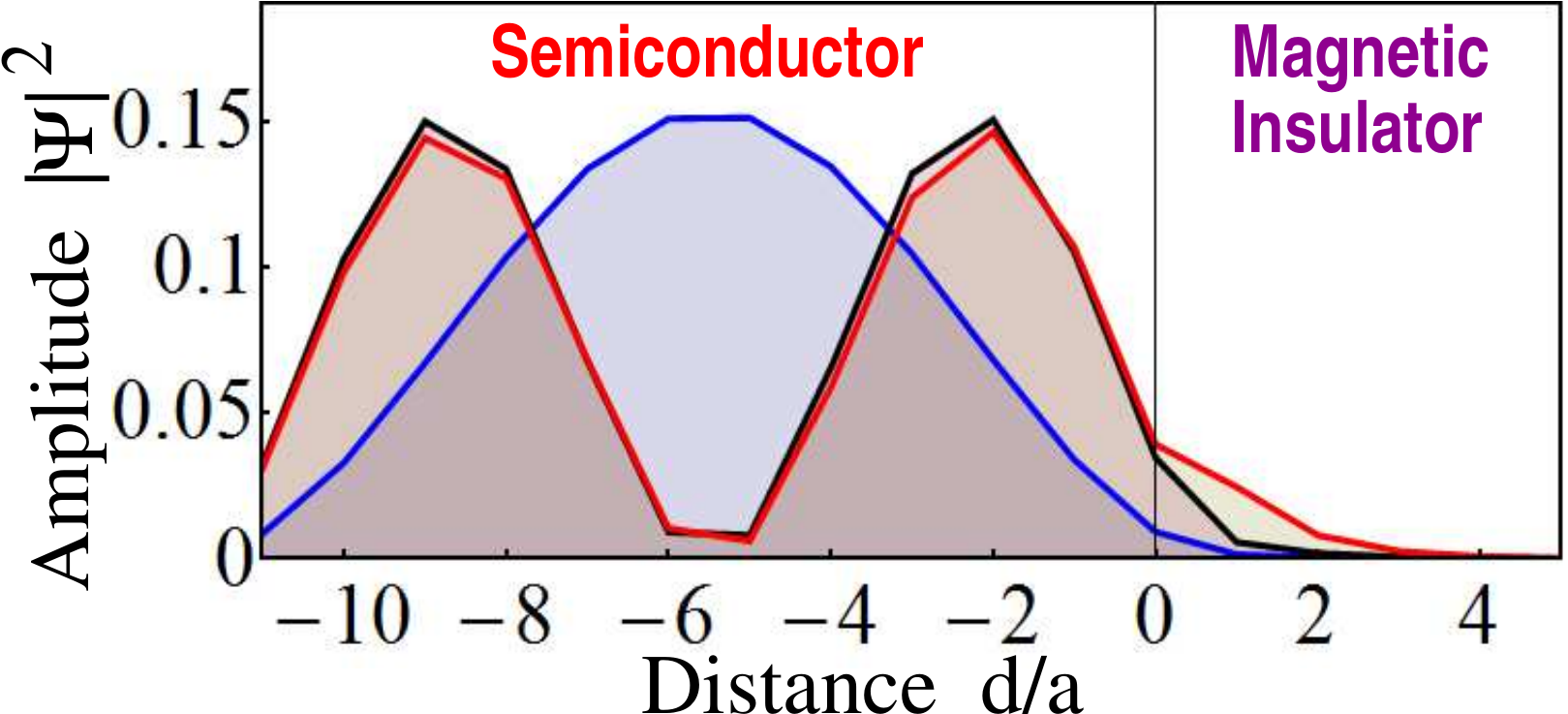}
\end{center}
\caption{(Color online) Dependence of the amplitude of low-energy states on the distance from the semiconductor - MI interface for three different sets of parameters. In order of increasing penetration depth into the MI: lowest semiconductor band ($n=1$) and $\tilde{t}_m=0.25$eV (blue), second semiconductor band ($n=2$) and $\tilde{t}_m=0.25$eV (black), and  $n=2$,  $\tilde{t}_m=0.5$eV (red). For a given set of parameters, a certain fraction of each low-energy wave function penetrates the insulator leading to ferromagnetic correlations in the semiconductor thin film,. i.e., to an effective Zeeman field induced by proximity effect.}
\label{FigSsM2}
\end{figure}

A natural question of practical importance concerns the dependence of the induced Zeeman splitting on the parameters of the model. To get a better intuition of the physics behind the proximity effect, we start with an analysis of the structure of the wave functions of the relevant states. Consider a semiconductor state with an energy within the insulating gap $\Gamma$. When the  semiconductor and the MI are coupled, the wave function describing this state will partly penetrate into the magnetic insulator, where it decays exponentially. These components of the wave functions inside the MI will acquire ferromagnetic correlations equivalent to an effective Zeeman field. Hence, the strength of the proximity effect is related to the fraction of the wave function that penetrates the MI. This qualitative picture of the proximity effect is illustrated in Fig. \ref{FigSsM2}. In general, the fraction of the wave function that penetrates the insulator depends on three types of factors: i) the properties of the quasi-two dimensional parent system, in this case the semiconductor film, ii) the properties of the insulating host system, and iii) the coupling strength at the interface.

The first and last types of factors are qualitatively illustrated in Fig. \ref{FigSsM2}. For example, by controlling the value of the on-site energy $\epsilon_0$, i.e., chemical potential of the semiconductor, one can bring the minima of different semiconductor bands within the insulating gap. The profiles of the wave functions in the transverse direction (i.e., perpendicular to the film) depend on which band is inside the gap, as shown in Fig. \ref{FigSsM2} for the first two bands, $n=1$ and $n=2$. Furthermore, the fraction of the wave function that penetrates the insulator depends on its amplitude at the interface. This amplitude is significantly different for $n=1$ and $n=2$, hence one expects significantly different strengths of the proximity effect. Of course, the amplitude at the interface can be also modified by changing the size of the system (N), or by applying a bias potential $V(z)$ that can tilt the spectral weight toward or away from the interface. In addition to the amplitude at the interface, which is determined by the parameters of the semiconductor, the fraction of the wave function that penetrates the MI also depends on the slope at the interface. In essence, this slope is controlled by the coupling parameter $\tilde{t}_m$, as illustrated in Fig. \ref{FigSsM2}.  Finally, the parameters of the exponential decay inside the insulator, as well as the strength of the ferromagnetic correlations depend on the properties of the insulating host.

\begin{figure}[tbp]
\begin{center}
\includegraphics[width=0.47\textwidth]{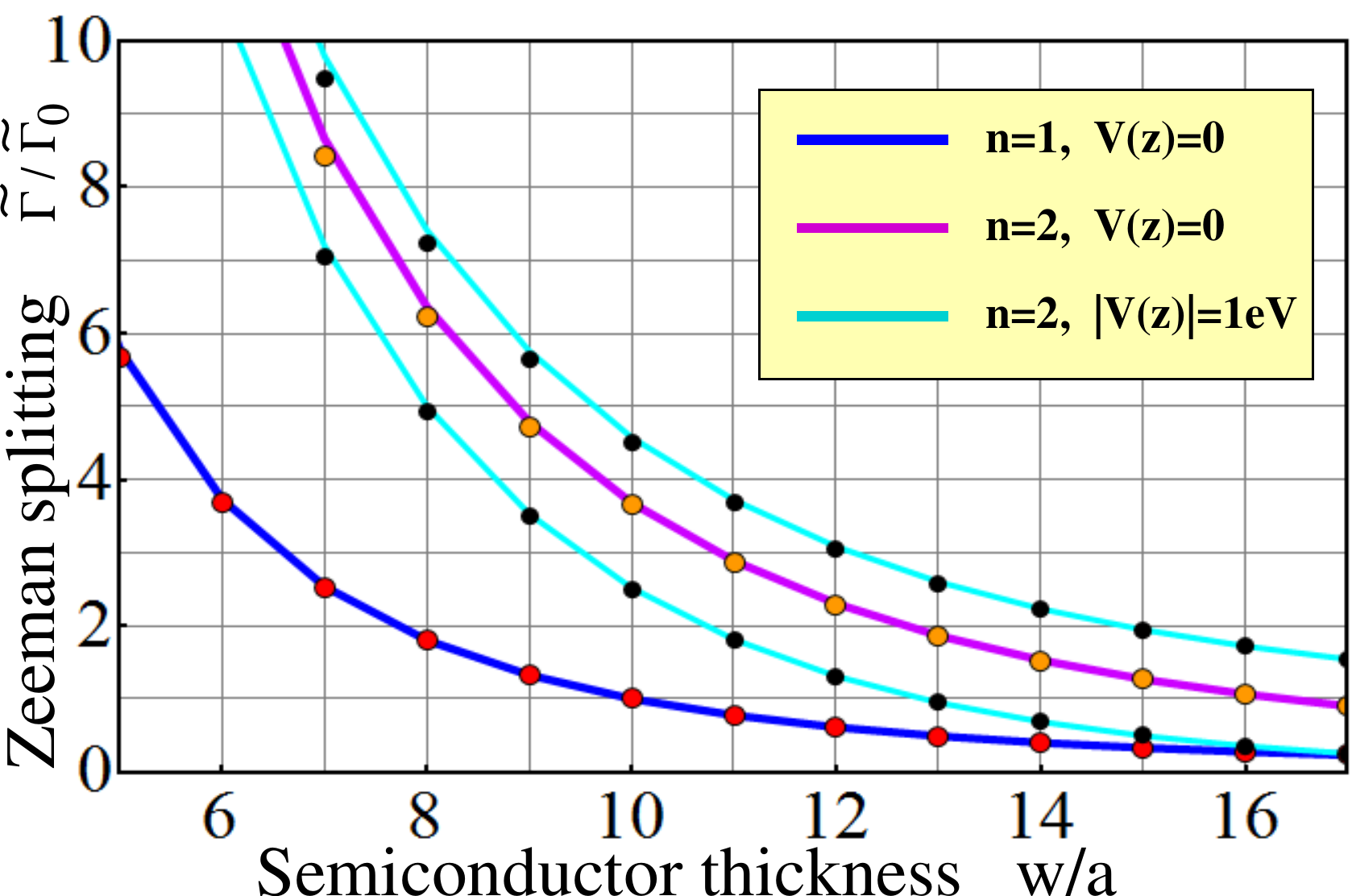}
\end{center}
\caption{(Color online) Dependence of the normalized effective Zeeman splitting on the semiconductor thickness for different sets of parameters (dots). The continuous lines represent the normalized wave function amplitude at the interface (the Nth layer) $\vert\psi_{n}(N)\vert^2/\vert\psi_{n}(10)\vert^2$. The reference gap is $\tilde{\Gamma}_0=2.955$ meV. The wave function amplitude in the absence of an applied bias, $V(z)=0$, is determined using Eq. (\ref{psi2}), while for a linear bias $V(z)=V_0 z/w$ with $V_0=\pm 1$ eV the amplitude is determined numerically.}
\label{FigSsM3}
\end{figure}

Next we proceed with a quantitative analysis of the induced Zeeman splitting  and derive an effective low-energy theory for the ferromagnetic proximity effect. We consider the situation when the states of the nth semiconductor band with wave vectors near ${\mathbf k}=0$ have energies inside the insulating gap. The states $\psi_{n{\mathbf k}\sigma}({\mathbf r}_i)$ corresponding to vanishing spin-orbit coupling, $\alpha=0$, form a convenient basis for the low-energy Hilbert subspace of interest. Projecting the Hamiltonian $H_0$ onto this subspace we obtain
\begin{equation}
H_{0{\rm eff}}({\mathbf k}) = \left(
\begin{array}{cc}
\frac{k^2}{2m^*} & \alpha(k_y-ik_x) \\
\alpha(k_y+ik_x) & \frac{k^2}{2m^*}
\end{array}
\right),  \label{H0eff}
\end{equation}
where $m^*$ is an effective mass with a value slightly different from the effective mass corresponding to the 3D semiconductor model. The difference stems from the quasi 2D geometry of the system. The insulating degrees of freedom can be integrated out and replaced by an interface self-energy. When projected onto the low-energy subspace, this contribution becomes
\begin{equation}
\Sigma_{\sigma\sigma^{\prime}}({\mathbf k}, \omega) = -\tilde{t}_{m}^2 \vert\psi_{n\mathbf k}(z_0)\vert^2 G_{\sigma\sigma^{\prime}}^{(m)}({\mathbf k}, \omega; z_0+\delta_z), \label{Sigmam}
\end{equation}
where $G_{\sigma\sigma^{\prime}}^{(m)}({\mathbf k}, \omega; z_0+\delta_z)$ is the Green function of the magnetic insulator and $\vert\psi_{\mathbf k}(z_0)\vert^2$ the amplitude of the semiconductor wave function, both at the interface. From Eq. (\ref{Sigmam}) one immediately notice that, neglecting dynamical effects, i.e., setting $\omega=0$, any induced Zeeman splitting has to be proportional with the amplitude of the wave function at the interface times the square of the interface transparency. For a vanishing bias potential, $V(z)=0$, the wave function amplitude on the j atomic layer is proportional to $\sin^2[\pi n j/(N+1)]$ and the amplitude at the interface becomes
\begin{equation}
\vert\psi_{n\mathbf k}(z_0)\vert^2 = \frac{2\sin^2\left(\frac{n \pi}{N+1}\right)}{N+1}. \label{psi2}
\end{equation}
Note that, for clarity, we dropped the spin label, as the amplitude is spin independent. Also note that for large N the amplitude becomes $\vert\psi_{n\mathbf k}(z_0)\vert^2 \approx 2n^2\pi^2/(N+1)^3$.
The dependence of the induced Zeeman splitting on the wave function amplitude is shown in Fig. \ref{FigSsM3}.
\begin{figure}[tbp]
\begin{center}
\includegraphics[width=0.47\textwidth]{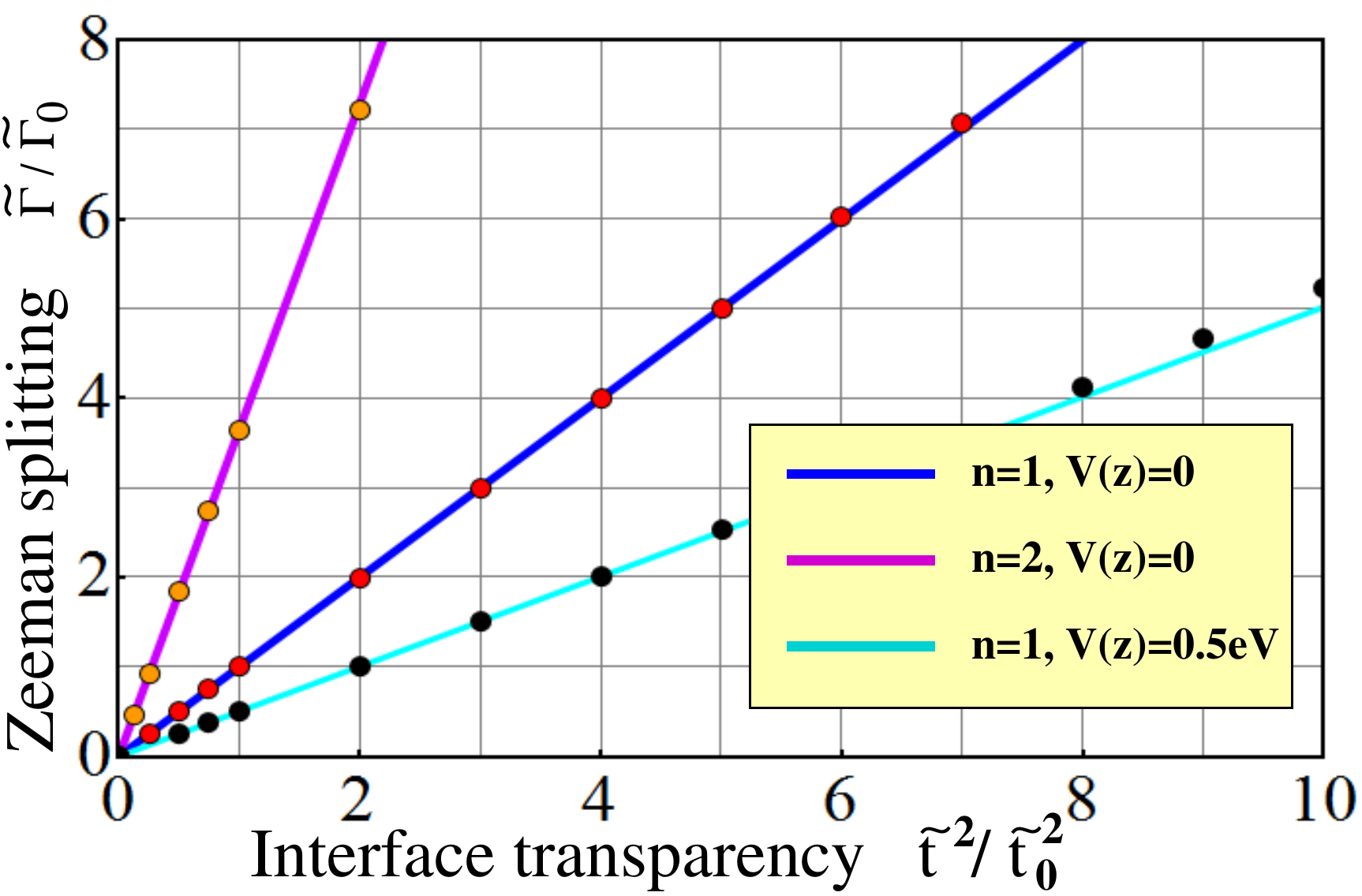}
\end{center}
\caption{(Color online) Dependence of the normalized Zeeman splitting on the interface transparency for a system with $N=12$ and different model parameters (dots). The straight lines are guides for the eye. The reference interface coupling is $\tilde{t}_{m0}=250$ meV. The rather small deviations from a linear dependence indicate that dynamical effects are negligible, i.e., neglecting the frequency dependence in Eq. (\ref{Sigmam}) is a good approximation.}
\label{FigSsM4}
\end{figure}
As the film thickness $w = N a$ is increased, the value of the of the wave function amplitude at the interface drops rapidly, as expected from Eq. (\ref{psi2}). Due to the proximity effect, an effective Zeeman splitting creates a gap in the nth semiconductor band at ${\mathbf k}=0$,
\begin{equation}
\tilde{\Gamma} =  E_{n2}(0) - E_{n2}(0),        \label{tildeG}
\end{equation}
where $E_{nj}({\mathbf k})$ are the energies of the nth semiconductor band within the magnetic insulator gap (see lower panel of Fig. \ref{FigSsM1}).
The proportionality between the induced Zeeman gap $\tilde{\Gamma}$ (dots in
Fig. \ref{FigSsM3}) and the amplitude of the wave function (lines in
Fig. \ref{FigSsM3}) reveal the absence of significant dynamical effects Eq. (\ref{Sigmam}). This conclusion is further supported by the linear dependence  $\tilde{\Gamma}$ on the square of the interface transparency, $\tilde{t}_m^2$, shown in Fig. \ref{FigSsM4}. We emphasize that the effective coupling constant that determines the strength of the ferromagnetic proximity effect, $g_m = 2\tilde{t}_m^2 \vert\psi_{n}(z_0)\vert^2 / \Lambda_m$, where $\Lambda_m$ is a characteristic bandwidth for the magnetic insulator, can be tuned by: a) modifying the semiconductor film thickness (see Fig. \ref{FigSsM3}),  b) applying a bias voltage (see Figs. \ref{FigSsM3} and \ref{FigSsM4}), and c) changing the semiconductor - MI coupling (Fig. \ref{FigSsM4}).

\begin{figure}[tbp]
\begin{center}
\includegraphics[width=0.47\textwidth]{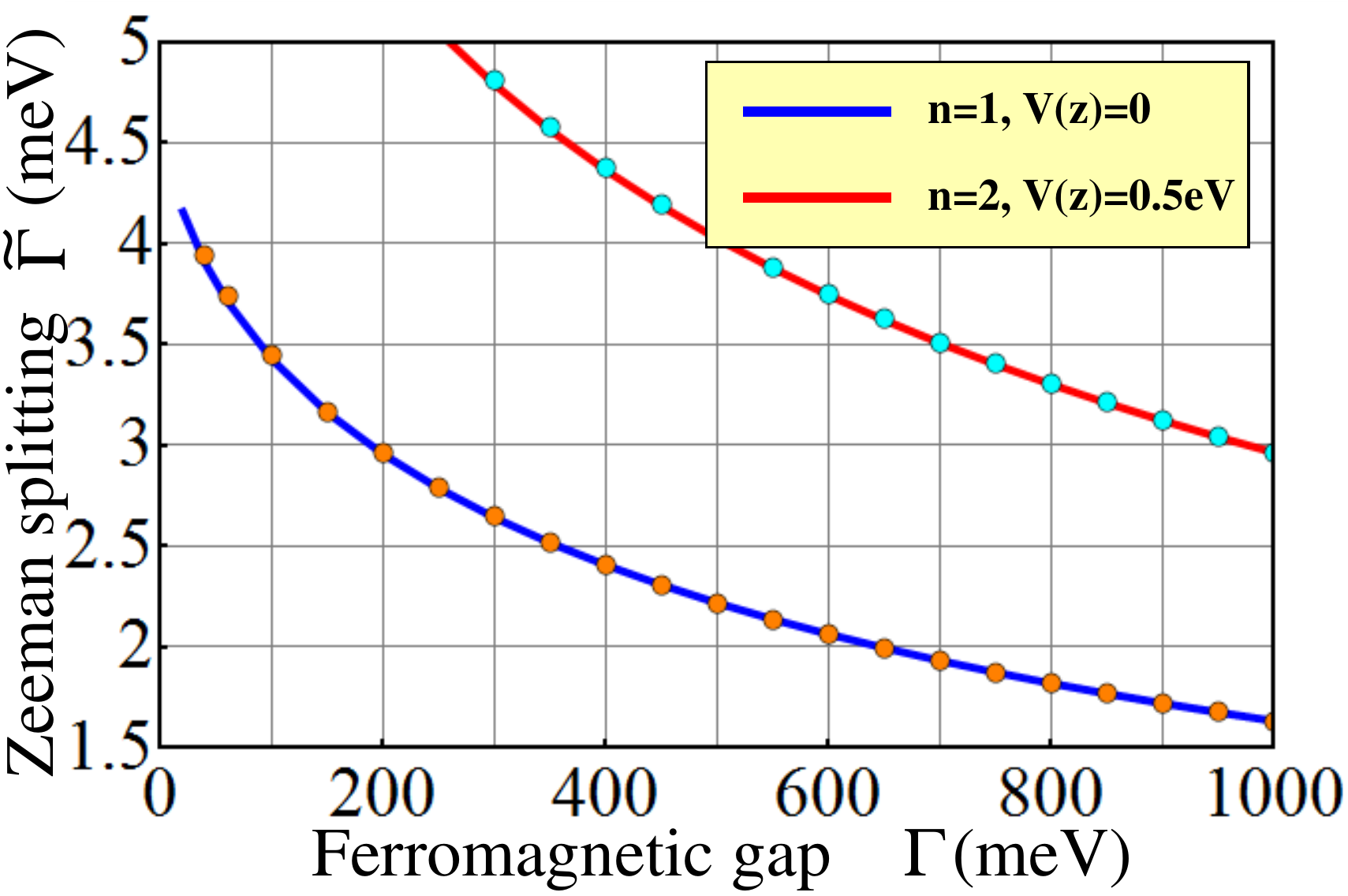}
\end{center}
\caption{(Color online) Dependence of the effective Zeeman splitting on the ferromagnetic insulator gap for two sets of semiconductor parameters (dots). The continuous lines are calculated using the effective low-energy theory described by Eq. (\ref{DelGtil}). Notice the remarkable agreement between the low-energy theory and the exact results and the difference in energy scales between the insulating gap and the induced gap.}
\label{FigSsM5}
\end{figure}

So far we have discussed the dependence of the proximity effect on the properties of the quasi-two dimensional parent system and on the coupling strength at the interface. Next we briefly investigate the role of the  insulating host system. We note that a full analysis would require a treatment of the magnetic insulator beyond the simple mean-field picture used here, but this would not change the results obtained so far. Within our mean-field approximation, the MI Green function from Eq. (\ref{Sigmam}) can be easily evaluated,
\begin{equation}
G_{\sigma\sigma^{\prime}}^{(m)}({\mathbf k}, \omega; z_0+\delta_z) = \delta_{\sigma\sigma^{\prime}}\sum_{\nu}\frac{1}{\omega-E_{\nu\sigma}({\mathbf k}) + i\eta} \vert\chi_{\nu {\mathbf k} \sigma}(z_0+\delta_z)\vert^2, \label{Gm}
\end{equation}
where $E_{\nu\sigma}({\mathbf k})$ are the energies of the insulator bands and $\chi_{\nu {\mathbf k} \sigma}(z_0+\delta_z$ the values of the  corresponding eigenstates at the interface. It is convenient to express the Green function in terms of the partial density of states $\rho_{{\mathbf k} \sigma}(\omega) = \sum_{\nu}\delta(\omega- E_{\nu\sigma}({\mathbf k}))\vert\chi_{\nu {\mathbf k} \sigma}(z_0+\delta_z)\vert^2$. Within our model, this partial density of states becomes
\begin{equation}
\rho_{{\mathbf k} \sigma}(\omega) = \frac{2}{\pi \Lambda_{\sigma}}\sqrt{1-\left(\frac{\omega-\bar{E}_{\sigma}({\mathbf k})}
{\Lambda_{\sigma}}\right)^2},  \label{rho}
\end{equation}
with $\Lambda_{\sigma} = 2\vert t_{m\sigma}\vert$ being half of the bandwidth in the insulating phase and $\bar{E}_{\sigma}({\mathbf k})$ the energy values at the middle of the valence and conduction bands for a given  wave vector parallel to the interface. At ${\mathbf k}=0$ we have $\bar{E}_{\sigma}(0)= -\mu-\sigma(\Gamma/2 +\Lambda_{\sigma})$. Note that $\rho_{{\mathbf k} \sigma}(\omega)$ vanishes for values of $\omega$ outside the bandwidth. Using Eq. (\ref{rho}), we obtain for the Green function the expression
\begin{equation}
G_{\sigma\sigma}^{(m)} = \frac{2}{\Lambda_{\sigma}}\left[\frac{\omega-\bar{E}_{\sigma}}{\Lambda_{\sigma}} -{\rm sign}(\omega-\bar{E}_{\sigma})\sqrt{\left( \frac{\omega-\bar{E}_{\sigma}}{\Lambda_{\sigma}}\right)^2-1} \right]. \label{Gss}
\end{equation}

\begin{figure}[tbp]
\begin{center}
\includegraphics[width=0.47\textwidth]{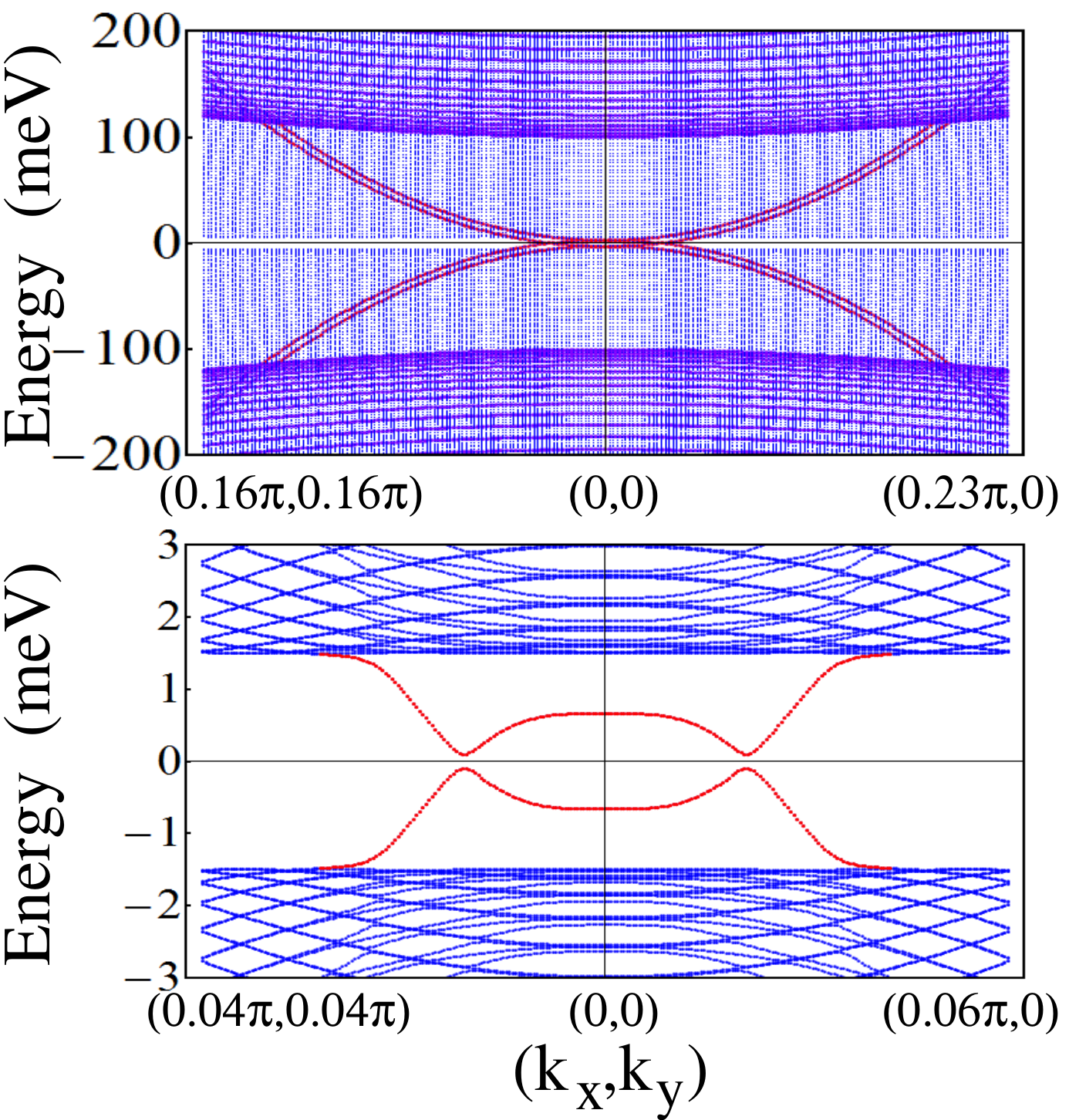}
\end{center}
\caption{(Color online) BdG spectrum of the full Hamiltonian (\ref{Htot}) for a semiconductor thin film with $N=10$ sandwiched between a MI ($\Gamma=0.2$eV, $\tilde{t}_m=250$meV, $\tilde{\Gamma}=2.95$meV) and an s-wave SC ($\Delta=1.5$meV, $\tilde{t}_s=130$meV). The red points represent states that reside (mostly) within the semiconductor, magenta designates the MI bands, while the SC states are blue. Due to the effective Zeeman filed, the semiconductor bands are split and only one crosses the chemical potential (see upper panel). As a result of the superconductor proximity effect, a small gap opens at the crossing points (see lower panel).}
\label{FigSsM6}
\end{figure}

\noindent Note that the imaginary part of the Green function vanishes for values of $\omega$ within the insulating gap. Also, because the energies of interest are much smaller than the insulator bandwidth, $\omega\ll\Lambda_{\sigma}$, we can neglect the frequency dependence in Eq. (\ref{Gss}). Finally, within the static approximation, we obtain for the induced Zeeman splitting the expression
\begin{eqnarray}
\tilde{\Gamma}&=&\Delta\tilde{\Gamma}_{\downarrow} - \Delta\tilde{\Gamma}_{\uparrow}, \label{DelGtil} \\
\Delta\tilde{\Gamma}_{\sigma} &=& g_{m\sigma}
\left[\frac{-\mu-\sigma(\frac{\Gamma}{2} +\Lambda_{\sigma})}{\Lambda_{\sigma}}+\sigma\sqrt{\left( \frac{-\mu-\sigma(\frac{\Gamma}{2} +\Lambda_{\sigma})}{\Lambda_{\sigma}}\right)^2-1} \right], \nonumber
\end{eqnarray}
where the effective coupling constant is $g_m = 2\tilde{t}_m^2 \vert\psi_{n}(z_0)\vert^2 / \Lambda_{\sigma}$. To test the accuracy of this effective low-energy theory, we compare the values of the induced Zeeman splitting predicted by Eq. (\ref{DelGtil}) with the numerical calculations. The results for various values of the insulating gap shown excellent agreement (see Fig. \ref{FigSsM5}). Finally, we note that away from ${\mathbf k}=0$ the dispersion of the low-energy bands $E_{nj}({\mathbf k})$ can be obtained by adding the self-energy contribution (\ref{Sigmam}) to the effective theory described by Eq. (\ref{H0eff}). Within the static approximation we have
\begin{equation}
H_{{\rm eff}}({\mathbf k}) = \left(
\begin{array}{cc}
\frac{k^2}{2m^*}-\frac{\tilde{\Gamma}}{2} & \alpha(k_y-ik_x) \\
\alpha(k_y+ik_x) & \frac{k^2}{2m^*}+ \frac{\tilde{\Gamma}}{2}
\end{array}
\right),  \label{Heff}
\end{equation}
with $\tilde{\Gamma}$ given by Eq. (\ref{DelGtil}). Explicit calculations for various sets of parameters show that this low-energy theory represents an excellent approximation for all ${\mathbf k}$ values of interest.

\begin{figure}[tbp]
\begin{center}
\includegraphics[width=0.47\textwidth]{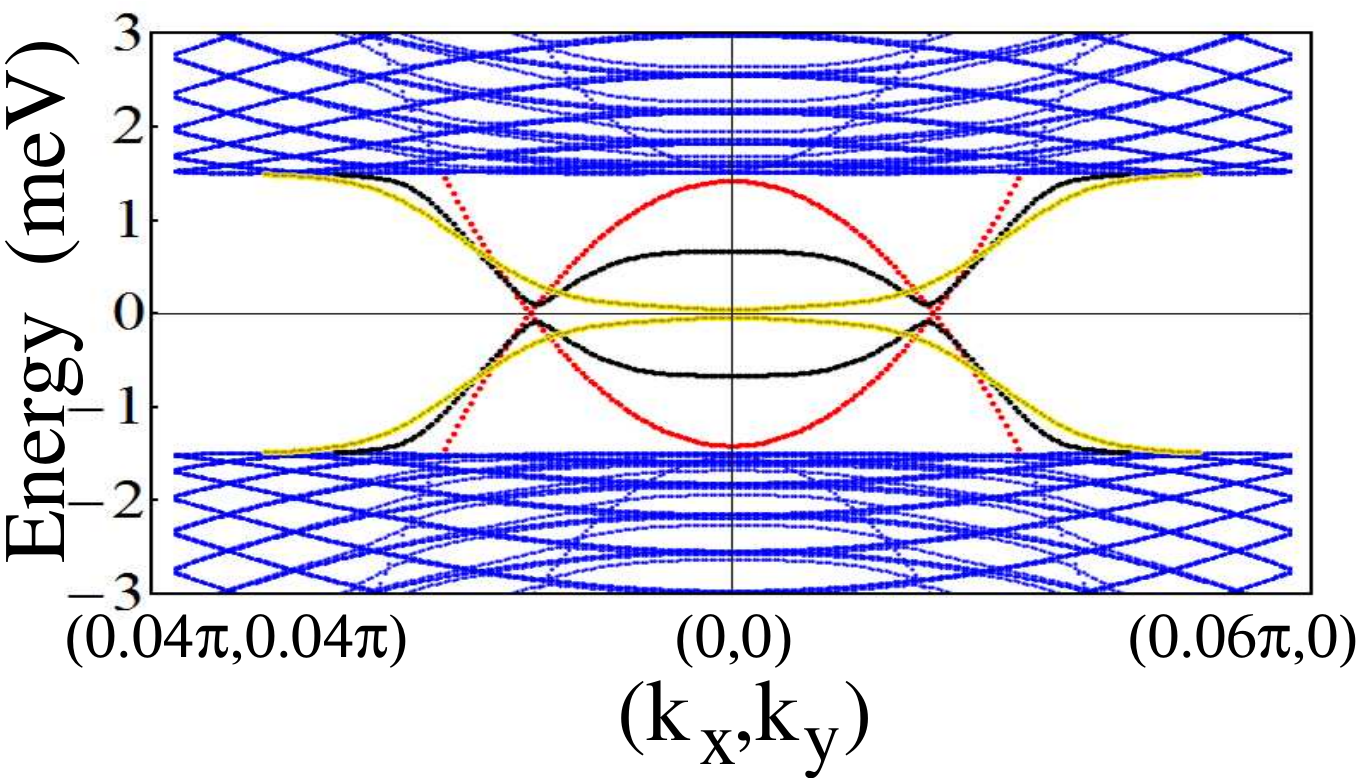}
\end{center}
\caption{(Color online) Dependence of the low-energy spectrum on the coupling between the semiconductor and the SC: $\tilde{t}_s=0$ (red), $\tilde{t}_s=130$meV (black), $\tilde{t}_s=190$ (yellow). A proximity effect - induced gap opens at finite ${\mathbf k}$. As $\tilde{t}_s$ is increased, the minimal gap shifts to lower wave vectors and first increases, then decreases and eventually vanishes at a critical coupling before opening again (see also Fig. \ref{FigSsM8}). }
\label{FigSsM7}
\end{figure}

\subsection{Superconducting proximity effect}

Next, we turn our attention to the effects induced by the proximity of an s-wave superconductor on the semiconductor-MI heterostructure. The parameters of the semiconductor-MI interface are fixed, with $\tilde{t}_m=250$ meV. We consider a semiconductor thin film with $N=10$ and create a new interface at the free surface of the semiconductor by coupling it to a SC with an s-wave gap $\Delta$, i.e., we add the terms given by equations (\ref{Hsc}) and (\ref{HtilS}) to the total Hamiltonian (\ref{Htot}). The corresponding spectrum is shown in Fig. \ref{FigSsM6}. The semiconductor band is split due to the ferromagnetic proximity effect and the chemical potential is tuned so that it  crosses only the lower energy mode. When the coupling to the SC is turned on, a small gap opens at low energies due to the SC proximity effect (see Fig. \ref{FigSsM6} lower panel).

 Before a quantitative analysis, let us illustrate qualitatively the behavior of the proximity-induced SC gap. Fig. \ref{FigSsM7} shows the low-energy spectrum for three different values of the coupling constant $\tilde{t}_s$. For $\tilde{t}_s=0$ (red line) the BdG spectrum is gapless. A non-vanishing interface coupling opens a small gap at the crossing points. The value of this finite wave vector gap increases with $\tilde{t}_s$, but the gap at ${\mathbf k}=0$ decreases. (black line). Eventually, at a critical value $\tilde{t}_{sc}$ the gap vanishes at ${\mathbf k}=0$, before opening again for larger couplings (yellow curve). This closing of the induced gap signals the presence of a quantum phase transition~\cite{Sau}.

\begin{figure}[tbp]
\begin{center}
\includegraphics[width=0.47\textwidth]{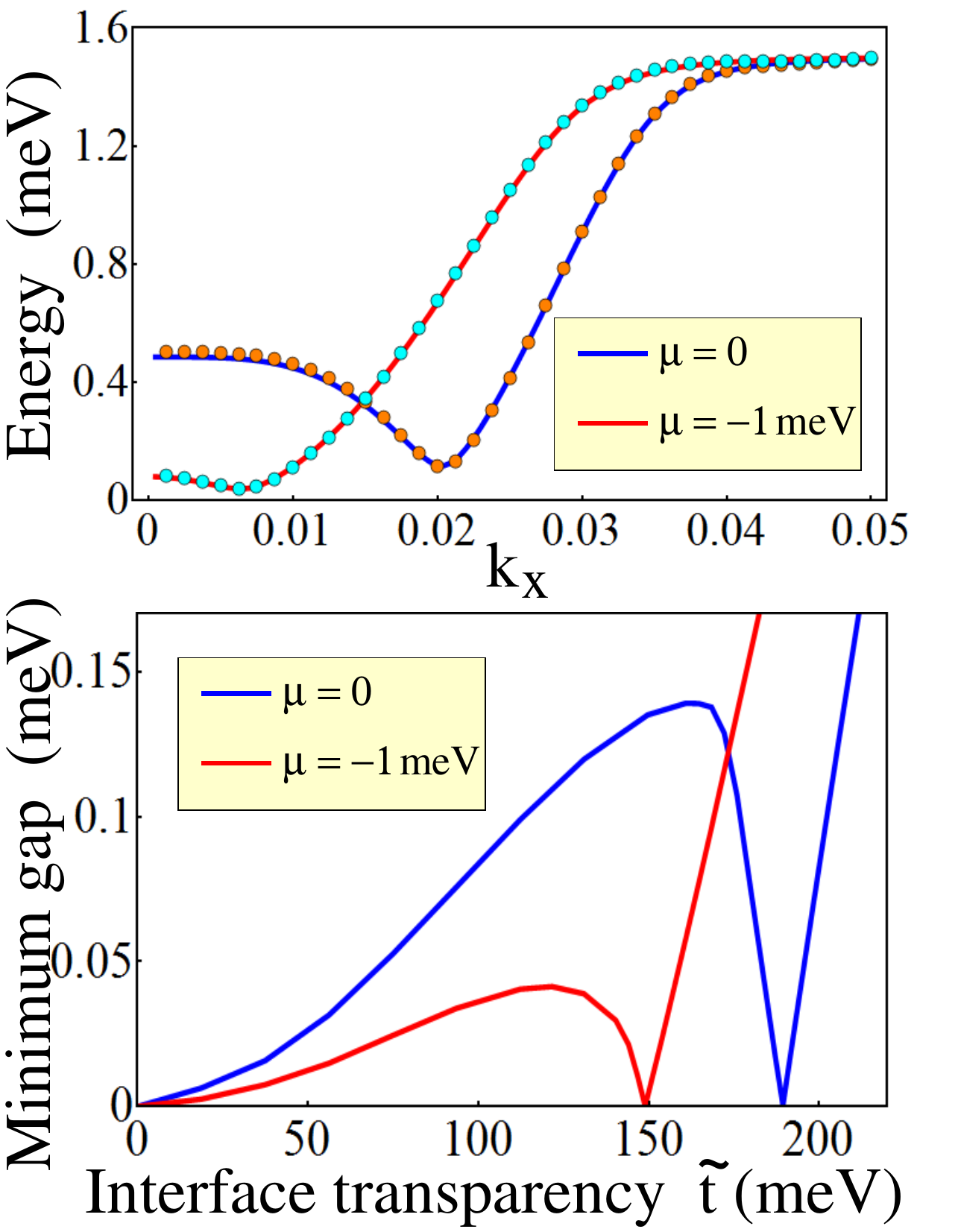}
\end{center}
\caption{(Color online) Upper panel: Comparison between the solution of the effective low-energy theory given by Eq. (\ref{effBdG}) (lines) and the numerical solution of Hamiltonian (\ref{Htot}) (dots) for two values of the chemical potential. Only positive energies are shown. Lower panel: Dependence of the induced minimum gap on the interface transparency. The vanishing of the gap at a critical value $\tilde{t}_{sc}(\mu)$  reflects a quantum phase transition between a topologically non-trivial SC (at small values of $\tilde{t}_{s}$) and a trivial s-wave SC (at large couplings) [reference]. }
\label{FigSsM8}
\end{figure}

In order to understand this behavior, it is useful to develop an effective low-energy theory for the SC proximity effect. As shown previously,~\cite{robustness,SSLdS} dynamical corrections are crucial in capturing the low-energy physics in this case, in contrast to the ferromagnetic proximity effect. Using the results obtained in Refs. [\onlinecite{SSLdS,robustness}], the Green function describing the low-energy physics of the heterostructure can be written as
\begin{equation}
-G^{-1}=(\xi_k+\lambda_k \sigma_++\lambda_k^*\sigma_-)\tau_z+\frac{\tilde{\Gamma}}{2}\sigma_z+\frac{g_s\Delta}{\sqrt{\Delta^2-\omega^2}}\tau_x- \omega\left(1+\frac{g_s}{\sqrt{\Delta^2-\omega^2}}\right)
\end{equation}
where $\xi_k = k^2/2m^* -\mu$, $\tilde{\Gamma}$ is given by Eq. (\ref{DelGtil}), $\lambda_k = \alpha(k_y-i kx)$, and the effective coupling is $g_s = 2\tilde{t}_s^2\vert\psi(z_s)\vert^2/\Lambda_s$. The wave function amplitude is evaluated at the semiconductor - SC interface and $\Lambda_s= 2t_s$ is half of the SC bandwidth. The low-energy spectrum can be obtained by solving the corresponding BdG equation, ${\rm Det}(G^{-1})=0$. Explicitly, we have
\begin{eqnarray}
&&\omega^2\left(1+\frac{g_s}{\sqrt{\Delta^2-\omega^2}}\right)^2 = \frac{\tilde{\Gamma}^2}{4}+\xi^2+\vert\lambda_k\vert^2+ \frac{g_s^2\Delta^2}{\Delta^2-\omega^2} \nonumber \\
&-& 2\sqrt{\xi_k^2\left(\frac{\tilde{\Gamma}^2}{4}+\vert\lambda_k\vert^2\right) + \frac{\tilde{\Gamma}^2}{4}\frac{g_s^2\Delta^2}{\Delta^2-\omega^2} }. \label{effBdG}
\end{eqnarray}
A comparison between the solution of Eq. (\ref{effBdG}) and the numerical calculations is shown in the upper panel of Fig. \ref{FigSsM8}. The good agreement between the two calculations indicates that all the relevant ingredients have been incorporated into the effective low-energy theory. The dependence of the minimum gap on the interface coupling is shown in the lower
panel of Fig. \ref{FigSsM8}. Note that the critical value of $\tilde{t}_{s}$ at which the gap vanishes can be obtained from Eq. (\ref{effBdG}) by setting $\omega=0$. Explicitly we have
\begin{equation}
\tilde{t}_{sc}(\mu) = \frac{\sqrt{t_s}}{\vert\psi(z_s)\vert}\left(\frac{\tilde{\Gamma}^2}{4} -\mu^2 \right)^{\frac{1}{4}}.
\end{equation}

\begin{figure}[tbp]
\begin{center}
\includegraphics[width=0.47\textwidth]{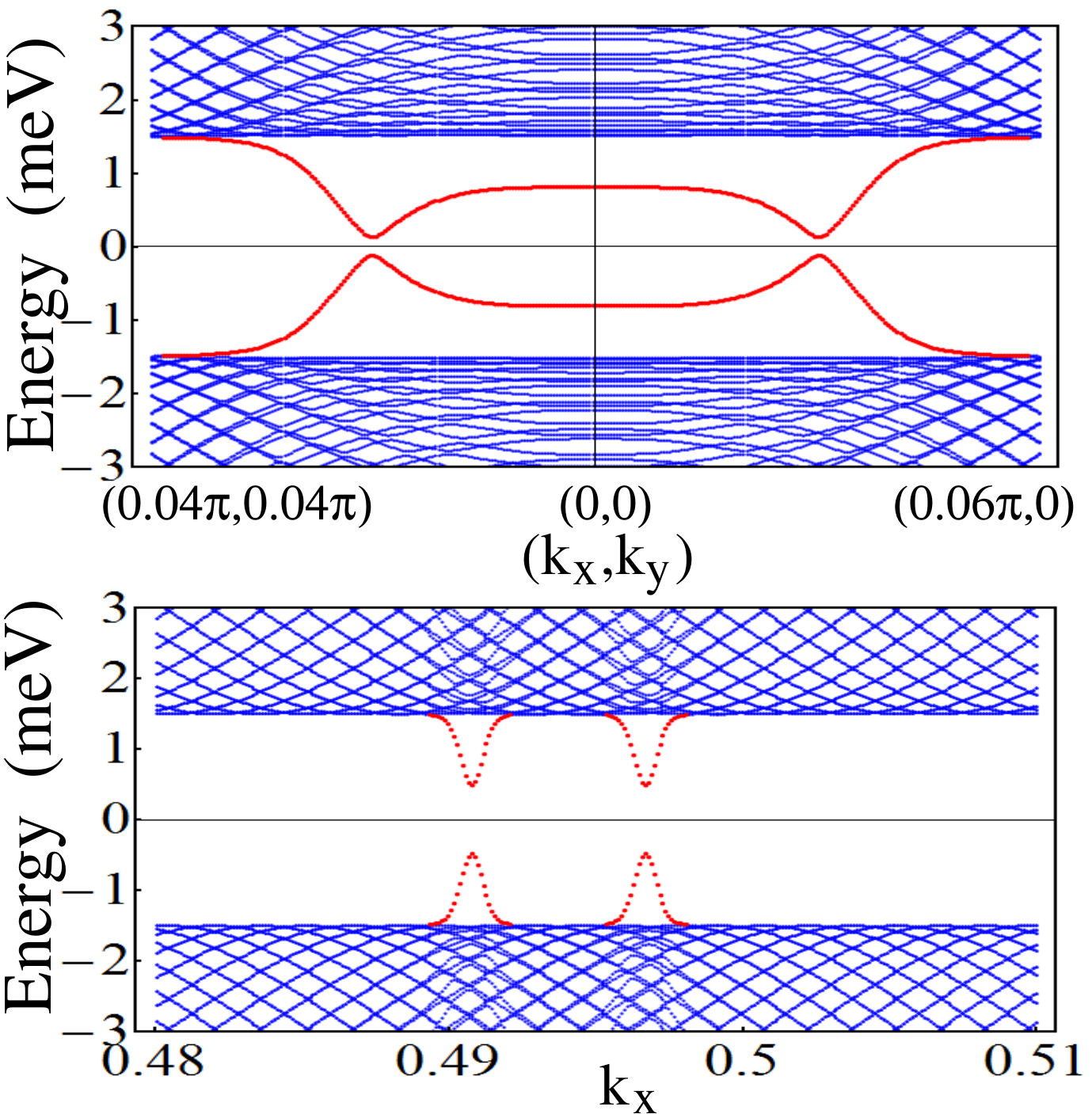}
\end{center}
\caption{(Color online) Details of the low-energy spectrum for the case when the $n=2$ semiconductor band has the minimum in the vicinity of the chemical potential. The low-energy physics in the vicinity of ${\mathbf k}=0$ (upper panel) is described by the effective theory given by Eq. (\ref{effBdG}) with a coupling constant $g_s$ that includes the amplitude of the $n=2$ state at the interface. The $n=1$ semiconductor bands cross the chemical potential at larger values of ${\mathbf k}$ (lower panel) and are gaped due SC proximity effect. Note that the gap at large wave vectors does not vanish at a critical coupling.}
\label{FigSsM9}
\end{figure}

The last question that we address in this section concerns the situation when a higher energy semiconductor band, $n>1$, has the minimum in the vicinity of the chemical potential. For concreteness, we consider the case $n=2$. The physics in the vicinity of ${\mathbf k}=0$ is similar with the case studied above. In addition, the $n=1$ bands cross the chemical potential at some large value of ${\mathbf k}$. Nonetheless, assuming that the partial density of states of the superconducting metal $\rho_{{\mathbf k}\sigma}(\omega)$ does not vary significantly with the wave vector, i.e., the effective mass of the metal is much grater than the effective mass of the semiconductor, the $n=1$ bands will be gaped and the induced gap is typically grater than the minimum gap near ${\mathbf k}=0$. Hence, the general conclusions of our analysis of the $n=1$ case hold for $n>1$. To illustrate this point, we show in Fig. \ref{FigSsM9} the relevant details of the low-energy spectrum for the case $n=2$.

\section{Majorana fermion modes in one dimensional nanowire.}
\begin{figure}[tbp]
\begin{center}
\includegraphics[width=0.47\textwidth]{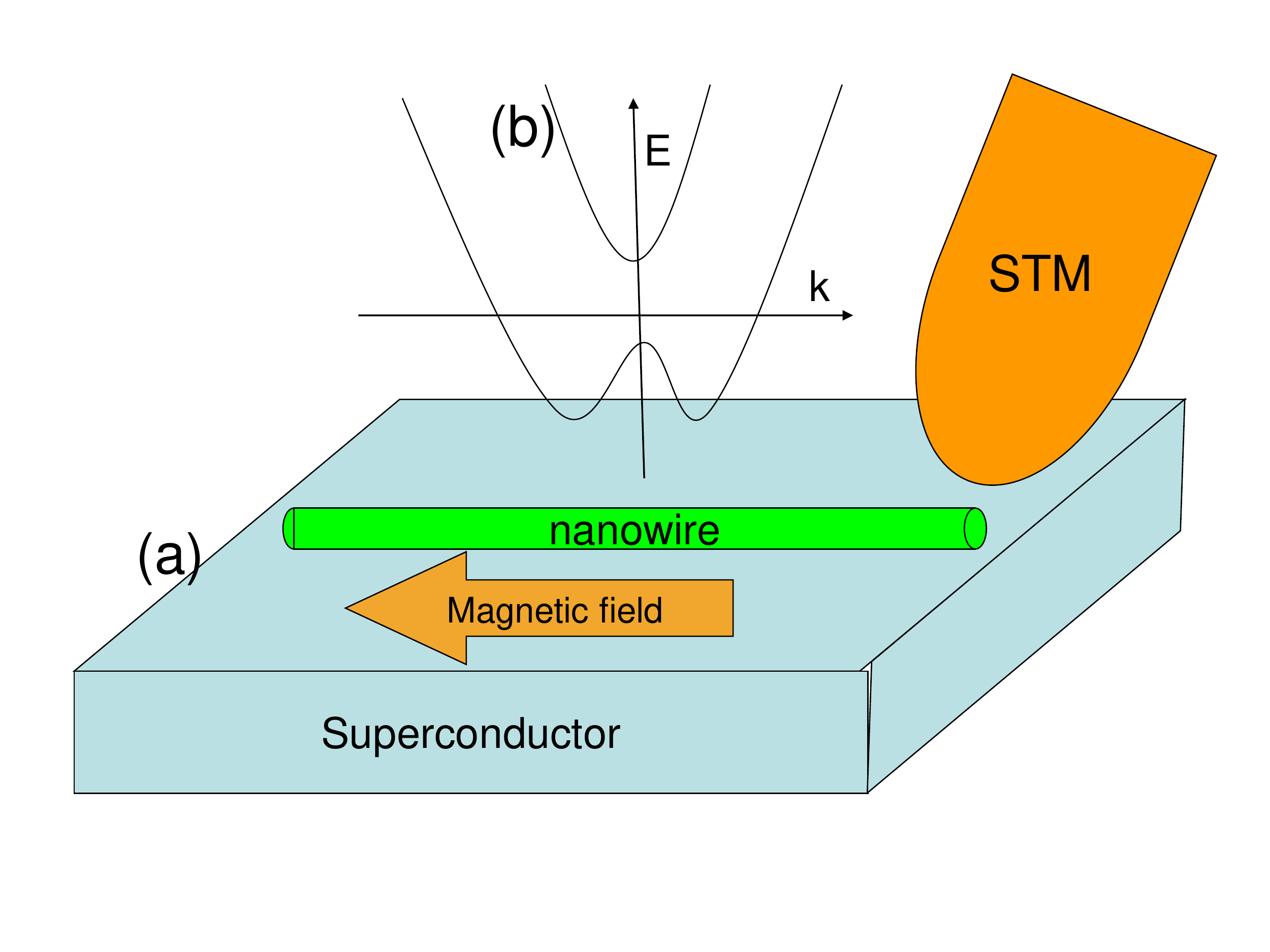}
\end{center}
\caption{(a): Geometry to detect zero-energy Majorana fermions using
STM spectroscopy on a semiconducting nanowire.  The Zeeman splitting is induced by a parallel magnetic field,
while the chemical potential is controlled by an external gate (not shown). The Majorana fermion
 mode localized at the end of the nanowire gives rise to a
zero-bias peak in the STM tunneling spectrum from the end. The
tunneling spectrum from the bulk of the nanowire is gapped.
  (b): Band-structure of the nanowire in the topological
 superconducting phase.}
\label{fig:nanowire}
\end{figure}

In the previous sections, we discussed how Majorana states may appear
at vortices and edges of various two dimensional spin-orbit
coupled semiconductor heterostructures. In this section, we show that
 Majorana fermions can also be realized in the much simpler
 one-dimensional nanowire set up (Fig. \ref{fig:nanowire}).
 In this set-up we propose to
study a semiconducting nanowire with a sizeable spin-orbit coupling
( \textit{e.g.}, InAs) placed on an $s$-wave superconductor
 (\textit{e.g.}, Al). An in-plane magnetic field is used to create a
 Zeeman splitting in the nanowire, \cite{roman, Gil} which gives rise to
  the band-structure shown in
 Fig. \ref{fig:nanowire}(b). The direction of the magnetic field
 (parallel or perpendicular to the
 length of the wire)  that is required to open a gap in the nanowire
 depends on the exact direction of the spin-orbit coupling in the wire.
 The chemical potential in the wire $\mu$ is assumed to be controlled
 by external gate voltages.

We argue below that, for the Zeeman splitting satisfying
 $V_Z>\sqrt{\mu^2+\Delta^2}$, a single non-degenerate
 zero energy state exists at the end of the wire as the only
 low-energy bound state. The second quantized operator for this
state is again a Majorana fermion operator. This mode can
be detected as a zero-bias conductance peak in the STM tunneling
 spectrum.  On reducing the Zeeman splitting so that
$V_Z<\sqrt{\mu^2+\Delta^2}$, the Majorana mode, and hence the
zero-bias conductance peak, should disappear from the tunneling
spectrum.

The BdG Hamiltonian for a one-dimensional single band semiconductor
with spin-orbit coupling (which is linear in the momentum) in proximity to
an $s$-wave superconductor (Fig. ~\ref{fig:nanowire}), can be
written as
\begin{equation}
H_{BdG}=(-\eta \partial_y^2-\mu(y))\tau_z + V_z\bm\sigma\cdot \hat{\bm B}+\imath\alpha\partial_y\hat{\bm \rho}\cdot\bm \sigma\tau_z+\Delta\cos{\phi}\tau_x+\Delta\sin{\phi}\tau_y.\label{eq:1DBdG}
\end{equation}
Here the unit vector $\hat{\bm B}$ gives the direction of the effective
 Zeeman field and the unit vector $\hat{\bm \rho}$ characterizes the
spin-orbit coupling.
From inspection it is now clear that this  Hamiltonian is formally identical to the Hamiltonian
 for the edge states, Eq.~\ref{eq:BdGedge}, at $k_x=0$. Therefore,
from the solution of the chiral edge state at $k_x=0$
for the semiconductor thin film (Sec. X B),
 we conclude that in the topological phase of the wire $(C_0<0)$, the ends
of the nanowire support localized zero-energy Majorana states.

\subsection{STM detection of Majorana end modes in the nanowire.}
In the previous section, we showed that the end of a nanowire in the
topological phase is characterized by a Majorana mode. The Majorana
modes at the ends of a one dimensional $p$-wave superconductor have been
shown to lead to distinct signatures in the STM tunneling spectrum.~\cite{sengupta,bolech}
 In what follows, we analytically and numerically
calculate the STM conductance spectrum from one end of our semiconducting
nanowire. We find that, for the Zeeman coupling satisfying $(V_Z>\sqrt{\Delta^2+\mu^2})$, this spectrum has a zero-bias conductance peak.
 The zero-bias peak disappears as the Zeeman splitting is reduced to
satisfy $(V_Z<\sqrt{\Delta^2+\mu^2})$.

The tunneling spectrum of a superconducting system, similar to the tunneling conductance of normal systems, can be calculated using the Keldysh
 formalism of non-equilibrium Green functions~\cite{bolech} where the superconducting system is coupled to a tip which is initially in thermal equilibrium at a chemical potential $\mu_t$, by an adiabatically increasing
tunneling amplitude perturbation $V$. To
 calculate the current,
we consider an STM tip state $\psi_\sigma^\dagger(x')$ where $x'$ is
restricted to the tip.
 The tunneling Hamiltonian between the
STM tip and the superconducting wire can be written as
\begin{equation}
H_{tunnel}=\int dx dx' \sum_\sigma [V(xx')\psi_\sigma^\dagger(x')\psi_\sigma(x)+V^*(xx')\psi_\sigma^\dagger(x)\psi_\sigma(x')].
\end{equation}
The tunneling current can be related to the Keldysh Green function of
the combined system as~\cite{bolech}
\begin{equation}
I=\int dx dx'\sum_\sigma  [V^*(xx')G^{(K)}_{\sigma\sigma}(xx')-V(xx')G^{(K)}_{\sigma\sigma}(x'x)]\label{eq:current0}.
\end{equation}

The Keldysh Green function $G^{(K)}$ can be evaluated using the
 Dyson equation
\begin{align}
& G^{(K)}_{\sigma\sigma}(xx')=-G^{(R)}_{\sigma\sigma}(xx_1)V(x_1x'_1)g^{(K)}_{\sigma\sigma}(x'_1,x')-G^{(K)}_{\sigma\sigma}(xx_1)V(x_1x'_1)g^{(A)}_{\sigma\sigma}(x'_1,x')\\
& G^{(K)}_{\sigma\sigma}(x'x)=-g^{(R)}_{\sigma\sigma}(x'x'_1)V^*(x_1x'_1)G^{(K)}_{\sigma\sigma}(x_1,x)-g^{(K)}_{\sigma\sigma}(x'x'_1)V^*(x_1x'_1)G^{(A)}_{\sigma\sigma}(x_1,x)
\end{align}
where $g=(H_0-\omega)$ are the unperturbed Green functions and $H_0$ is the unperturbed Hamiltonian. Since the initial systems
are in equilibrium, $g^{(K)}=(g^{(R)}-g^{(A)})\tanh{\frac{\epsilon-\mu}{2 T}}$. Substituting these into Eq.~(\ref{eq:current0}), we get
\begin{equation}
I=\imath\int dx dx_1 d\omega\sum_\sigma \Gamma_{\sigma}(xx_1;\omega)[\tanh{\frac{\omega-\mu_t}{2 T}}\{G^{(R)}_{\sigma\sigma}(xx_1;\omega)-G^{(A)}_{\sigma\sigma}(x,x_1;\omega)\}+G^{(K)}_{\sigma\sigma}(xx_1;\omega)]
\end{equation}
where
\begin{equation}
\Gamma_{\sigma}(xx_1;\omega)=\int dx' dx'_1 V^*(xx')V(x_1x'_1)\rho_{\sigma}(xx_1;\omega)
\end{equation}
and the tip spectral function at energy $\omega$ is given
\begin{equation}
\rho_{\sigma}(xx_1;\omega)=\imath[g^{(A)}_{\sigma\sigma}(x'_1,x')-g^{(R)}_{\sigma\sigma}(x'_1x')].
\end{equation}

The exact Green functions in the superconductor can be calculated by integrating out the tip using Dyson's equation through the relation
\begin{equation}
G(xx_1;\omega)=g(xx_1;\omega)+g(xx_2;\omega)\Sigma(x_2x_3;\omega)G(x_3x_1;\omega)
\end{equation}
where the tip induced self-energy is given by
\begin{align}
&\Sigma(xx_1;\omega)=\int_{-\infty}^{\infty} d\omega'\frac{\Gamma(xx_1;\omega')}{\omega-\omega'}\\
&\Sigma^{K}(xx_1;\omega)=\tanh{\frac{\epsilon-\mu}{2 T}}\Gamma(xx_1;\omega).
\end{align}
The self-energy discussed above is for the normal Green function, which is
relevant for the tip. However to describe the superconductor, we need to
consider the Nambu spinor Green function at
complex frequency $\omega$. To be consistent with
particle-hole symmetry, the self-energy in Nambu spinor
notation is given by
\begin{equation}
\Sigma_{Nambu}(\omega)=\left(\begin{array}{cc}\Sigma(\omega) &0\\0&-\Sigma^*(-\omega^*)\end{array}\right).
\end{equation}
The retarded and advanced self-energy are given at frequencies $\omega\pm\imath\delta$ respectively.
The Dyson equations for the full Green function $G=(H_{BdG}-\omega)^{-1}$ in terms of the self-energy can be decomposed component-wise~\cite{haug} as
\begin{align}
&G(\omega)=(1-g(\omega)\Sigma(\omega))^{-1}g(\omega)\label{eq:dyson}\\
&G^{(K)}=(1+G^{(R)}\Sigma^{(R)})g^{(K)}(1+\Sigma^{(A)}G^{(A)})+G^{(R)}\Sigma^{(K)}G^{(A)}.
\end{align}
Since the starting system is in equilibrium, $g^{(K)}=\tanh{\frac{\epsilon-\mu}{2 T}}[g^{(R)}-g^{(A)}]$ and writing $(1+\Sigma G)=G H_0$ where $H_0$
is the unperturbed Hamiltonian, the latter equation can be reduced to
\begin{equation}
G^{(K)}=G^{(R)}\Sigma^{(K)}G^{(A)}.
\end{equation}
Finally using the equilibrium constraint on the tip, one obtains the
expression for the current~\cite{bolech}
\begin{equation}
I=\imath\int dx dx_1 d\omega\sum_\sigma \Gamma_{\sigma}(xx_1;\omega)\tanh{\frac{\omega-\mu_t}{2 T}}\{G^{(R)}_{\sigma\sigma}(xx_1;\omega)-G^{(A)}_{\sigma\sigma}(x,x_1;\omega)\}.
\end{equation}

To proceed further we must make assumptions about the Green function for the STM tip. For a simple STM tip tunneling to the end of the nanowire, the tunneling can be assumed to
 occur between one point on the tip and the end of the wire. Furthermore, we can assume that the local density of states of the tip is a Lorentzian function 
$((\omega-\mu_t)^2/W^2+1)^{-1}$ where $W$ is the band-width about the tip. With these assumptions $\Gamma$ is given by 
\begin{equation}
\Gamma_{\sigma}(xx_1;\omega)=\frac{\Gamma }{(\omega-\mu_t)^2/W^2+1}\delta(x)\delta(x_1)
\end{equation}
where $\mu_t$ is the chemical potential of the tip.
The corresponding self-energy is given by 
\begin{equation}
\Sigma(xx_1;\omega)=\frac{\Gamma W}{\omega-\imath \sgn{Im(\omega)} W}
\end{equation}
where $\sgn{Im(\omega)}$ is the sign of the imaginary part of $\omega$.
The expression for the current with the above spectral density is given by   
\begin{equation}
I=\imath\Gamma \int d\omega\sum_\sigma \tanh{\frac{\omega-\mu_t}{2 T}}\frac{1}{\omega^2/W^2+1}\{G^{(R)}_{\sigma\sigma}(00;\omega)-G^{(A)}_{\sigma\sigma}(00;\omega)\}.
\end{equation}
Using the identity $\tanh(\frac{\omega}{2 T})=4\omega T\sum_{n\geq 0}\frac{1}{\omega^2+(2 n+1)^2\pi^2 T^2}$ and the fact $G^{(R)}$ and $G^{(A)}$ are analytic in the upper and lower half
complex frequency planes respectively, the $\omega$ integral can be
replaced by a discrete sum over imaginary Matsubara frequencies as
\begin{align}
&I=T\Gamma \sum_{\sigma,n\geq 0}\frac{W^2}{W^2-(2 n+1)^2\pi^2 T^2}Re\{G_{\sigma\sigma}(00;\mu_t+\imath (2 n+1)\pi T)\}\nonumber\\
&-T\Gamma W\sum_\sigma \tan{\frac{W}{2 T}}\frac{1}{\omega^2/W^2+1}Re\{G_{\sigma\sigma}(00;\mu_t+\imath W)\}\label{eq:Imat_full}.
\end{align}
 The second term in the expression for $I$ regulates the singularity in the first term for $W\sim (2 n+1)\pi T$.
Using the Dyson equations (Eq.~\ref{eq:dyson}), the expression for the Green function $G$ can be written in terms of the unperturbed nanowire Green function $g(00;\omega)$  at complex frequency $\omega$ as 
\begin{equation}
G_{\sigma\sigma}(00;\omega)=g(00;\mu_t+\imath (2 n+1)\pi T)\left[1-\Sigma(00;\mu_t+\imath (2 n+1)\pi T)) g(00;\mu_t+\imath (2 n+1)\pi T)\right]^{-1}|_{\sigma\sigma}\label{eq:Gfull}.
\end{equation}

 The unperturbed nanowire Green  $g(x,x'=0;\omega)$ at $x'=0$ satisfies the BdG equation
\begin{equation}
[\{-\eta \partial_x^2-\mu-V_g(x)-\imath\xi\sigma_y\partial_x-\}\tau_z+V_z \sigma_z+\Delta \tau_x-\omega]g(x,x'=a;\omega)=\delta(x-a)
\end{equation}
for $x>-a$ with the boundary condition $g(x=-a,x'=0;\omega)=0$. Here the
end of the wire has been taken to be at $x=-a$.
Away from the boundary $x=-a$ and the contact point of the tip, $x=x'=0$, the Green function can be expanded
\begin{align}
&g(xx'=0;\omega)=\sum_n \Psi_n C_{n,-}^\dagger e^{z_n x}\textrm{ for }x<0\\
&g(xx'=0;\omega)=\sum_{Re(z_n)<0} \Psi_n C_{n,+}^\dagger e^{z_n x}\textrm{ for }x>0
\end{align}
where
\begin{equation}
[\{-\eta z_n^2-\mu+\imath \xi \sigma_y z_n\}\tau_z+V_z \sigma_z+\Delta \tau_x-\omega]\Psi_n=0\label{eq:QEP}
\end{equation}
and $C$ are a set of vectors that are determined from boundary conditions.
The quadratic eigenvalue problem in Eq.~(\ref{eq:QEP}) can be reduced to a linear
generalized eigenvalue problem by defining $\Phi=z\Psi$ as
\begin{equation}
z\Phi=-\frac{\xi}{\alpha}\sigma_y \Phi-\frac{1}{\alpha}[\{-\mu+V_z \sigma_z\tau_z\}+\imath\Delta \tau_y-E\tau_z]\Psi.
\end{equation}
The coefficient vectors $C$ are determined numerically by solving the boundary conditions
\begin{align}
&g(x=-a,x'=0;\omega)=0\label{eq:BC}\\
&-\tau_z[\partial_x g(x,x'=0;\omega)|_{x=0_+}-\partial_x g(x,x'=0;\omega)|_{x=0_-}]=\bm 1\\
&g(x=0_-,x'=0;\omega)=g(x=0_+,x'=0;\omega).
\end{align}

\begin{figure}[tbp]
\begin{center}
\includegraphics[width=0.8\textwidth,angle=270]{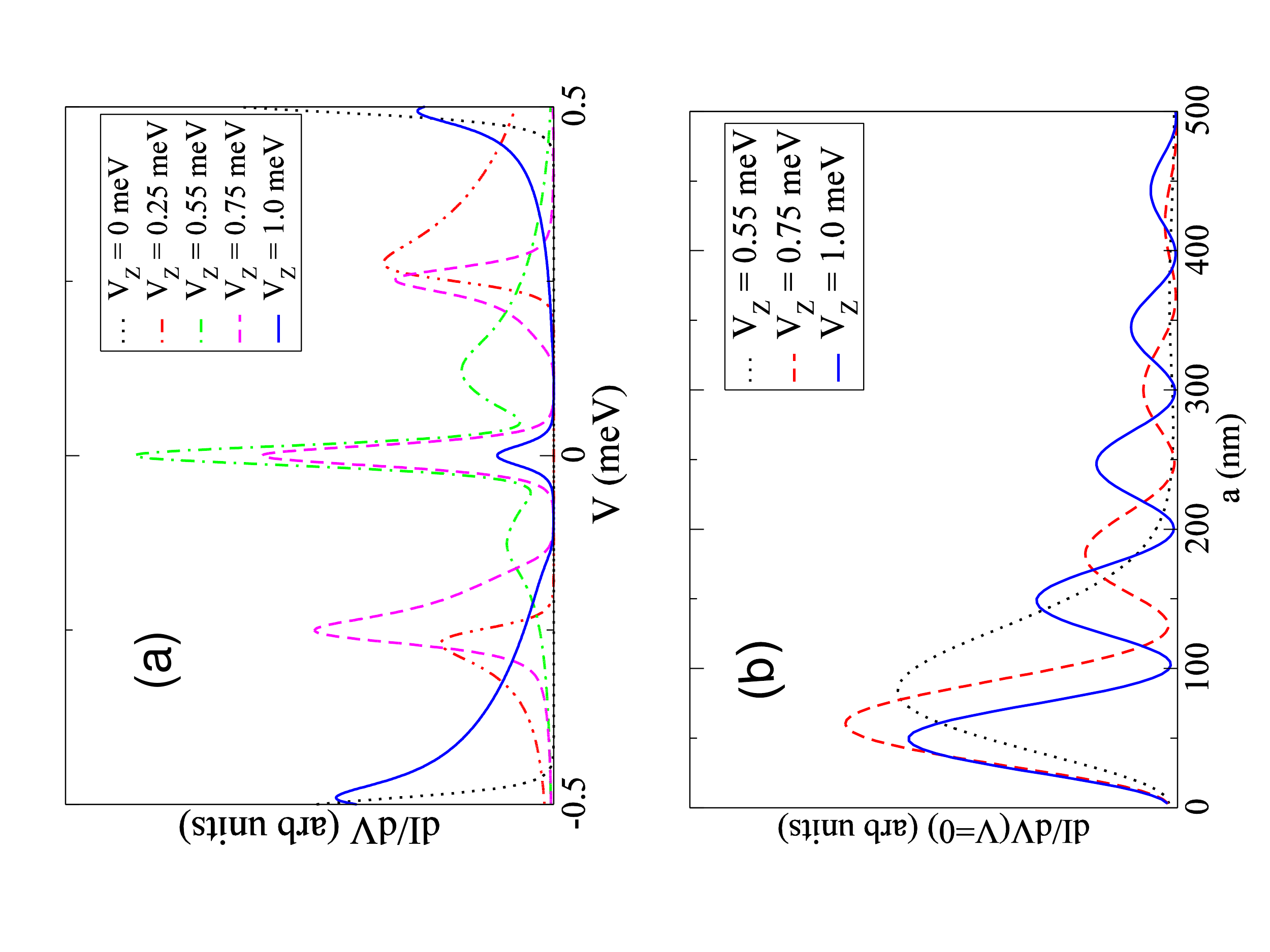}
\end{center}
\caption{(Color online) (a) Conductance $dI/dV$ as a function of voltage $\mu_t=V$. For the plot we have taken $\eta=\hbar^2/2 m^*$ where $m^*=0.04 m_e$ for InAs, $\Delta=0.5$ meV for Nb, $\mu=0$, $\alpha=0.1$ eV-\AA, and $T=100$ mK. The different values of $V_Z$ in meV are given in the inset. The distance of the STM tip from the end of the nanowire has been taken as $a=100$ nm. The plots for $V_Z > 0.5$ meV (corresponding to $B_x\sim 0.5$ T for InAs with $g_{InAs}\sim 35$) are in the topological phase and show a zero bias peak while the plots for $V_Z < 0.5$ meV do not. (b) Position dependence of the zero-bias conductance in the topological phase. The amplitude
of the zero-bias conductance is seen to be localized near the end of the wire and corresponds to the localized wave-function of the Majorana mode at the end of the
wire. }\label{fig:conductance}
\end{figure}

The current $I(\mu_t=V)$ is calculated by using the Green function  $g(x=0,x'=0;\omega)$  in Eqs.~(\ref{eq:Imat_full}) and ~(\ref{eq:Gfull}). The conductance
 $\frac{dI}{dV}$ obtained by numerical differentiation of the current
 is shown in Fig.~(\ref{fig:conductance}). As can be seen from panel (a)
 of the figure, the Majorana fermion mode at the end of the wire (in
 the non-Abelian phase, \textit{i.e.} $V_Z > \sqrt{\Delta^2+\mu^2}=0.5$ meV)
  gives rise to a peak in the conductance at zero-bias $(V=0)$.
 Apart from the zero-bias peak,  the bulk
 states in the nanowire also contribute to the STM conductance at
 bias voltages above the bulk energy gap. It is important to emphasize that,
 unlike  $p$-wave superconducting nanowires,  the tunneling spectrum of the present
  system (Fig. ~(\ref{fig:conductance})) has no
 state (other than the zero-energy Majorana state at the end) below the bulk energy gap $\Delta$. Therefore the effective
 mini-gap at the ends is order the bulk energy gap $(\Delta)$.
 As the Zeeman splitting is lowered towards the critical value
 $V_{Z,c}=\sqrt{\Delta^2+\mu^2}=0.5$ meV, the bulk energy gap closes before
 re-opening for $V_Z<\sqrt{\Delta^2+\mu^2}=0.5$ meV. For $V_Z$ in this regime, the semiconductor is in a regular $s$-wave superconducting
phase and there is no peak at
 zero bias. Instead there are 2 peaks
which are associated with van Hove singularities of the bulk
quasiparticle spectrum. These peaks
become sharper as the Zeeman potential is gradually lowered to zero.

STM spectroscopy can be used to not only verify the existence of the
zero energy Majorana states localized around the ends but also to
study the spatial structure of the zero energy wave-functions.
 Panel (b) of Fig.~(\ref{fig:conductance}) shows the
zero-bias tunneling conductance as a function of distance $a$
 from one end of the
wire. The calculation shows that the conductance vanishes at the end of
 the wire (the Majorana wave-function must vanish at the end of the wire
to satisfy the physical boundary condition). The zero-bias conductance
then rises to a peak and  oscillates with an envelope that decays away from the
 end of the wire.  The decay length  becomes longer as the
in-plane magnetic field is tuned towards the critical value $V_{Z,c}=0.5$ meV.
No such zero-bias conductance should be observable for $V_Z< V_{Z,c}$.
 Thus, the STM experiment from the semiconducting nanowire
 provides one of the most experimentally
 feasible probes of Majorana fermions in condensed matter systems.

The numerical results discussed in the previous paragraphs (Fig.~(\ref{fig:conductance}), can be understood 
analytically in the limit that the STM tunneling strenght is smaller than
 the thermal broadening $(\Gamma \ll T)$.
In this limit the current can be approximated as
\begin{equation}
I(\mu_t)\approx T\Gamma \sum_\sigma\sum_{n\geq 0}Re[g(00;\mu_t+\imath (2 n+1)\pi T)_{\sigma\sigma}]\label{eq:Imat0}
\end{equation}
which can be cast into the more familiar expression for STM current 
\begin{equation}
I(\mu_t)\approx \Gamma \sum_\sigma \int d\omega' A_{\sigma\sigma}(\omega')\tanh{\left(\frac{(\omega'-\mu_t)}{2 T}\right)}\label{eq:Imat}
\end{equation}
where the STM spectral function is given by $A_{\sigma\sigma'}(\omega)=Im(g_{\sigma\sigma'}(00;\omega))$\cite{sumanta_stm}.
The corresponding conductivity 
\begin{equation}
G(\mu_t)=\frac{d I(\mu_t)}{d \mu_t}=  \Gamma  \sum_\sigma \int d\omega' A_{\sigma\sigma}(\omega')\textrm{sech}^2\left(\frac{\mu_t-\omega'}{T}\right)\label{eq:conductance}.
\end{equation}
From the previous subsection, it is clear that the topological
phase is characterized by a single Majorana state localized at the end.
This zero energy Majorana mode is expected to lead to a zero-bias
peak in the tunneling spectrum. To see this we consider the
  $4\times 4$ Nambu Green function near $\omega=0$ where it is
dominated by the zero energy pole as
\begin{equation}
g(00;\omega)\approx\frac{\Psi \Psi^\dagger}{\omega}
\end{equation}
where $\Psi^T=(u_\uparrow,u_\downarrow,u^*_\uparrow,u^*_\downarrow)$. The corresponding spectral function is 
\begin{equation}
A_\sigma(\omega)\approx u_\sigma u^*_{\sigma'}\delta(\omega).
\end{equation}
Using this in Eq.~\ref{eq:conductance}, the contribution of the zero-energy state to the conductance 
becomes 
\begin{equation}
G(\mu_t)=\frac{d I(\mu_t)}{d \mu_t}=\tilde{\Gamma} \textrm{sech}^2\left(\frac{\mu_t}{T}\right)
\end{equation}
where $\tilde{\Gamma}=\Gamma\sum_\sigma |u_\sigma|^2$.

Therfore STM spectroscopy provides a multi-facted tool to study the 
properties of the topological quantum phase transition in the wire as
 the magnetic field is tuned from $B_x\sim 0$ to $B_x\sim 0.5$ T (which is significantly 
below the parallel critical field of thin-film Nb \cite{quateman}).
Firstly the STM spectra away from the ends provides information about the 
induced superconducting gap in the wire, which diminishes as a function 
of applied magnetic field, goes to 0 at the transition and then 
increases. For magnetic fields above the critical value, a zero-bias 
peak appears in the STM spectrum near the ends of the wire. Finally, 
STM can also be used to probe the spatial structure of the 
 wave-function of this Majorana mode.
 Thus, the STM experiment from the semiconducting nanowire
 provides one of the most experimentally 
 feasible probes of Majorana fermions in condensed matter systems.

\section{Summary and Conclusion}
Let us first recapitulate the most important results contained in each section in this paper.
In Secs.~[II, III] we analyze in detail the Hamiltonian and the BdG equations for a spin-orbit coupled semiconductor in the proposed heterostructure geometry in the presence of a vortex. Here we provide all the mathematical details, left out in Ref.~[\onlinecite{Sau}], which are necessary to conclusively establish the presence of a non-degenerate Majorana mode at the vortex core. In Sec.~[IV], we use the same formalism to establish the existence of a Majorana mode localized in a ``vortex-like'' defect in the spin-orbit coupling that can be artificially created in a spin-orbit coupled atomic system potentially realizable in an optical lattice. Here we find that, contrary to a previous treatment of the same problem, ~\cite{Sato} the decay length of the zero-energy Majorana mode localized in the spin-orbit vortex is inversely proportional to the superconducting gap. Therefore, in the limit
of vanishing superconducting gap, the Majorana mode will delocalize
over the entire system. We then show in Sec.~[V] how the same formalism can be used to demonstrate the existence of a Majorana fermion mode in a vortex on the surface of a $3D$ topological insulator, even though in this simpler case an exact solution of the BdG equations is already available. ~\cite{fu_prl'08}

In Sec.~[VI], we confirm our approximate analytical calculations, which are indicative of the existence of the zero-energy modes in the vortices in the spin-orbit coupled semiconductor, by a full numerical solution of the BdG equations set up on a sphere.~~\cite{kraus'08} Here we obtain the full bound-state excitation spectrum for the BdG Hamiltonian relevant for a vortex-antivortex pair placed at the poles of a sphere. ~\cite{kraus'08} In the non-Abelian phase of the semiconductor film, this calculation produces a pair of lowest energy states whose energy eigenvalues approach zero exponentially with the radius of the sphere. This indicates the presence of one exact zero-energy state on each vortex in the limit of infinite inter-vortex separation.  The numerical calculations also show
that the excitation gap above the zero energy state, the mini-gap, can be made comparable to the induced $s$-wave gap $\Delta$ in the semiconductor film. This surprising result, which was obtained previously for the proximity-induced superconducting state on the surface of a TI,~\cite{robustness} is now extended to the spin-orbit coupled semiconductor in this paper. The enhancement of the mini-gap increases the regime of temperature in which any non-Abelian quasiparticle is accessible in experiments by many folds (from $T\sim \frac{\Delta^2}{E_F}$ to $T\sim \Delta$).
In Sec.~[VII] we briefly discuss the parameter regime in which the non-Abelian topological state is the ground state in the semiconductor film. The associated topological quantum phase transition (TQPT), which can be accessed by varying any one of the three parameters -- Zeeman coupling ($V_z$), chemical potential ($\mu$), and the proximity-induced $s$-wave gap ($\Delta$) -- is a transition at which the excitation gap vanishes at wavevector $k=0$. In Sec.~[VIII], we analyze the interplay of the Zeeman coupling and the proximity-induced $s$-wave superconductivity in the presence of spin-orbit coupling. We show that, even though the Zeeman coupling can eliminate $s$-wave superconductivity in the absence of spin-orbit coupling, the latter can give rise to a \emph{re-entrant} non-Abelian superconducting state despite the fact that $|V_z| > |\Delta|$.
Apart from the zero-energy Majorana states in order parameter defects such as vortices, the non-Abelian topological state in the semiconductor film is also characterized by gapless Majorana modes at the sample edges. In Sec.~[IX], we use the same techniques employed in the earlier sections to demonstrate the existence of these edge modes, which turn out to be \emph{chiral} Majorana modes because of the explicit breakdown of the time reversal symmetry.

In Sec.~[X] we study the proximity effects in superconductor - semiconductor - magnetic insulator heterostructures starting from a microscopic tight-binding model. The superconductor and the magnetic insulator are described at the mean-field level. We determine the excitation spectrum of a slab containing a semiconductor thin film sandwiched between an $s$-wave superconductor and a ferromagnetic insulator and identify the dependence of the induced gaps on the parameters of the model. In particular, we study the dependence of the effective Zeeman splitting and induced superconducting gap on the thickness of the semiconductor film, the applied bias potential and the strength of the coupling at the interfaces.

Finally, in Sec.~[XI] we demonstrate the existence of zero-energy Majorana modes at the ends of a one dimensional version of the spin-orbit coupled semiconductor system -- a semiconducting nanowire. It has been shown that it
 may be far simpler to experimentally realize Majorana fermions in the
 one dimensional nanowire system because the Zeeman splitting can be induced by a modest in-plane magnetic field, obviating the need for a proximate magnetic insulator. \cite{roman, Gil} We find that the Majorana modes at the two ends of the nanowire can be probed by scanning tunneling microscope experiments. We show by explicit calculations that the Majorana modes at the ends of the nanowire give rise to zero-bias conductance peaks in the tunneling spectrum at the ends. These peaks disappear on lowering the Zeeman coupling so that the system settles into the non-topological superconducting state. Furthermore these zero-bias conductance peaks are found to be the only features at bias voltages below the induced superconducting gap in the 
nanowire. We believe that the observation of this zero-bias tunneling peak in the semiconductor nanowire is the simplest experiment proposed so far to unambiguously detect a Majorana fermion state in a condensed matter system.

We note here that the Majorana fermions, being non-Abelian particles 
belonging to the $(SU_2)_2$ conformal field theory (i.e. the so-called
 "Ising anyon" universality class), cannot directly be used for universal
 fault-tolerant topological quantum computation (TQC).
 \cite{nayak_RevModPhys'08}  They
 can serve as topologically protected quantum memory or can be used in
 quantum computation along with supplementary unprotected quantum gates
 requiring only small amounts of error corrections. \cite{bravyi}
 Since the topological protection for the semiconductor heterostructures
 in our work is very robust, with the energy gap providing the protection
 being of the order of the superconducting gap (~1-10K) itself, our
 proposed system could serve as an excellent quantum memory in TQC
 applications.  In a recent development, Bonderson et al. have shown
 \cite{bonderson} that certain dynamic-topology-changing operations,
 which are feasible in our proposed semiconductor heterostructures,
 dubbed Ising Sandwich Heterostructures (ISH)in ref. \cite{bonderson},
 could
 allow fully fault-tolerant TQC to be carried out using our proosed
 systems.  Thus, in addition to the fundamental significance of the
 possible non-Abelian quantum order and the existence of topological
 Majorana excitations, these semiconductor-superconductor structures may
 have future application as the basic components of a topological quantum
 computer.

Our proposed non-Abelian system is possibly one of the simplest to study experimentally since it involves neither special materials nor exceptional purity nor the application of a high magnetic field. It is encouraging to note that proximity-induced $s$-wave superconductivity has already been realized in a host InAs semiconductor film,\cite{merkt,giazotto} which additionally has a sizable spin-orbit coupling. Experimentally, the only new effect that must be introduced to the system is a strong enough Zeeman splitting of the spins. We argue how this can be achieved also via the proximity effect due to a nearby magnetic insulator. It is important to emphasize that when the spin-orbit coupling is of the Rashba type, we require a Zeeman splitting which is
 perpendicular to the plane of the film. This is because a Zeeman splitting parallel to the film does not produce a gap in the one-electron band-structure. \cite{Sau, index-sm} Inducing a perpendicular splitting by applying a strong perpendicular magnetic field is not convenient, because the magnetic field will give rise to unwanted order parameter defects such as vortices. It is for this reason that we propose to induce the Zeeman splitting by the exchange proximity effect of an adjacent magnetic insulator (we ignore the small coupling of the spins in the semiconductor with the actual magnetic field of the magnetic insulator).  More recently, it has been shown that, when the spin-orbit coupling also has a component which is of the Dresselhaus type, the appropriate Zeeman splitting can also be induced by applying an in-plane magnetic field.~\cite{alicea} Note that an in-plane magnetic field does not produce unwanted vortex excitations. The one dimensional version of our system -- a semiconducting nanowire -- is also a non-Abelian system
in the presence of proximity induced \emph{s}-wave superconductivity and
a Zeeman coupling.  It is quite exciting that the superconducting
 proximity effect on an InAs nanowire has already been realized in experiments.\cite{doh} In this case, the Zeeman coupling can be more easily
 introduced by applying an external magnetic field parallel to the
 length of the wire because such a field does produce a gap in the one-electron band structure without producing unwanted excitations in the adjacent superconductor. This obviates the need for a nearby magnetic insulator. \cite{roman} In the topological superconducting state of the nanowire (\emph{i.e.,} the Zeeman coupling is above a critical value), there are zero-energy Majorana states localized around the two ends. Such zero energy states can be revealed by zero-bias conductance peaks in STM tunneling experiments at the ends of the wire. All other contributions to the
 conductance occur at energies higher than the bulk gap $\Delta$. There will be no such zero-bias peak when the wire is in the topologically trivial \emph{s}-wave superconducting state (\emph{i.e.,} the Zeeman coupling is below the critical value). Such an STM experiment from the semiconducting nanowire will serve as an unambiguous probe for the condensed matter manifestation of a Majorana fermion mode.

\section{Acknowledgement} This work is supported by DARPA-QuEST, JQI-NSF-PFC, and LPS-NSA.
ST acknowledges DOE/EPSCoR Grant \# DE-FG02-04ER-46139 and Clemson University start up funds for support.

\appendix
\section{Power series for the Rashba model at $r>R$}
Even though we found an analytic solution for the null vectors of the matrix for $\Delta(r)=0$ in the region $r<R$, we could not find
such a solution for $\Delta(r)=\Delta>0$ for $r>R$. In a previous section we claimed without explicit proof that the solution to this equation
can be written in terms of a power-series expansion for $\rho(1/r)$. Since we are interested
in the solution at large $R$ we expect a power-series in $1/R$ to converge. To generate the equation for the power-series for $\rho$ it is
convenient to shift to a basis where the $1/r=0$ part of the matrix is diagonal.
\begin{align}
&\left(\begin{array}{cc}\eta  (-\partial_r^2-\frac{1}{4 r^2}+2 z\partial_r-z^2) +V_z-\mu & \pm\Delta+\alpha  (\partial_r+\frac{1}{2 r} -z)\\ \mp \Delta-\alpha  (\partial_r-\frac{1}{2 r}-z)  & \eta  (-\partial_r^2+\frac{3}{4 r^2}+2 z\partial_r-z^2)-V_z-\mu \end{array}\right)=A+B
\end{align}
where 
\begin{align}
&A=\left(\begin{array}{cc}-\eta z^2 +V_z-\mu & \pm\Delta-z\alpha  \\ \mp \Delta+z\alpha    & -\eta z^2-V_z-\mu \end{array}\right)\\
&B=\eta  (-\partial_r^2-\frac{1}{4 r^2}+2 z\partial_r)+\imath\alpha \sigma_y \partial_r +\alpha \sigma_x/2 r-\frac{\eta}{2 r^2}\sigma_z.
\end{align}
If $S$ is the matrix of eigenvectors of $A$ then $A=S D S^{-1}$ where $D$ is a diagonal matrix such that $D_{22}=0$ and $D_{11}=Trace(A)=\eta$. Thus
the relevant system of differential equations becomes
\begin{align}
& [S D S^{-1}+\eta  (-\partial_r^2+\frac{1}{4 r^2}-\frac{1}{2 r^2}\sigma_z+2 z\partial_r)+\imath\alpha \sigma_y \partial_r +\alpha \sigma_x/2 r-\frac{1}{2 r^2}\sigma_z]\left(\begin{array}{c}\rho_{\uparrow}(1/r)\\\rho_{\downarrow}(1/r)\end{array}\right)=0\\
&\left(\begin{array}{c}\rho_{\uparrow,1}(1/r)\\\rho_{\downarrow,1}(1/r)\end{array}\right)=S^{-1}\left(\begin{array}{c}\rho_{\uparrow}(1/r)\\\rho_{\downarrow}(1/r)\end{array}\right)\\
&[\left(\begin{array}{cc} \eta & 0\\ 0  & 0\end{array}\right) +\eta  (-\partial_r^2+\frac{1}{4 r^2}-\frac{1}{2 r^2}S^{-1}\sigma_z S+2 z\partial_r)+\alpha S^{-1}(\imath \sigma_y)S \partial_r +\frac{\alpha}{2 r}S^{-1}\sigma_x S]\left(\begin{array}{c}\rho_{\uparrow,1}(1/r)\\\rho_{\downarrow,1}(1/r)\end{array}\right)=0
\end{align}
where in the above we note that it is sufficient to provide a series expansion for $\rho_1(1/r)$ since $S$ is independent of r.
The matrix $S$ may be written  explicitly as 
\begin{align}
& S=\left(\begin{array}{cc}\sqrt{b^2-c^2}+b & \sqrt{b^2-c^2}-b\\ -c  & c\end{array}\right)
\end{align}
where $a=\eta z^2-\mu$, $b= V_z$ and $c= \pm \Delta -\alpha z$.

It is easy to see that our equation is satisfied to order $1/r^2$ by the choice
\begin{align}
&\left(\begin{array}{c}\rho_{\uparrow,1}(1/r)\\\rho_{\downarrow,1}(1/r)\end{array}\right)=\left(\begin{array}{c}\alpha\frac{\sqrt{b^2-c^2}-b}{c\eta}(1/r)\\1\end{array}\right)
\end{align}
The upper and lower components of the above solutions are different orders 
in $1/r$.
Therefore we redefine our spinor as 
\begin{equation}
\left(\begin{array}{c}\rho_{\uparrow,2}(1/r)\\\rho_{\downarrow,2}(1/r)\end{array}\right)=\left(\begin{array}{cc}1&0\\0&\frac{1}{r}\end{array}\right)\left(\begin{array}{c}\rho_{\uparrow,1}(1/r)\\\rho_{\downarrow,1}(1/r)\end{array}\right)
\end{equation}
the above solution at lowest order motivates us to modify our ansatz so that the upper and lower component are of the same order in $1/r$
as below
\begin{equation}
M\left(\begin{array}{c}\rho_{\uparrow,2}(1/r)\\\rho_{\downarrow,2}(1/r)\end{array}\right)=0
\end{equation}
where the matrix differential operator $M$ is given by 
\begin{align}
&M=\left(\begin{array}{cc}1&0\\0&\frac{1}{r}\end{array}\right)[\left(\begin{array}{cc} \eta & 0\\ 0  & 0\end{array}\right) +\eta  (-\partial_r^2+\frac{1}{4 r^2}-\frac{1}{2 r^2}S^{-1}\sigma_z S+2 z\partial_r)+\alpha S^{-1}(\imath \sigma_y)S \partial_r +\frac{\alpha}{2 r}S^{-1}\sigma_x S]\nonumber\\
&\left(\begin{array}{cc}1&0\\0&r\end{array}\right).
\end{align}

The terms in the matrix part of the above equation can be separated into 2 categories. Those that preserve the order of a term $1/r^n$ and those that increase the
order of a term to $1/r^{n+1}$. The terms that preserve the order are contained within the matrix below
\begin{align}
&\left(\begin{array}{cc}1&0\\0&\frac{1}{r}\end{array}\right)[\left(\begin{array}{cc} \eta & 0\\ 0  & 0\end{array}\right) +\eta  (2 z\partial_r)+\alpha S^{-1}(\imath \sigma_y)S \partial_r +\frac{\alpha}{2 r}S^{-1}\sigma_x S]\left(\begin{array}{cc}1&0\\0&r\end{array}\right)+\textrm{higher order}\\
&=\left(\begin{array}{cc} \eta & 0\\ 0  & 0\end{array}\right)+(2\eta z +\alpha S^{-1}(\imath \sigma_y)S)\left(\begin{array}{cc} 0 & 0\\ 0  & 1\end{array}\right)+\alpha\frac{\sqrt{b^2-c^2}-b}{ c}\left(\begin{array}{cc} 0 & \frac{1}{2}+\frac{b r}{\sqrt{b^2-c^2}}\partial_r \\ 0  & 0\end{array}\right)+\textrm{higher order}.
\end{align}
The matrix written explicitly here preserves the order of $1/r^n$ while the rest of the terms generate terms of order $1/r^{n+1}$ or higher.
We can check this by applying  the above matrix  to a spinor proportional to $1/r^n$. The resulting spinor is 
\begin{align}
&[\left(\begin{array}{cc} \eta & 0\\ 0  & 0\end{array}\right)+(2\eta z +\alpha S^{-1}(\imath \sigma_y)S)\left(\begin{array}{cc} 0 & 0\\ 0  & 1\end{array}\right)+\alpha\frac{\sqrt{b^2-c^2}-b}{ c}\left(\begin{array}{cc} 0 & \frac{1}{2}+\frac{b r}{\sqrt{b^2-c^2}}\partial_r \\ 0  & 0\end{array}\right)]\frac{1}{r^n}\left(\begin{array}{c}\rho_{\uparrow}\\\rho_{\downarrow}\end{array}\right)\\
&=\frac{Q_n}{r^n}\left(\begin{array}{c}\rho_{\uparrow}\\\rho_{\downarrow}\end{array}\right)
\end{align}
where 
\begin{equation}
Q_n=[\left(\begin{array}{cc} \eta & 0\\ 0  & 0\end{array}\right)+(2\eta z +\alpha S^{-1}(\imath \sigma_y)S)\left(\begin{array}{cc} 0 & 0\\ 0  & 1\end{array}\right)+\alpha\frac{\sqrt{b^2-c^2}-b}{ c}\left(\begin{array}{cc} 0 & \frac{1}{2}-\frac{b (n+1)}{\sqrt{b^2-c^2}} \\ 0  & 0\end{array}\right)].
\end{equation}
From this it is clear that the action of the matrix differential operator $M$ 
on the spinor $(\rho_{\uparrow},\rho_{\downarrow})r^{-n}$ is the non-singular matrix that acts on the spinor $(\rho_{\uparrow},\rho_{\downarrow})$ in 
the above equation. 

This fact can be used to create a procedure for generating the power 
series for $\rho_2(1/r)$ iteratively. To see how this is the case 
consider the stage after the $(n-1)^{th}$ iteration, where the power series has been 
approximated to 
\begin{equation}
\rho_2^{(n-1)}(1/r)=\sum_{j=0}^{n-1} \frac{\rho_2^{(j)}}{r^j}.
\end{equation}
To calculate $\rho_2^{(n)}$, we note that 
\begin{equation}
0=M\rho_2^{(n)}(1/r)=M \frac{\rho_2^{(n)}}{r^n}+M\rho_2^{(n-1)}(1/r)= \frac{Q_n\rho_2^{(n)}}{r^n}+M\rho_2^{(n-1)}(1/r)+o(r^{-(n+1)}).
\end{equation}
Solving the equation to $o(r^{-n})$ yields the iterative relation 
\begin{equation}
\rho_2^{(n)}=-Q_n^{-1}\lim_{r\rightarrow \infty}r^n M\rho_2^{(n-1)}(1/r).
\end{equation}
As discussed previously, we already have a starting guess for the solution
for $n=1$ so that the residue is of order $1/r^2$. Using the non-singular matrix $Q_n$ we can continue to solve the for the higher-order
coefficient by inverting the matrix over the residue.
The above argument provides a procedure for how to construct the power-series for $\rho(1/r)$ referred to in Eq. 22 of the main body of the text.

\end{document}